\documentclass[preprint,3p,times]{elsarticle}

\usepackage{graphicx}
\usepackage{epstopdf, epsfig}
\newcommand{\bx}{{\mathbf x}}
\newcommand{\by}{{\mathbf y}}
\newcommand{\Ca}{\text{Ca}}
\usepackage{caption}
\usepackage{amsmath}
\usepackage{graphicx}  
\usepackage{amssymb}
\usepackage{multirow}
\usepackage{amsmath}
\usepackage{floatrow}
\usepackage[label font=bf,labelformat=simple]{subfig}% <-- changed
\usepackage{caption}
\usepackage{xcolor}

\journal{Journal of Computational Physics}

\begin{document}

\begin{frontmatter}



\title{Non-local model for surface tension in fluid-fluid simulations}

% Use letters for affiliations, numbers to show equal authorship (if applicable) and to indicate the corresponding author

\author[label1]{Amanda A. Howard}
\author[label1]{Alexandre M. Tartakovsky\corref{cor1}}
\address[label1]{Pacific Northwest National Laboratory, Richland, WA 99354, USA}
\ead{alexandre.tartakovsky@pnnl.gov}
\cortext[cor1]{Corresponding author.}
\address{}

\begin{abstract}
%Please provide an abstract of no more than 250 words in a single paragraph. Abstracts should explain to the general reader the major contributions of the article. References in the abstract must be cited in full within the abstract itself and cited in the text.
We propose a non-local model for surface tension obtained in the form of an integral of a molecular-force-like function with support $3.5\varepsilon$ added to the Navier-Stokes momentum conservation equation. We demonstrate analytically and numerically that with the non-local model interfaces with a radius of curvature larger than the support length behave macroscopically and microscopically, otherwise. For static droplets, the pressure difference $P_{\varepsilon, in} -  P_{\varepsilon, out}$ satisfies the Young-Laplace law for droplet radius greater than $3.5\varepsilon$ and otherwise deviates from the Young-Laplace law. The latter indicates that the surface tension in the proposed model decreases with decreasing radius of curvature, which agrees with molecular dynamics and experimental studies of nanodroplets. Using the non-local model we perform numerical simulations of droplets under dynamic conditions, including a rising droplet, a droplet in shear flow, and two colliding droplets in shear flow, and compare results with a standard Navier-Stokes model subject to the Young-Laplace boundary condition at the fluid-fluid interface implemented via the Conservative Level Set (CLS) method. We find good agreement with existing numerical methods and analytical results for a rising macroscopic droplet and a droplet in a shear flow. For colliding droplets in shear flow, the non-local model converges (with respect to the grid size) to the correct behavior, including sliding, coalescence, and merging and breaking of two droplets depending on the capillary number. In contrast, we find that the results of the CLS model are highly grid-size dependent. 
\end{abstract}


\end{frontmatter}

\section{Introduction}

Since the work of Young, Laplace, and Gauss, multiphase flow at the continuum scale has been modeled almost exclusively by the Young-Laplace (YL) law, imposed as a boundary condition for the Navier-Stokes (NS) equations at the fluid-fluid interface. In this work, we propose an alternative continuum description of the multiphase flow in the form of an integral of a molecular-force-like function with support $\varepsilon$ added to the Navier-Stokes momentum conservation equation. 

At the nanoscale, multiphase flow is traditionally described by the equations of molecular dynamics (MD). MD simulations show a sharp density and pressure drop across the fluid-fluid interface in a region of around ten nanometers \cite{Masuda2011, Nakamura2011}. Outside of this region the pressure satisfies the YL law, i.e., the pressure difference across the interface is linearly proportional to the interface curvature. 

Another molecular-scale feature of the fluid-fluid interface is the surface tension dependence on the curvature radius for curvature radii smaller than 10 nanometers \cite{Kashchiev2003}. As the curvature radius increases, surface tension asymptotically approaches its ``macroscopic'' value $\sigma_0$. Therefore, one can conclude that the NS-YL model is suitable for describing interfacial dynamics on scales larger than 10 nanometers given that the thermal fluctuations are properly accounted for \cite{Landau1987}. Comparisons between continuum and MD simulations have shown disagreement below a droplet diameter of 36 nanometers, with drastic differences at 10 nanometers \cite{Bardia2016}.  

{ One example of a physical system that cannot be described by the YL law are nanobubbles.  Bulk nanobubbles, also known as ultrafine bubbles, show extraordinarily long-term stability compared to that which would be expected from the Young-Laplace Law with macroscopic surface tension \cite{Jadhav2020}, up to three months in a laboratory setting \cite{Michailidi2020}. This long-term stability leads to a number of applications, for example long term stability of bulk nanobubbles is an advantage when used as an ultrasound contrast agents over ultrasound contrast agents with larger bubbles, which have a limited half-life \cite{Perera2017}. Additionally, nanobubbles have a higher surface area than macro-scale bubbles for the same volume fraction of bubbles, and change the physiochemical properties of the fluid-bubble system \cite{Ohgaki2010, Liu2013}. Bulk nanobubbles show strong promise for biomedical applications in ultrasound \cite{Hernandez2018, Abenojar2019}, radiofrequency ablation of tumors \cite{Perera2014}, and for mediating drug delivery \cite{Hernandez2017}, as well as water and waste treatment \cite{Agarwal2011, Uchida2011, Temesgen2017, Wang2019, Hu2018, Atkinson2019}, prevention of membrane and surface fouling and cleaning \cite{Chen2009, Zhu2016, Ghadimkhani2016}, froth flotation \cite{Fan2010}, and plant and animal growth \cite{Zhou2019, Ebina2013}. Numerical simulations of a large number of nanobubbles over large timescales are essential for these and other important applications.  
}

%Here, we propose a continuum nonlocal surface tension model that when combined with the NS equation is capable of capturing molecular-scale features, such as the pressure drop across the interface and the surface tension dependence on the curvature radius. 

On the molecular scale, surface tension results from the broken symmetry in the molecular interactions near the interface, i.e., the molecular forces acting between like molecules differ from forces acting between unlike molecules.  In a similar manner, the molecular-like-forces generate surface tension in the non-local model.

%Droplet collisions have incredibly important and ubiquitous applications across a wide range of systems, including drug delivery, waste treatment, and industrial products such as ink jet printers.
Many interfacial phenomena are multiscale in nature. For example, 
modeling colliding macroscale droplets requires resolving a large range of relevant length scales and topology changes that occur at the droplet interfaces and a thin fluid film forming between two colliding droplets, and is another example where the YL law fails under certain capillary numbers \cite{Jiang2007}. The interfaces of droplets become locally flat as they approach each other, so the surface tension force due to the YL law becomes zero.  
%One specific challenge is that in the course of the bubble collision the film between them must necessarily thin to a thickness less than the grid spacing \cite{Zhang2011}. 
The film of surrounding fluid forms between droplets and reaches $100-1000\AA$  before rupturing. This film drains as the droplets come closer to each other, eventually rupturing due to the inter-molecular van der Waals forces, which are absent in the macroscale models \cite{Jiang2007}. 

In traditional local front capturing Level Set and volume-of-fluid methods, droplet coalescence will occur if the droplets are separated by less than about one grid point, a phenomenon known as numerical (or artificial) coalescence \cite{Coyajee2009}. Simulations of binary head-on collisions by \cite{Pan2005} using the Level Set method were unable to capture the small deformation regime of bubble merging because they could not resolve the film drainage and rupturing. The Coupled Level-Set/Volume-of-Fluid (CLSVOF) method was proposed to overcome numerical coalescence, at the expense of prohibiting even physically-feasible coalescence \cite{Sussman2000}. In the simulations by \cite{Pan2008}, coalescence occurred at timing determined by experiments or by a van der Waals force with augmented range such that the length scale was large enough to be resolved by the simulations, implemented by adding a surface force at the interface. This study, and others, showed that the behavior of the droplets after coalescence is sensitive to the timing of the front rupture, and experimental data may not be available in all cases to determine the coalescence time  \cite{Pan2008, Mason2012}. To resolve this challenge,  sub-grid-scale (SGS) models have been proposed, where a semi-analytical model for the thin film dynamics is coupled with the local model at the droplet lengthscale to determine if and when the front will rupture \cite{Mason2012, Liu2018}. This results in a predictive method, removing the necessity of prescribing the time for the fronts to merge. SGS models were considered in the CLSVOF method by \cite{Coyajee2009} and \cite{Kwakkel2013} and in the front tracking method by \cite{Tryggvason2010}. These models require coupling knowledge of the dynamics of the nano-scale gap with the macro-scale bubble dynamics as well as computationally expensive periodic searching through the domain to find interfaces close to collision \cite{Chan2018}. Another approach by \cite{Jiang2007} included van der Waals forces in the momentum conservation equation in addition to the YL surface tension force.  
%which are physically necessary for coalescence but act on a length scale much smaller than those resolved in simulations of macroscopic bubbles. 
This approach was limited to the head-on collisions with equal sized droplets. 

%Therefore, numerical models based on the Young-Laplace boundary condition, fluid-fluid interfaces are merged ``artificially.'' In interface tracking methods, interfaces are ``forced'' to merge when the distance between droplets becomes smaller than a specified value. {Needs references.} It was shown that the resulting behavior strongly depends on threshold value, which in turn can depend on capillary number. In diffused interface models such as volume of fluid, interfaces merge when the colored functions overlap.  \textcolor{red}{Please provide references and describe shortcomings of such treatment. Also explain how interfaces merge in the Level Set method and provide references.}    

To demonstrate the multiscale nature of the non-local model, we present a semi-analytical steady-state solution for the fluid pressure across a fluid-fluid interface. This solution shows nanoscale behavior for the radius of curvature smaller than $3.5\varepsilon$ and macroscopic behavior (i.e., the solution follows the YL law) for the radius of curvature larger than $3.5\varepsilon$. 

Using the non-local model, we perform numerical simulations of droplets under dynamic conditions, including a rising droplet, a droplet in a shear flow, and two colliding droplets in a shear flow, and compare results with standard NS model subject to the YL boundary condition at the fluid-fluid interface implemented via the Conservative Level Set (CLS) method. We find good agreement with the CLS method for a rising macroscopic droplet and a droplet in a shear flow. For a small capillary number (Ca = 0.24), we find that the non-local model predicts a microscopic droplet in shear flow to form ``ears''. The CLS method predicts the same (macroscopic) shape of the droplet regardless of the droplet size. 

Finally, for colliding droplets in shear flow we find that the non-local model converges (with respect to the grid size) to the correct behavior, including (depending on capillary number)  sliding, coalescing, and temporary bridging (coalescing and then separating) of two droplets. On the other hand, in our simulations the CLS method results are highly grid-size dependent.

\section{Non-local surface tension model}

We consider flow of two incompressible Newtonian fluids, denoted $\alpha$ and $\beta$ in a fixed domain $\Omega = \Omega_\alpha(t) \cup \Omega_\beta(t)$. The macroscopic (hydrodynamic) model for two-phase flow includes the continuity equation for phase $i = \alpha$ or $\beta$
\begin{equation}
\nabla \cdot \mathbf{u}_i = 0, \quad\mathbf{x}\in\Omega_i(t), 
\end{equation}
and the momentum conservation equation 
\begin{equation}
\rho_i \left( \frac{\partial \mathbf{u}_i }{\partial t} +\mathbf{u}_i \cdot \nabla \mathbf{u}_i  \right) = -\nabla P_i + \nabla \cdot \boldsymbol{\tau}_i + \rho_i \mathbf{g}, \quad\mathbf{x}\in\Omega_i(t),
\label{Eq-Mom}
\end{equation}
 subject to the YL boundary condition at the fluid-fluid interface
\begin{equation}\label{Young-Laplace}
(P_\alpha-P_\beta)\mathbf{n}= (\boldsymbol{\tau}_\alpha-\boldsymbol{\tau}_\beta)\cdot\mathbf{n} +\kappa \sigma_0 \mathbf{n},
\quad\mathbf{x}\in\Omega_\alpha \cap \Omega_\beta,
\end{equation}
where the subscript $i$ denotes phase $i$, $\rho_i$ is the density, $\mathbf{u}_i$ is the velocity,  
$P_i$ is the pressure, $\mathbf{g}$ is the gravitational acceleration, $\tau_i = [\mu_i (\nabla \mathbf{u}_i + \nabla \mathbf{u}_i^T)]$ is the viscous stress tensor with the dynamic viscosity $\mu_i$,  $\sigma_0$ is the macroscopic surface tension, and $\mathbf{n}$ is the normal vector.

To simplify numerical treatment of these equations, it is common to replace eqs. \ref{Eq-Mom} and \ref{Young-Laplace} with \cite{Brackbill1992}
\begin{equation}
\rho \left( \frac{\partial \mathbf{u} }{\partial t} +\mathbf{u} \cdot \nabla \mathbf{u} \right) = -\nabla P + \nabla \cdot \boldsymbol{\tau} + \rho \mathbf{g} + \mathbf{F}, \quad\mathbf{x}\in\Omega,
\label{Eq-Mom-Brackbill}
\end{equation}
where the surface force $\mathbf{F}$ is given by the YL law as 
\begin{equation}
\mathbf{F} = \sigma_0 \kappa \hat{\bf{n}}
\label{eq:YLlaw}
\end{equation}
where $\kappa$ is the interface curvature. 
We introduce the color function $\phi$:
\begin{equation}
\phi(\mathbf{x}) = \left\{ \begin{array}{ll}
      0, & \mathbf{x} \in \Omega_\alpha, \\
       1, & \mathbf{x} \in \Omega_\beta.
\end{array}
\right. 
\label{equ:color_func}
\end{equation}
The color function is advected with the velocity $\mathbf{u}$ as  $\frac {\partial \phi}{\partial t} + \mathbf{u} \cdot \nabla \phi = 0.$
Then, from \cite{Brackbill1992}, the force 
\begin{equation}
\mathbf{F} =\sigma_0 \left(-\nabla \cdot \frac{\nabla \phi}{|\nabla \phi|}\right) \nabla \phi
\label{eq:YLlawvolume}
\end{equation}
gives the same total force as eq. \ref{eq:YLlaw}, but spread over the interface width.

Here, we propose to replace the YL definition of the surface tension force from eq. \ref{eq:YLlaw} with the non-local model 
\begin{equation}
\mathbf{F} = -\int_\Omega s(\mathbf{x}, \mathbf{y}) f_\varepsilon(|\mathbf{x}-\mathbf{y}|) \frac{\mathbf{x}-\mathbf{y}}{|\mathbf{x}-\mathbf{y}|} \; d\mathbf{y}, \; \; \mathbf{x} \in \Omega,
\label{eq:Nonlocal_force}
\end{equation}
where $ f_\varepsilon(|\mathbf{x}-\mathbf{y}|)$ is the force shape function and $s(\mathbf{x}, \mathbf{y})$ is the force strength. We obtain eq. \ref{eq:Nonlocal_force} as the continuous limit of the so-called pairwise surface tension force that is used in multiphase Smoothed Particle Hydrodynamics \cite{Tartakovsky2005,Tartakovsky2016}, a fully Lagrangian  particle method. 
The force strength is given by 
\begin{equation}
s(\mathbf{x}, \mathbf{y})= \left\{ \begin{array}{ll}
      s_{\alpha\alpha}, & \mathbf{x} \in \Omega_\alpha, \; \mathbf{y} \in \Omega_\alpha, \\
            s_{\alpha\beta}, & \mathbf{x} \in \Omega_\alpha, \; \mathbf{y} \in \Omega_\beta, \\
      s_{\beta\beta}, & \mathbf{x} \in \Omega_\beta, \; \mathbf{y} \in \Omega_\beta. \\
\end{array}
\right. 
\end{equation}
To ensure that $\sigma_0$ is positive, the coefficients must satisfy $  s_{\alpha\alpha} +   s_{\beta\beta} >   2s_{\alpha\beta}$. For convenience, we take
$  s_{\alpha\alpha}  =   s_{\beta\beta}  =   10^k s_{\alpha\beta}$ with $k = 3$. Then,  $  s_{\alpha\alpha}$ and $  s_{\beta\beta}$ can be found as a function of $\sigma_0$:
\begin{equation}\label{s11}
 s_{\alpha \alpha }   = {s}_{ \beta \beta} =\frac1{2(1-10^{-k})}  \frac{\sigma_0}{ \lambda},
\end{equation}
where
$ \lambda = \frac18 \pi  \int \limits_0^\infty z^4 
   f_\varepsilon(z)dz
$
and
$
  \lambda = \frac13  \int \limits_0^\infty z^3  f_\varepsilon(z)dz
$
in three and two spatial dimensions, respectively.

As in molecular dynamics,  $f_\varepsilon(|\mathbf{x}-\mathbf{y}|)$ must be repulsive for  small $|\mathbf{x}-\mathbf{y}|$, attractive for large $|\mathbf{x}-\mathbf{y}|$, and, for computational efficiency, should have compact support $h=O(\varepsilon)$ or become negligibly small for $|\mathbf{x}-\mathbf{y}| \geq h$.
Several forms of $f_\varepsilon$ have been proposed \cite{Tartakovsky2016}. Here, we use 
\begin{equation}
f_\varepsilon(|\mathbf{x}-\mathbf{y}|) = |\mathbf{x}-\mathbf{y}| \left[-A e^{-\frac{|\mathbf{x}-\mathbf{y}|^2}{2\varepsilon_0^2}}+ e^{-\frac{|\mathbf{x}-\mathbf{y}|^2}{2\varepsilon^2}}\right]
\label{eq:forceshapefunction}
\end{equation}
and $\lambda = \frac{1}{2}\sqrt{2 \pi}(-A \varepsilon_0^5 + \varepsilon^5).$ 
%The parameter $\varepsilon$ is chosen to be on the order of the grid size $dx$.

A fundamental difference between the non-local model in eq. \ref{eq:Nonlocal_force} and the YL law is that the former has an internal length scale $\varepsilon$, while the latter does not have any internal length scale.  Because of this, the YL law predicts the same behavior (for the same dimensionless numbers) regardless of the problem's length scale. In the following, we obtain an analytical solution for pressure that demonstrates that  the non-local model behaves ``macroscopically'' (follows the YL law) on the scale larger than  $\varepsilon$ and ``microscopically'' (deviates from the YL law in a way consistent with molecular dynamics simulations of droplets), otherwise.  

Under static conditions, eqs. \ref{Eq-Mom-Brackbill} and \ref{eq:Nonlocal_force} can be solved analytically for a circular interface separating two fluids in two dimensions. 
The pressure as a function of the distance $r$ from the droplet center can be found as \cite{Howard2019}: 
\begin{equation}
{P(r)-P(\infty) 
= -4 \pi( s_{aa} -s_{ab}) \left(\varepsilon^4 G(r, \varepsilon) - A\varepsilon_0^4 G(r, \varepsilon_0)\right)}
\label{eq:Pexact}
\end{equation}
where $G(r, \varepsilon) $ is given by eq. \ref{eqn:G}:
\begin{align}
G(r, \varepsilon)  &= \begin{cases} e^{-\frac{a^2}{2\varepsilon^2}} \sum_{l = 0}^\infty \frac{1}{l!(l+1)!} \left( \frac{a^2}{2\varepsilon^2} \right)^{l+1}  \left[\Gamma\left(l+1,  \frac{a^2}{2\varepsilon^2}\right) -\frac{1}{2}\Gamma\left(l+1,  \frac{r^2}{2\varepsilon^2}\right) \right] & r < a  \\
 e^{-\frac{a^2}{2\varepsilon^2}} \sum_{l = 0}^\infty \frac{1}{l!(l+1)!} \left( \frac{a^2}{2\varepsilon^2} \right)^{l+1}  \left[\frac{1}{2}\Gamma\left(l+1,  \frac{r^2}{2\varepsilon^2}\right) \right] & r \geq a. \end{cases}  \label{eqn:G}
  \end{align}

\begin{figure}
\centering
 \sidesubfloat[]{   
    \includegraphics[width=0.35\linewidth]{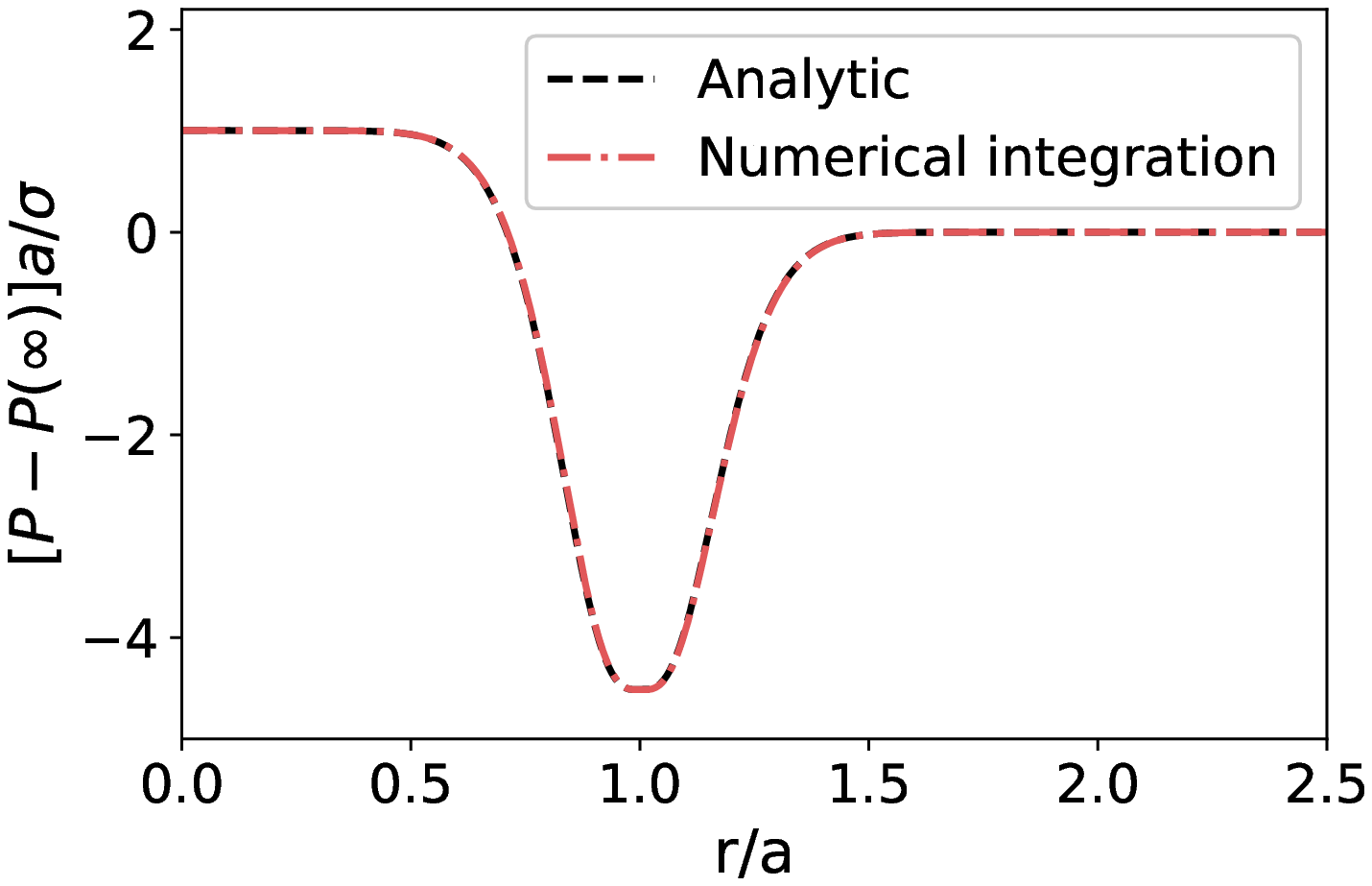}} \sidesubfloat[]{ 
   \includegraphics[width=0.35\linewidth]{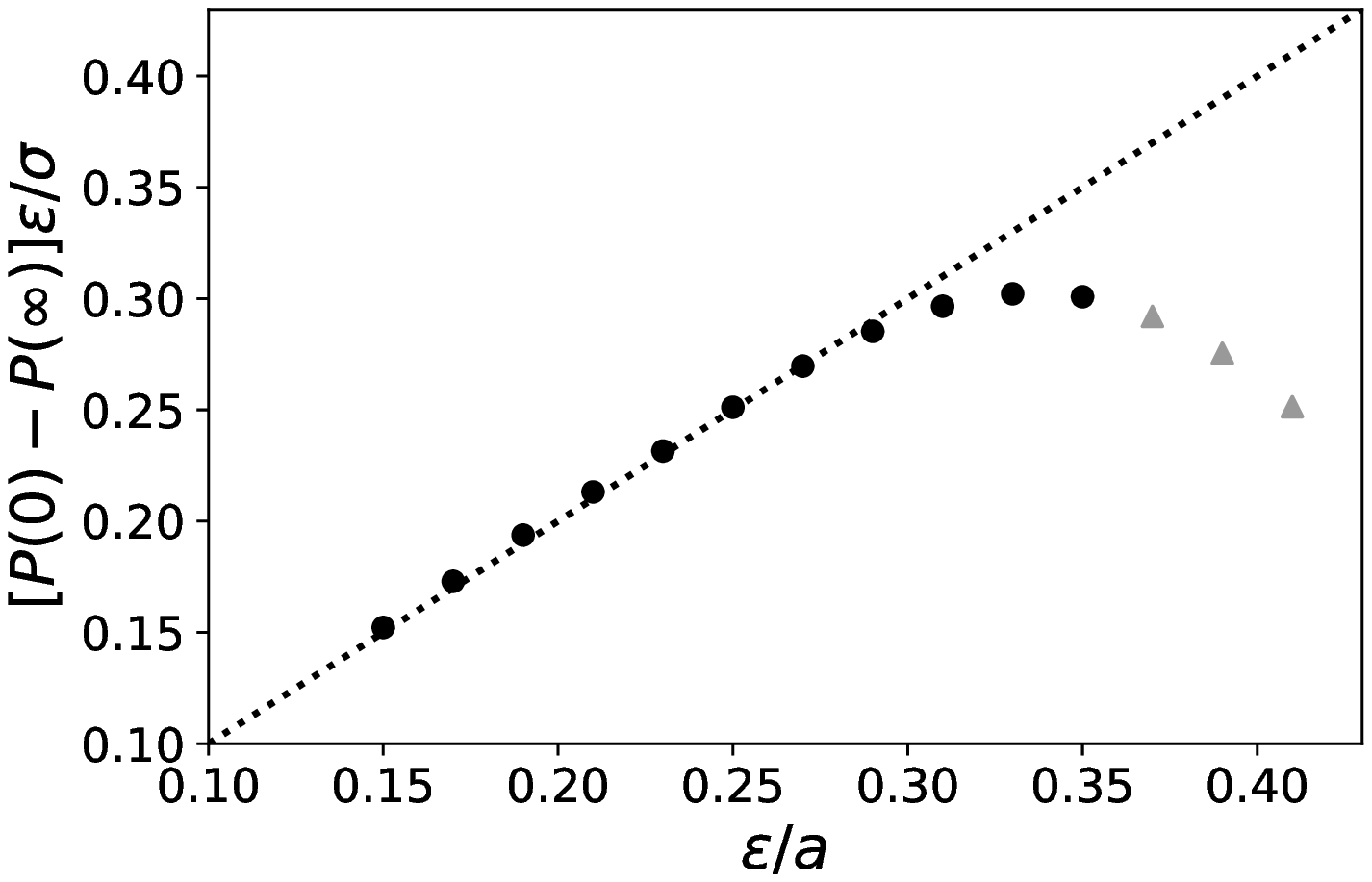}}
   \caption{Comparison of numerical integration and the analytical solution with $a/\varepsilon = 6$. (a) Pressure profile from eq. \ref{eq:Pexact}. (b) Pressure jump at the center of a circular droplet calculated from the analytic solution in eq. \ref{eq:Pexact}. The dashed line shows the linear value from the YL equation, eq. \ref{Young-Laplace}. The pressure jump agrees with the YL equation when the value of $\varepsilon$ is much smaller than the radius $a$, $\varepsilon \ll a$. As the curvature increases relative to $\varepsilon$ the pressure jump decreases relative to the one predicted by the YL law. The triangles represent values for which the non-local model gives non-physical values.}
       \label{fig:analytic_profiles}
\end{figure}

The pressure in eq. \ref{eq:Pexact} as a function of $r$ is plotted in Fig. \ref{fig:analytic_profiles}$A$ for a droplet with the radius $a$. The pressure profile qualitatively agrees with behavior observed in molecular dynamics simulations \cite{Park2001, Masuda2011, Nakamura2011, Malek2018}. The pressure is zero far from the droplet, then at the interface the pressure drops and becomes negative, as seen in MD simulations. At the center of the droplet the pressure reaches a constant value. The pressure jump from inside to outside the interface is given by the YL law (eq. \ref{eq:YLlaw}). 

Fig. \ref{fig:analytic_profiles}$B$ demonstrates that  the pressure difference $P_{\varepsilon, in} -  P_{\varepsilon, out}$  as a function of curvature
 agrees with the YL law (eq. \ref{Young-Laplace}) for droplet radius $a > 3.5 \varepsilon$ ($P_{\varepsilon, in}$ and $P_{\varepsilon, out}$ are pressures inside and outside of the droplet at the distance greater than $3.5\varepsilon$ from the interface). For smaller droplets with $a< 3.5\varepsilon$, the pressure jump begins to deviate from the YL law.
 %This corresponds with very small bubbles, where the YL law does not hold. 
For ``tiny'' droplets with $a< 0.3\varepsilon$, the pressure difference begins to decrease (the triangle symbols in Fig. \ref{fig:analytic_profiles}$B$). This indicates the limit of incompressible treatment of small (nano) droplets.

%Further studies will look at refining the form of the force shape function in eq. \ref{eq:forceshapefunction} and consider non-zero compressibility of fluids  to improve the behavior of the non-local model for large values of $\varepsilon \kappa$. 

%For the radius of curvature larger than $\varepsilon$, the components  of the isotropic stress tensor satisfy the YL equation.  Note that for the considered problems (two fluids under static conditions separated by a flat or spherical interface), eq. \ref{Eq-Mom-Brackbill} will predict an isotropic stress tensor.

%The described above analytical solutions and numerical results in the following section  show that the nonlocal model behavior depends on the ratio of its internal length scale  to the characteristic length of the problem. 

%\subsection{Local surface tension model}

%To provide a comparison with the non-local model, we also implement the conservative level set method from \cite{Olsson2005} and \cite{Harten1977}. The volume force due to the surface tension is found by 
%\begin{equation}
%\mathbf{F} = \sigma \left(-\nabla \cdot \frac{\nabla \phi}{|\nabla \phi|}\right) \nabla \phi.
%\end{equation} This force spreads the surface tension force per interfacial area over the finite interface width given by the color function $\phi$, while maintaining the same total force. 

\section{Numerical implementation of the non-local model}
\label{sec:numerical}

Eq.  \ref{Eq-Mom} is discretized with a prediction-correction scheme based on the finite volume method presented in  \cite{Tryggvason2012}. A temporary velocity is first calculated without the pressure, 
\begin{equation}
\frac{\mathbf{u}^*-\mathbf{u}^n}{\Delta t} = \nabla_h \cdot \mathbf{u}^n\mathbf{u}^n + \mathbf{g} 
+\frac{1}{\rho^n} \mathbf{F}^n
+ \frac{1}{\rho^n}\nabla_h^2\mathbf{u}^n
\end{equation}
where $\mathbf{u}^n$ is the velocity field at time step $n$, $\mathbf{u}^*$ is the temporary velocity, $\mathbf{g}$ represents external forces such as gravity, and $\rho^n$ is the density field. The subscript $h$ denotes discrete finite volume operators. The pressure $p^n$ is then calculated so that $\mathbf{u}^{n+1}$ satisfies the incompressibility condition, $\nabla_h \cdot \mathbf{u}^{n+1} =0$, which gives
\begin{equation}
\nabla_h \cdot \left( \frac{1}{\rho^n} \nabla_h p^n \right) = \frac{1}{\Delta t} \nabla_h \cdot \mathbf{u}^*
\end{equation}
for the pressure and 
\begin{equation}
\frac{\mathbf{u}^{n+1}-\mathbf{u}^*}{\Delta t} = \frac{-\nabla_h p^n}{\rho^n}
\end{equation}
for the updated velocity at time step $n+1$. We use a staggered mesh, with the velocity calculated on the mesh edges and pressure, density, viscosity, and the color function updated on the cell centers. 
The pressure $p^n$ is solved using a successive over relaxation (SOR) scheme \cite{Tryggvason2012}. Following \cite{Tryggvason2012}, we set the time step to $\Delta t = \frac{\Delta x^2\min(\rho_\alpha, \rho_\beta)}{4.25\max(\mu_\alpha, \mu_\beta)}$.

We use a linear interpolation within elements and the color function at time step $n$, $\phi^n$, to discretize eq. \ref{eq:Nonlocal_force}:
\begin{align}\label{Cont-SPH-Mom-Sum-phi}
\mathbf{F}^n(\mathbf{x}) &=  \int_{\Omega_\alpha^n\cup\Omega_\beta^n } s(\mathbf{x}, \mathbf{y}) f_\varepsilon(|\mathbf{x}-\mathbf{y}|) \frac{\bx-\by}{|\bx-\by|} d \by  \nonumber \\
&= \sum_j  f_\varepsilon(|\bx - \by|) \frac{\bx-\by}{|\bx-\by|}  V_j    \left(s_{\alpha \alpha}[\phi^n(\bx)\phi^n(\by) +(1-\phi^n(\bx) )(1-\phi^n(\by) )] + s_{\alpha \beta}[\phi^n(\bx)(1-\phi^n(\by))+ (1-\phi^n(\bx) )\phi^n(\by) ]
 \right) , 
\end{align}
where $V_j$ is the size of element $j$. In the numerical simulations,   $\varepsilon/\varepsilon_0 = 2$,   
 $A =  (\varepsilon/\varepsilon_0)^3$, and $s_{\alpha \alpha }   = {s}_{ \beta \beta} = 10^5 {s}_{ \alpha \beta} $.  Unless otherwise noted we take 
$\varepsilon = 4\Delta x /3.5$ where $\Delta x$ is the grid resolution and $V_j = \Delta x^2$.

When advecting the color function, $\phi$, it is important to use an accurate method that is conservative and also preserves the sharpness of the front. As noted in \cite{Sethian2003}, implementation of high order upwind schemes is difficult on a staggered mesh. We use the method presented in \cite{Olsson2005} and updated in \cite{Olsson2007} for a conservative advection scheme of the color function. To update the color function, we first solve 
\begin{equation}
\phi_t + \textbf{u}\cdot \nabla \phi = 0 \label{eq:advection}
\end{equation}
using the conservative Total Variation Diminishing (TVD) method with Superbee limiter from \cite{Olsson2005} with a second-order Runge-Kutta scheme in time. 
As noted in \cite{Olsson2005}, with eq. \ref{eq:advection} alone the interface width and profile will not remain constant over the course of a simulation. To remedy this, we solve the compression-diffusion equation from \cite{Olsson2005} after eq. \ref{eq:advection}. The compression-diffusion equation is given by
\begin{equation}
\phi_\tau + \nabla \cdot (\phi(1-\phi) \mathbf{n}) = \kappa \Delta \phi \label{eq:CDEquation}
\end{equation}
where $\mathbf{n}$ is the normal of the interface and $\kappa = \frac{1}{2}(\Delta x)^{1-d}$. $d$ is a parameter chosen so that $d \ll 1$ \cite{Olsson2005}. In the paper we take $d = 0.$ The time parameter is denoted by $\tau$ to distinguish it from the simulation time step, and is chosen as $\Delta \tau = \frac{1}{2}(\Delta x)^{1+d}$.  Eq. \ref{eq:CDEquation} is solved until the solution reaches steady state, defined by $\int | \phi^{n+1}-\phi^n| \leq \text{TOL}\cdot\Delta \tau $ for a given tolerance $\text{TOL}$, where $\phi^n$ denotes the solution after $n$ iterations.  To update $\phi^{n+1}$ we use a second-order Runge-Kutta scheme. The intermediate compression step provides an artificial compression normal to the front interface to maintain a sharp front, with constant thickness proportional to $\kappa$, preventing diffusion as the front is advected. The added viscosity term, $\kappa \Delta \phi $, prevents discontinuities at the interface. The compression flux $(\phi(1-\phi) \mathbf{n}) $ acts in regions where $0 < \phi < 1$ and provides compression normal to the interface, sharpening the interface profile. 

Because the area each phase is defined as the area inside and outside of the $\phi = 0.5$ contour, conservation of $\phi$ does not necessarily imply conservation of each phase. However, our results in sec. \ref{sec:rising_bubble}  indicate that the area loss is minimal. Once the color function is calculated, the density and viscosity are found directly by 
\begin{equation}
\mu^{n+1} =\mu_1 + (\mu_2-\mu_1) \phi^{n+1},
\end{equation}
\begin{equation}
\rho^{n+1} =\rho_1 + (\rho_2-\rho_1) \phi^{n+1}.
\end{equation}

The numerical algorithm described in this section is implemented in a parallel  C++ code with OpenMP. In this paper, simulations with the non-local methods are compared with simulations calculated with the CLS method. To provide a direct comparison, the simulations using the CLS method are completed with the same code, with the surface tension force in eq. \ref{Cont-SPH-Mom-Sum-phi} replaced by eq. \ref{eq:YLlaw}.

\section{Numerical results}

In this section, we validate the numerical implementation of the non-local model against analytical solutions, solutions obtained with the CLS \cite{Osher1988, Sussman1998, Olsson2005,Harten1977, Olsson2007} finite-volume discretization of the Navier-Stokes equations subject to the YL boundary condition, and published numerical simulations. In the following, we refer to CLS simulations of the Navier-Stokes equations subject to the YL boundary condition as a local model. Comparisons with the results of the local model will be used to illustrate advantages of the proposed here non-local model. 

\subsection{Static pressure in a droplet}
\begin{figure}
\centering
 \sidesubfloat[]{ \includegraphics[width=0.35\columnwidth]{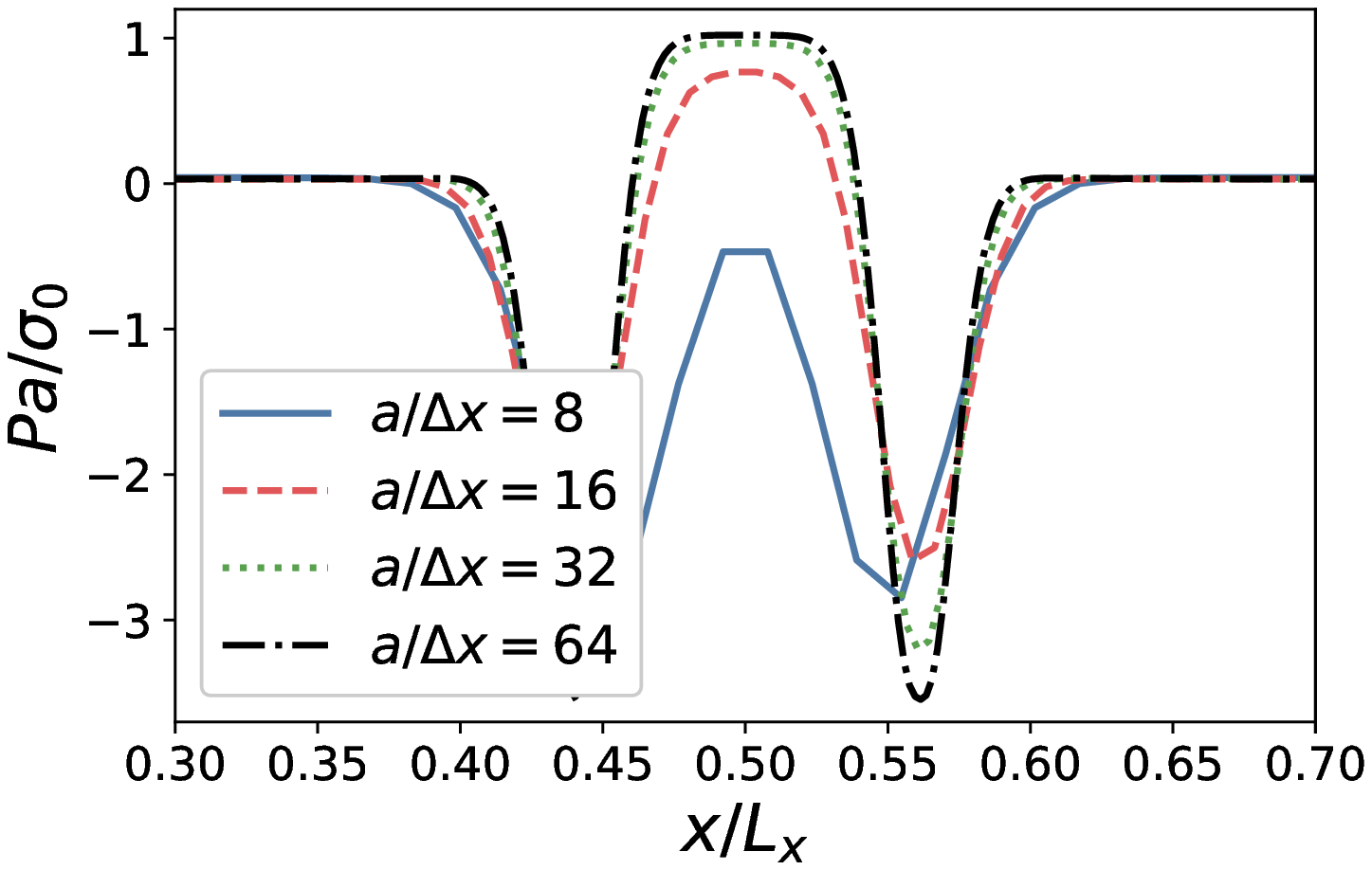}}
 \sidesubfloat[]{  
\includegraphics[width=0.35\columnwidth]{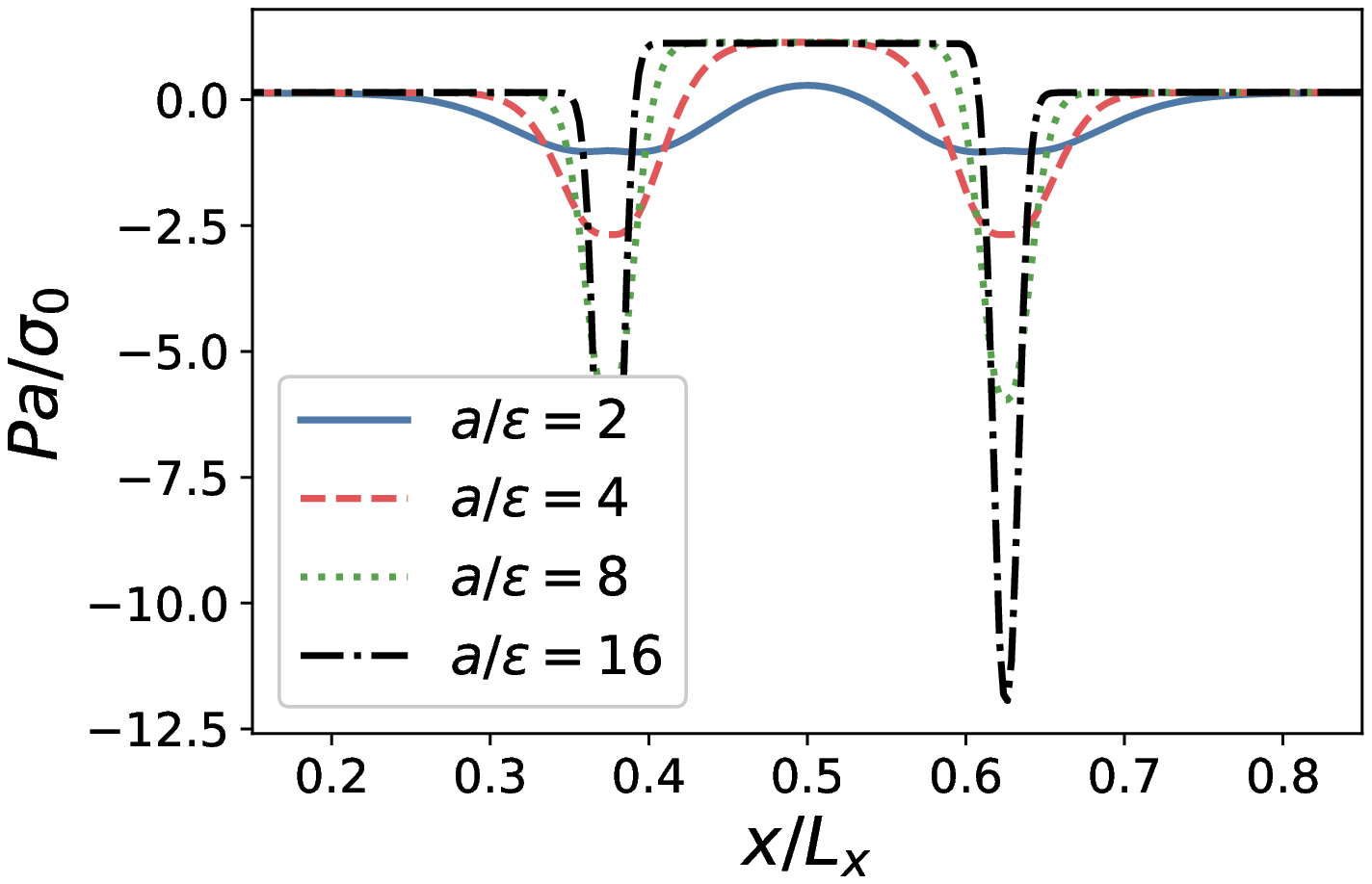}}
\caption{(a) Pressure across the center of a circular droplet with varying resolution. The droplet radius $a$ is fixed as $a = 0.0625L_x$, where $L_x = L_y = 4.0$ are the dimensions of the domain. The parameter $\varepsilon$ in the force shape function is fixed at $\varepsilon = 0.2a$. (b) Pressure across the center of a circular droplet with varying radius of support for the force shape function. The droplet radius $a$ is fixed as $a = 0.125L_x$, where $L_x = L_y = 2.0$ are the dimensions of the domain, with resolution $\Delta x = \Delta y = 1/200$.  In both cases the fluid parameters are set to $\mu_f = \mu_d = 0.15$, $\rho_f = 1.0$, $\rho_d = 2.0$, and $\sigma_0 = 5.0$. }
\label{fig:Pressure}
\end{figure}

We first test the code against analytical non-local and YL solutions for a static droplet with radius $a$.  
%
%A static droplet satisfies the Young-Laplace law
%\begin{equation}
%(P_{in}-P_{out}) = \frac{\sigma_0}{\kappa}.
%\end{equation}
%For a circular bubble we have the the curvature is the inverse of the radius, $\kappa = \frac{1}{a}$. $P_{in}$ and $P_{out}$ are the pressure inside and outside the droplet, respectively. 
 In Fig. \ref{fig:Pressure}$A$, the pressure across the center of the droplet is plotted for varying resolution $\Delta x$. As the resolution across the domain increases, $P_{\varepsilon, in} -  P_{\varepsilon, out}$ converges to the value expected by the YL law, with an error of less than $1\%$ for $a / \Delta x = 64$. In Fig. \ref{fig:Pressure}$B$, the resolution remains constant while the parameter $\varepsilon$ is varied relative to $a$.  Here, $P_{\varepsilon, in} -  P_{\varepsilon, out}$ is in excellent agreement with the YL law when $a/\varepsilon > 3.5$. 

\subsection{Rising bubble}
\label{sec:rising_bubble}
 %NOTE: The bubble and droplet terminology in this section has been standardized to agree with Hysing, so I do not want to change it. 

\begin{table}
\centering
\caption{Parameters for benchmark test cases 1 and 2 from \cite{Hysing2009}.}
\begin{tabular}{ccccccc}
\hline
Test case & $\rho_1$ & $\rho_2$ & $\mu_1$ & $\mu_2$ & $g$ & $\sigma$  \\
\hline 
1 & 1000 & 100 & 10 & 1 & 0.98 & 24.5 \\ 
2 & 1000 & 1 & 10 & 0.1 & 0.98 & 1.96  \\ 
\hline 
\end{tabular}
\label{tab:Hysingparams}
\end{table}

A set of quantitative benchmarks for a rising bubble was proposed in \cite{Hysing2009} for validation and comparison between numerical methods for interfacial flow. \cite{Hysing2009} considered a bubble rising due to buoyancy effects in a two-dimensional domain with $L_x = 1$, $L_y =2$, no-slip boundary conditions at the top and bottom walls, and free slip conditions on the vertical walls. The droplet has an initial radius $a = 0.25$ and is centered at $(x_c, y_c) = (0.5, 0.5)$. Values of the viscosity, density, surface tension, and gravitational acceleration are given in table \ref{tab:Hysingparams} for the two test cases, denoted Case 1 and Case 2. In \cite{Hysing2009}, three numerical discretizations of local models were compared,   two Eulerian Level Set finite-element codes (TP2D, denoted by Group 1, and FreeLIFE, denoted by Group 2) and an arbitrary Lagrangian-Eulerian moving grid method (MooNMD, denoted by Group 3.) The results of \cite{Hysing2009} have been compared with other numerical local models by many subsequent authors, including a conservative Level Set method with front sharpening \cite{Strubelj2009}, the volume of fluid method implemented in OpenFOAM\textsuperscript{\textregistered} \cite{Klostermann2013}, a diffuse interface model  \cite{Aland2012}, and a finite element based Level Set method  \cite{Zahedi2012}. In \cite{Strubelj2009} and \cite{Klostermann2013}, the results from \cite{Hysing2009} were matched to within $2\%$ and $4\%$, respectively, while in \cite{Zahedi2012}, it was found that higher resolution in their model is needed to achieve the accuracy in \cite{Hysing2009}.

We run the non-local model for Case 1 and Case 2 until the dimensionless time $t=3$  with the parameters given in table \ref{tab:Hysingparams} and $\varepsilon = \frac{8}{3.5}\Delta x$, where $\Delta x = 1/80$, $1/160$, or $1/320$  is the grid spacing. The characteristic length scale is $L = 2a$, where $a$ is the bubble radius, and time scale $L/\sqrt{2ga}$ where $g$ is the gravitational velocity. 
In \cite{Hysing2009}, data are provided for the bubble area $A(t)$, the center of mass $\mathbf{x}_c$
\begin{equation}
\mathbf{x}_c =  (x_c(t), y_c(t)) = \frac{\int_{\Omega_2(t)} \mathbf{x} \; d\mathbf{x}}{\int_{\Omega_2(t)} 1 \; d\mathbf{x}}, 
 \label{eq:cm}
\end{equation}  
where $\Omega_2(t)$ is the domain occupied by the bubble, the rise velocity $\mathbf{v}_c$ 
\begin{equation}
   \mathbf{v}_c =   (u_c(t), {v}_c(t)) = \frac{\int_{\Omega_2(t)} \mathbf{v}(\mathbf{x},t)  \; d\mathbf{x}}{\int_{\Omega_2(t)} 1 \; d\mathbf{x}},
      \label{eq:vc}
\end{equation} 
where $\mathbf{v} = (u, v)$ is the fluid velocity,  and the circularity $\hat{c}(t)$
\begin{equation}
   \hat c(t) = \frac{2 \pi a_{circle}(t)}{P_b (t)},% = \frac{\text{perimeter of area-equivalent bubble}}{\text{perimeter of bubble}}.
    \label{eq:hatc}
\end{equation} 
where $a_{circle}(t)=\sqrt{A(t)/\pi}$ and $P_b (t)$ is the bubble perimeter.

\begin{figure}
\centering
\begin{minipage}{.3\textwidth}
	 \sidesubfloat[]{ \includegraphics[width=\linewidth]{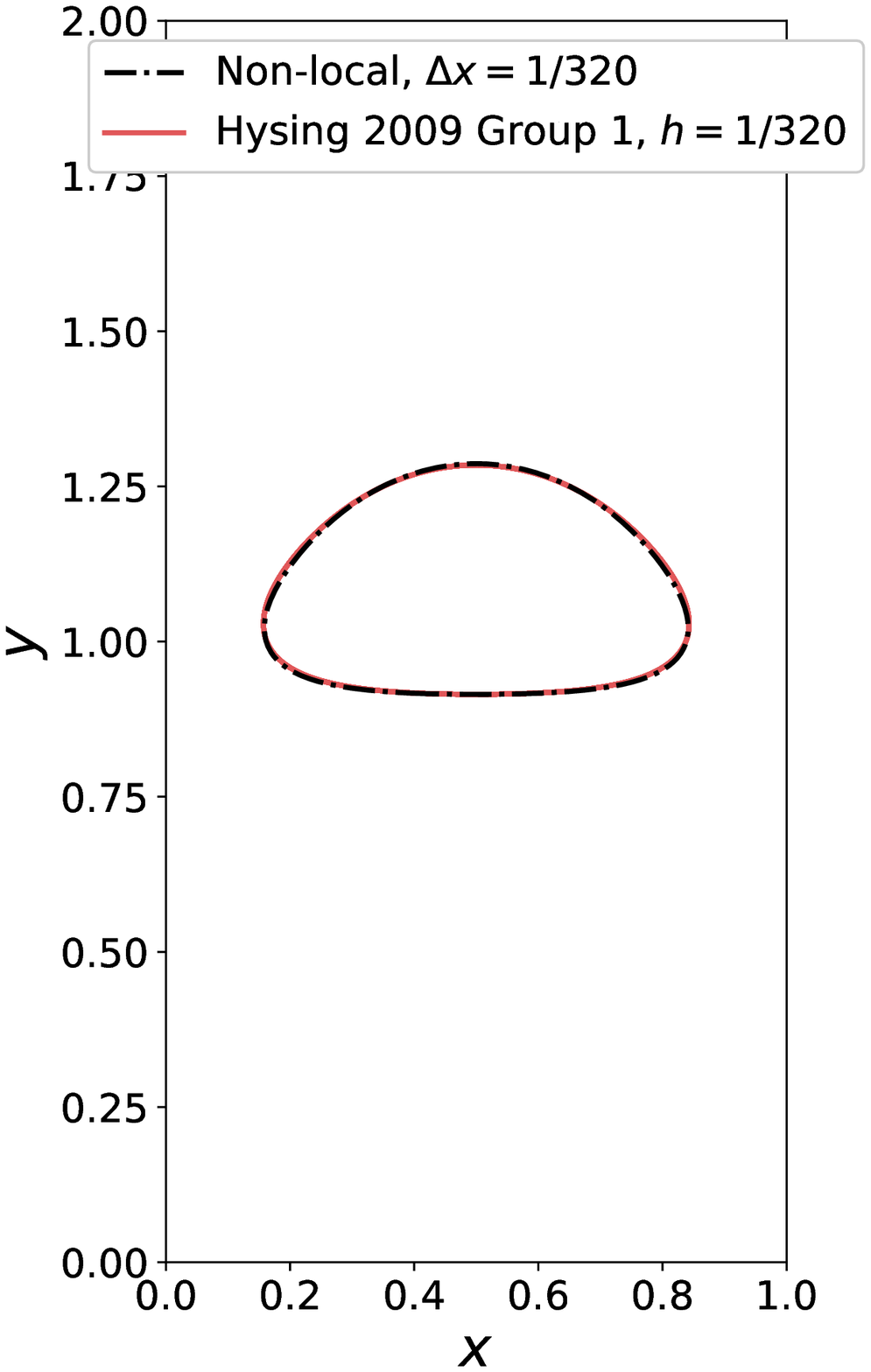}}
\end{minipage}
\begin{minipage}{0.65\textwidth}
  \sidesubfloat[]{    \includegraphics[width=0.45\columnwidth]{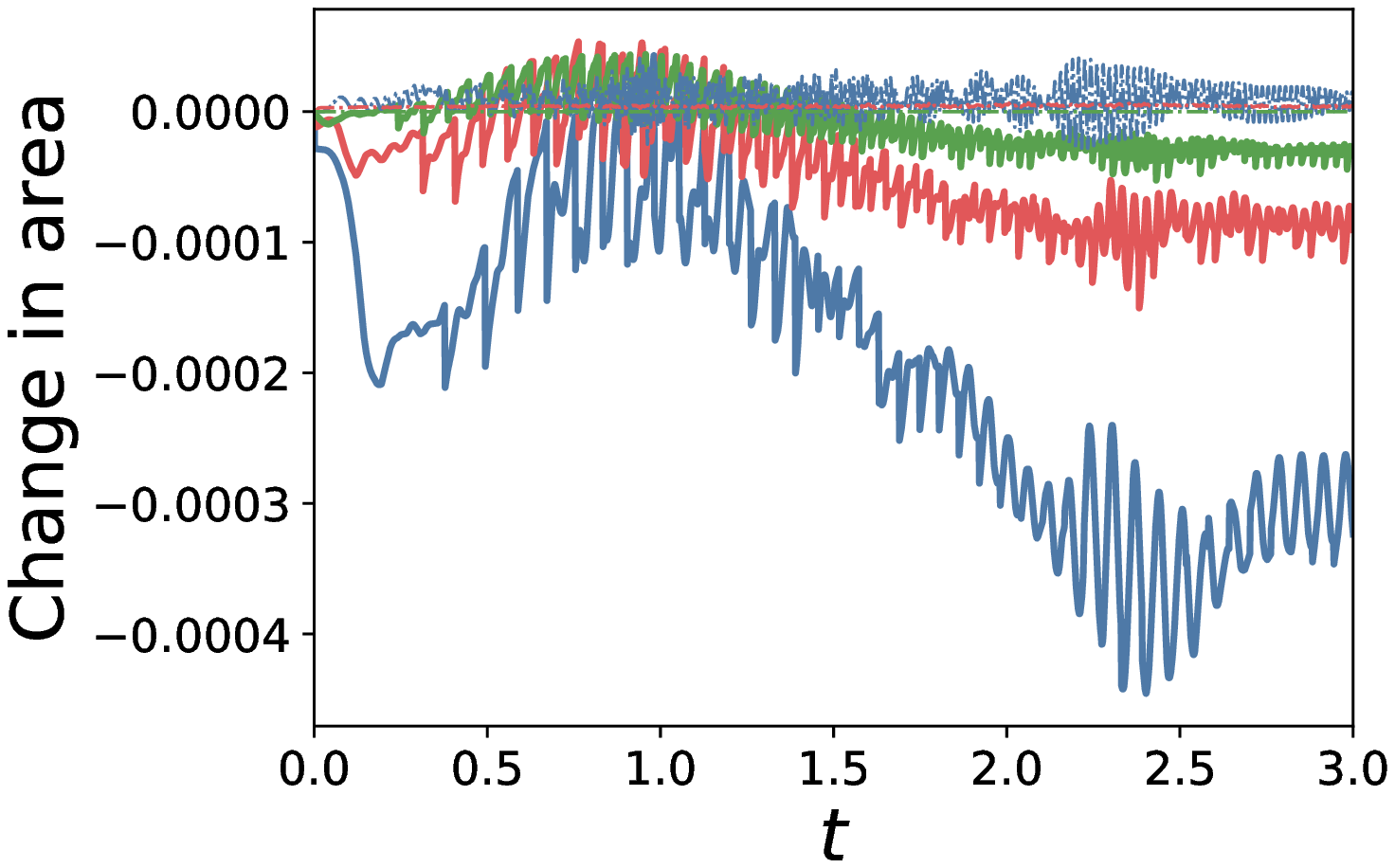}}
 \sidesubfloat[]{     \includegraphics[width=0.45\columnwidth]{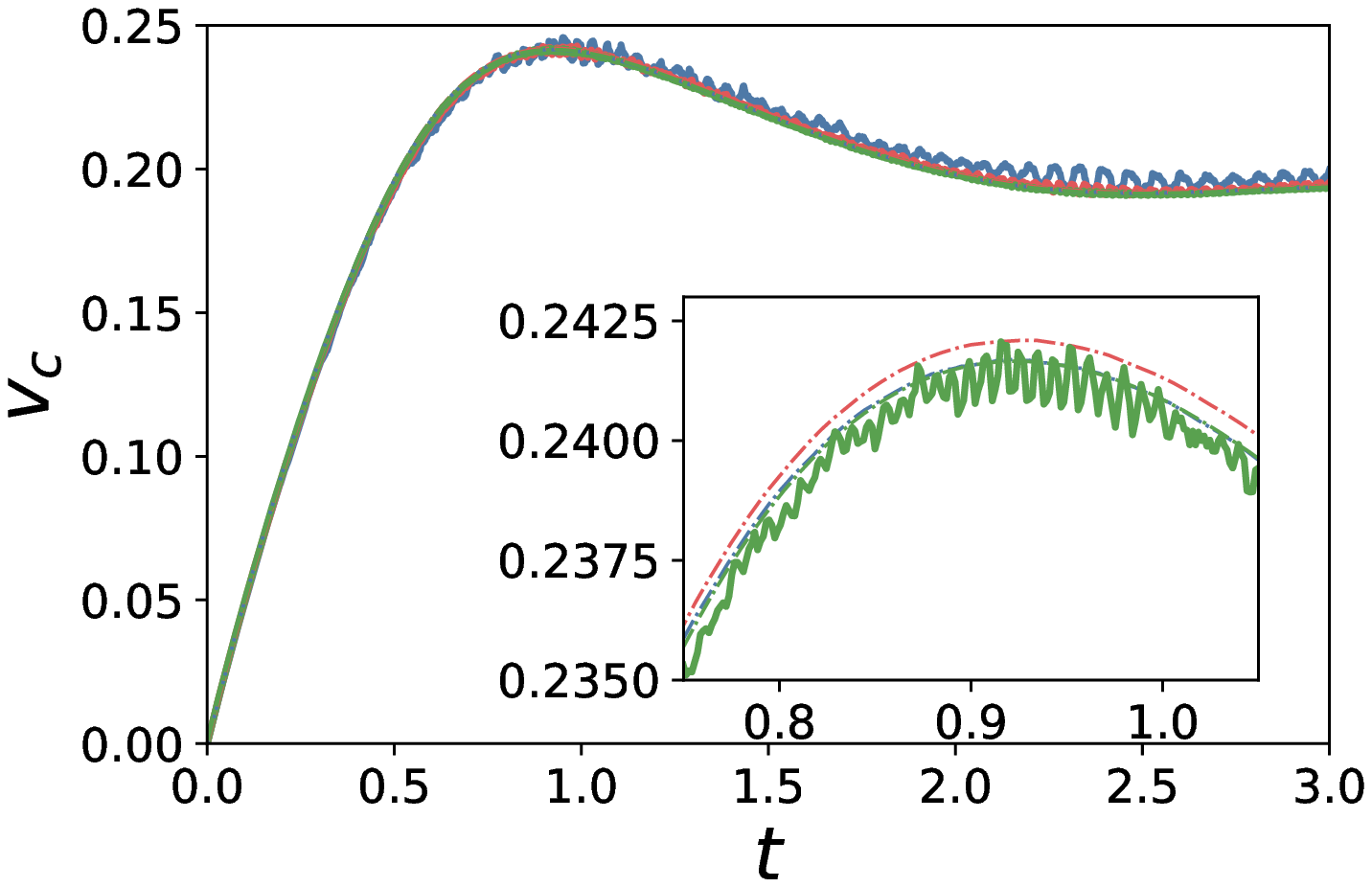}}\\
  \sidesubfloat[]{    \includegraphics[width=0.45\columnwidth]{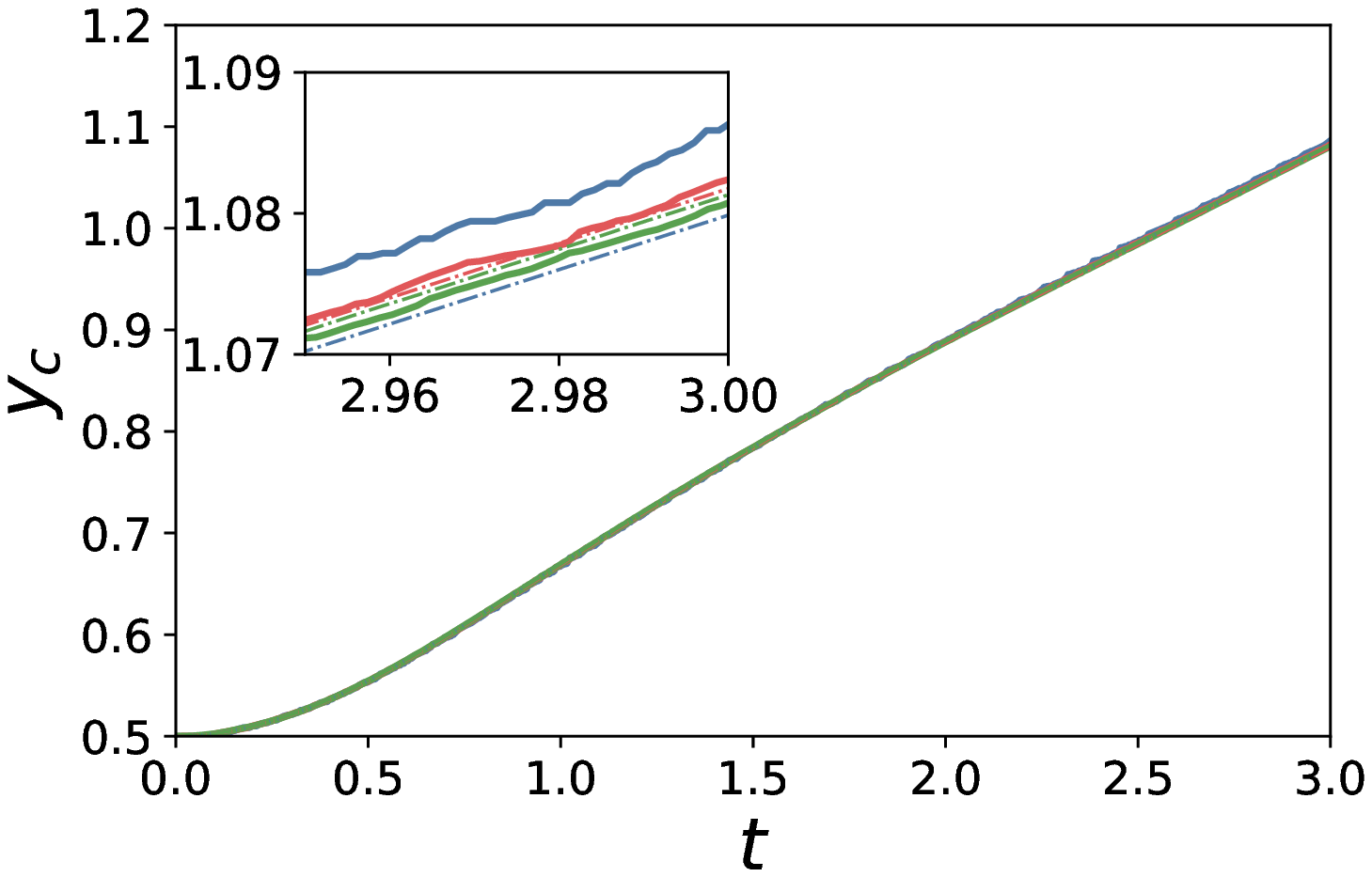}}
  \sidesubfloat[]{    \includegraphics[width=0.45\columnwidth]{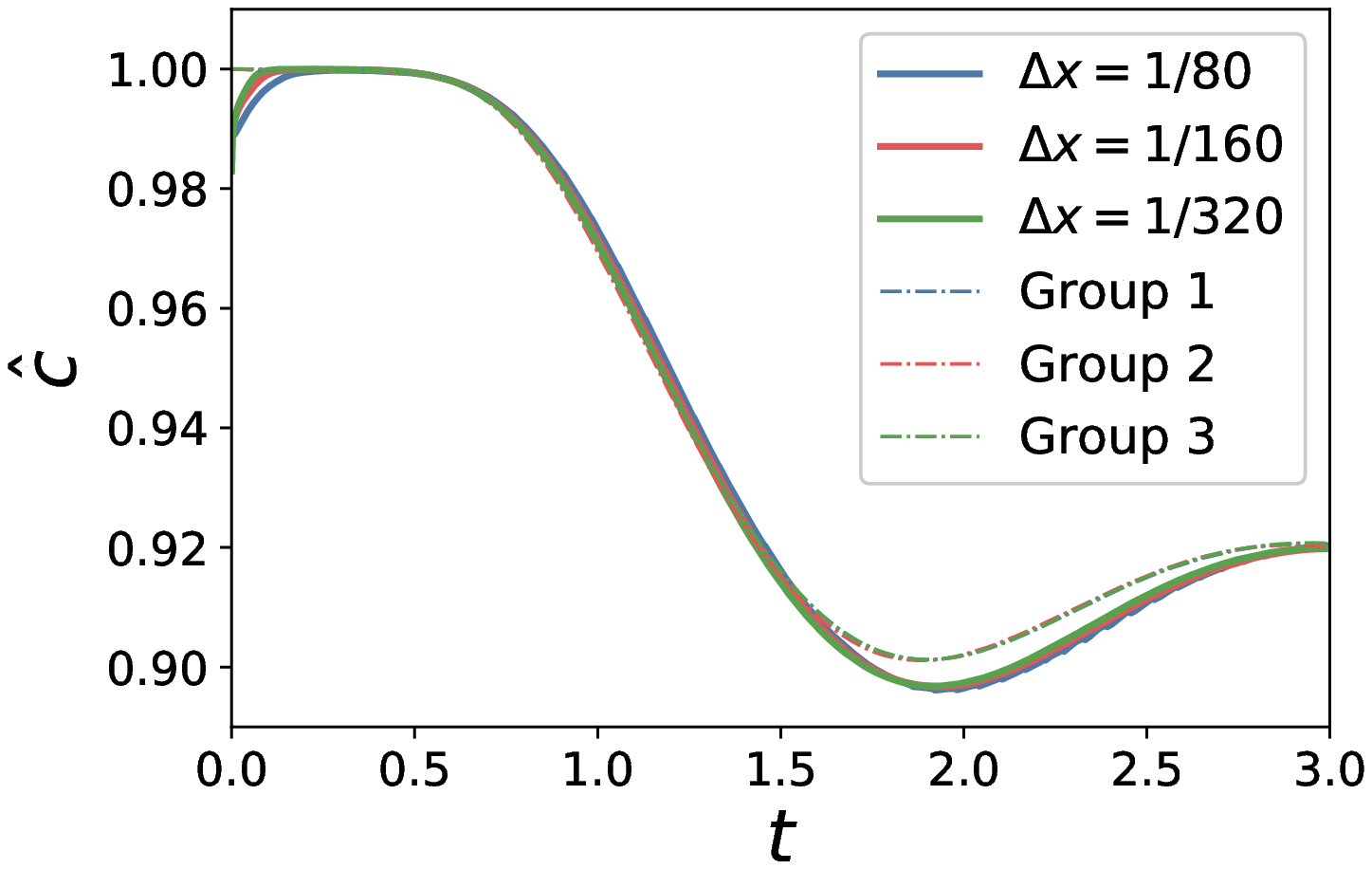}}
    \end{minipage}

\caption{ Case 1. (a) Location of the $\phi = 0.5$ contour at $t = 3.0$ for the non-local model with grid spacing $\Delta x = 1/320$ compared with the front location from \cite{Hysing2009} Group 1 for cell size $h = 1/320$. (b-e) Comparison for Case 1 between the results of the non-local model with grid spacing $\Delta x = 1/80, 1/160$, and $1/320$ and the results from \cite{Hysing2009} Group 1 (cell size $h = 1/320$), Group 2 (element mesh size $h = 1/160$), and Group 3 (900 degrees of freedom on the interface). (b) Change in bubble area $A(t) - A(t = 0)$, (c) rise velocity $v_c$ given by eq. \ref{eq:vc}, (d) center of mass $y_c$ given by eq. \ref{eq:cm}, and (e) bubble circularity $\hat c$ given by eq. \ref{eq:hatc}. The parameters for Case 1 and 2 are given in table \ref{tab:Hysingparams}. (Color online.) 
}
\label{fig:benchmark1_front}
\end{figure}

\begin{figure}
\centering
\begin{minipage}{.3\textwidth}
	 \sidesubfloat[]{ \includegraphics[width=\linewidth]{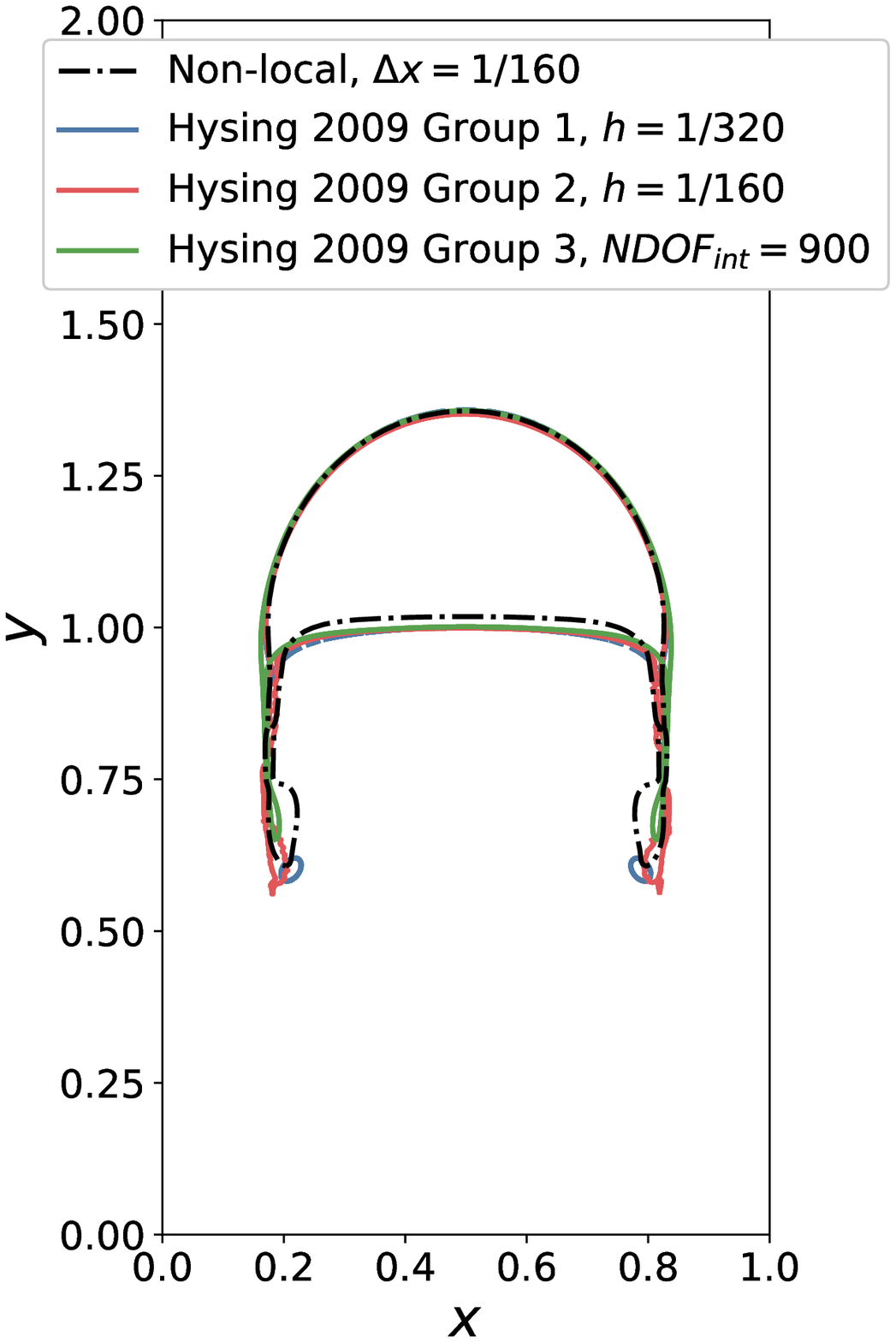}}
\end{minipage}
\begin{minipage}{0.65\textwidth}
  \sidesubfloat[]{    \includegraphics[width=0.45\columnwidth]{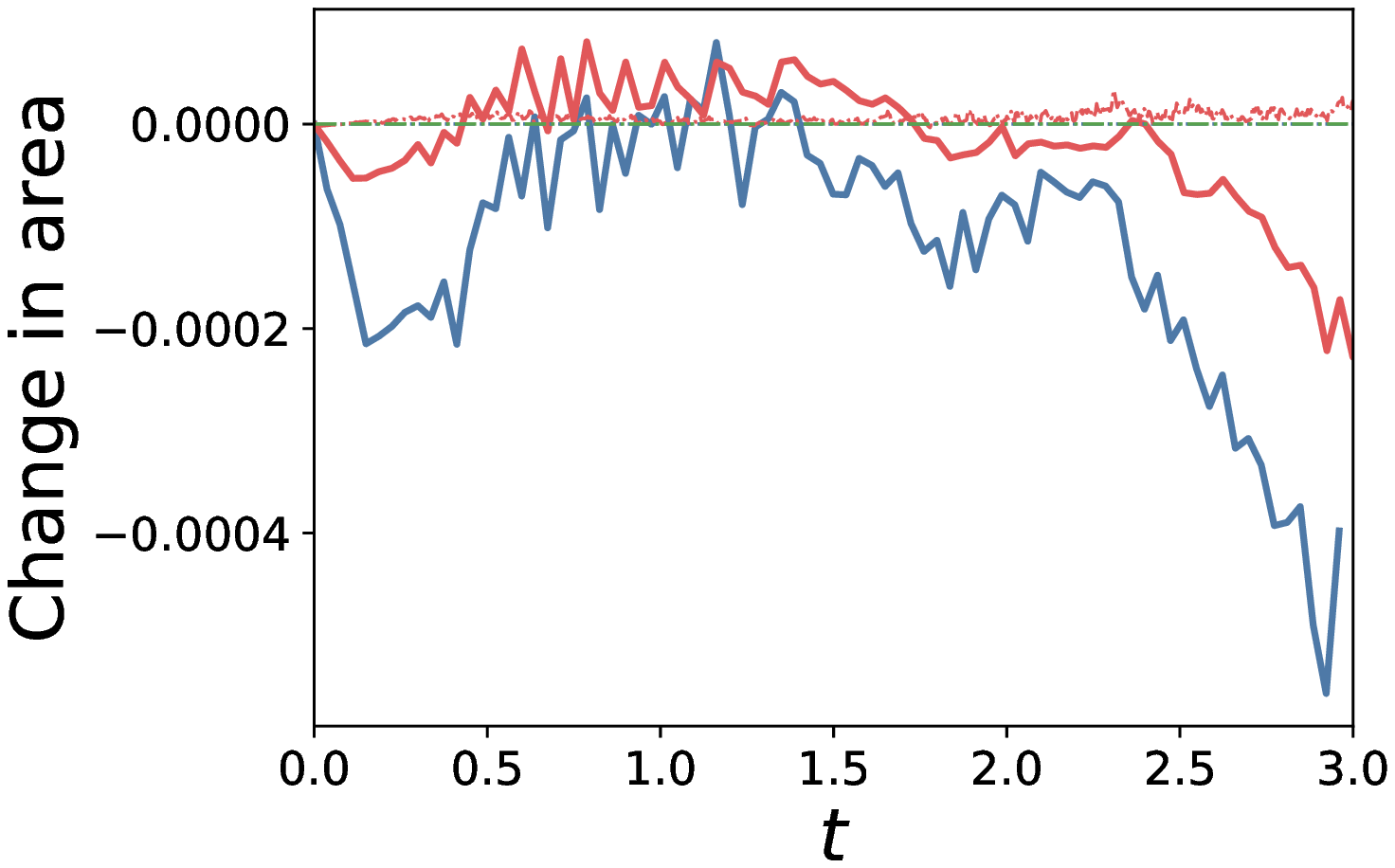}}
 \sidesubfloat[]{     \includegraphics[width=0.45\columnwidth]{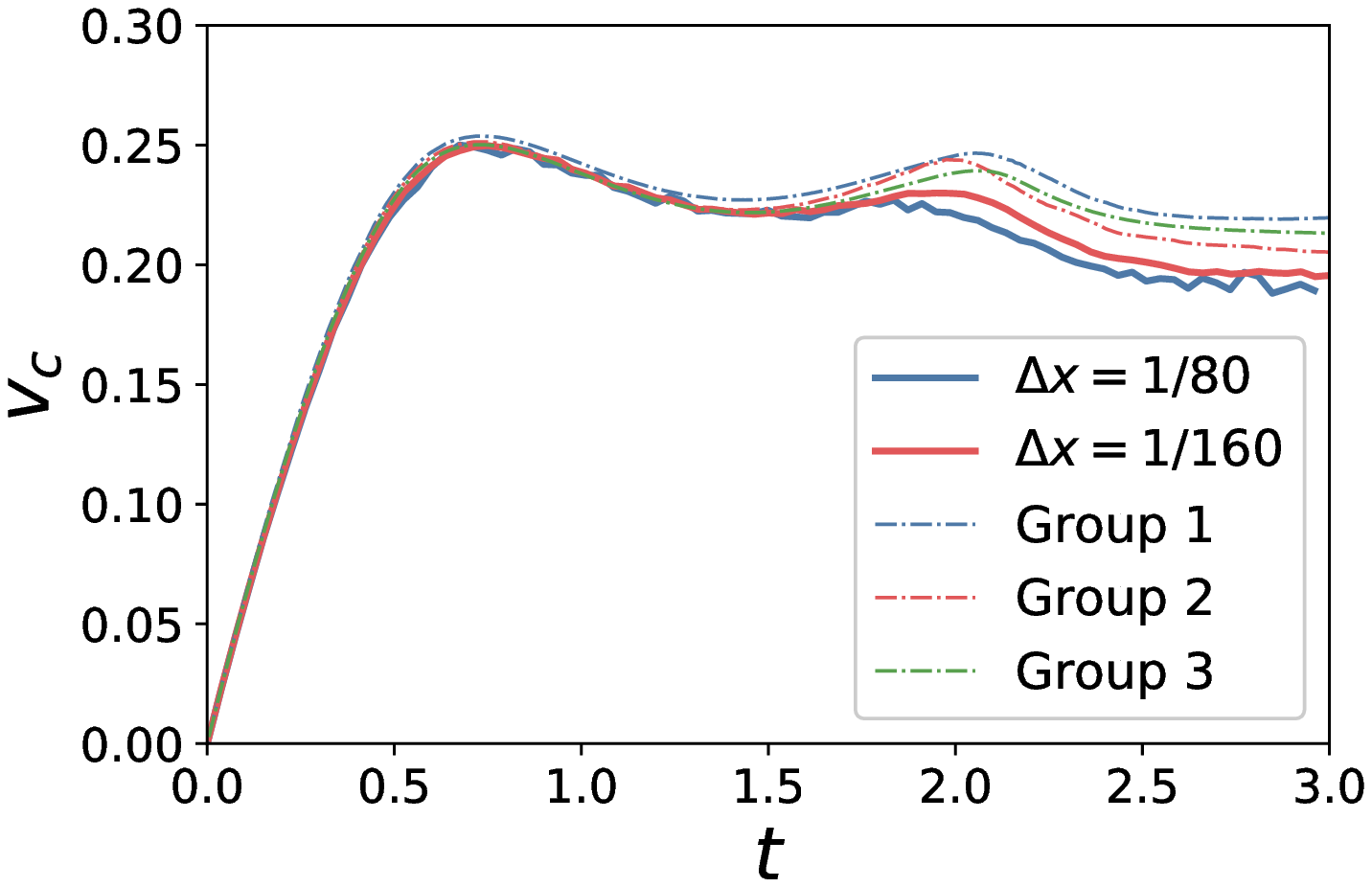}}\\
  \sidesubfloat[]{    \includegraphics[width=0.45\columnwidth]{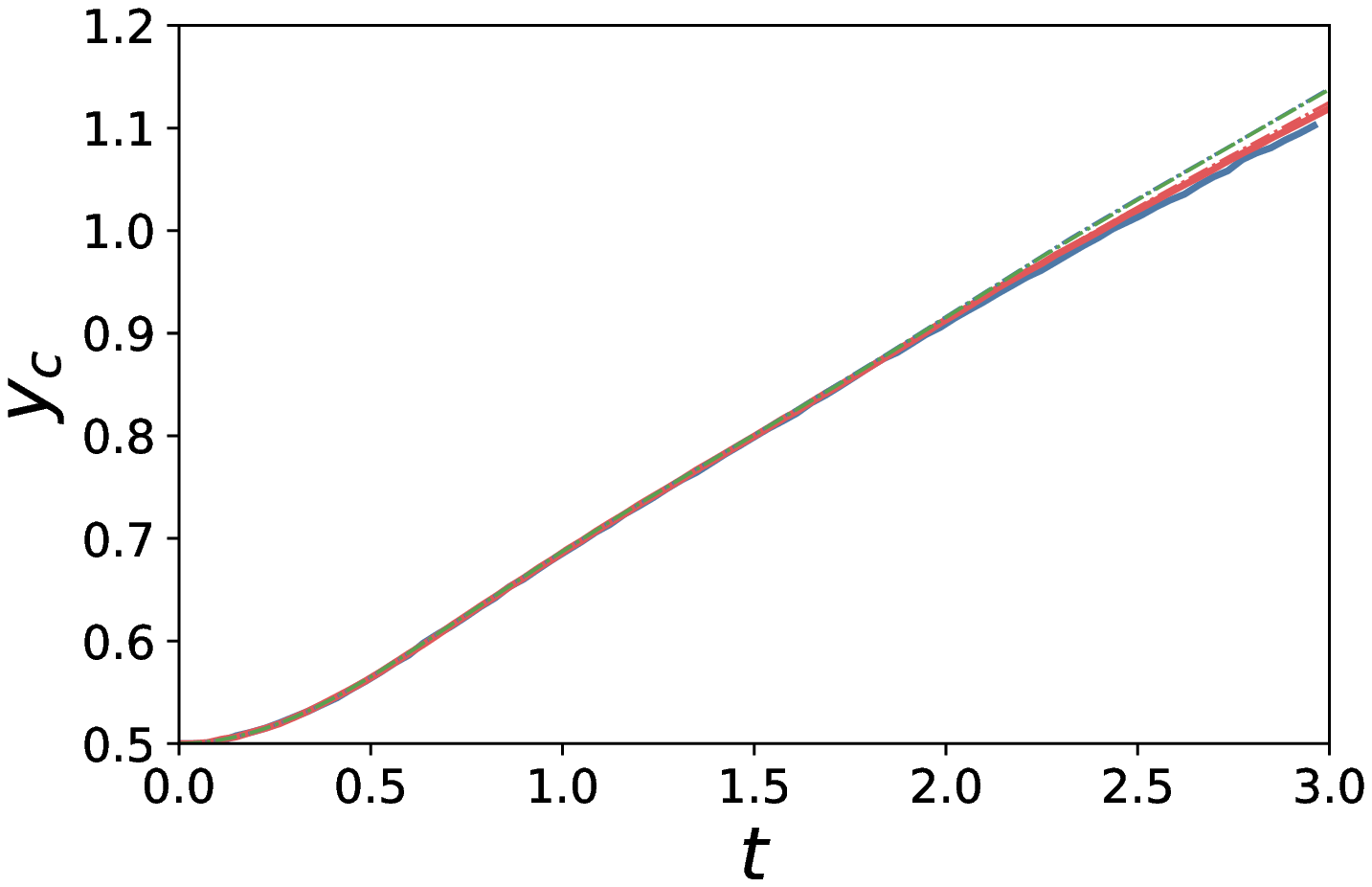}}
  \sidesubfloat[]{    \includegraphics[width=0.45\columnwidth]{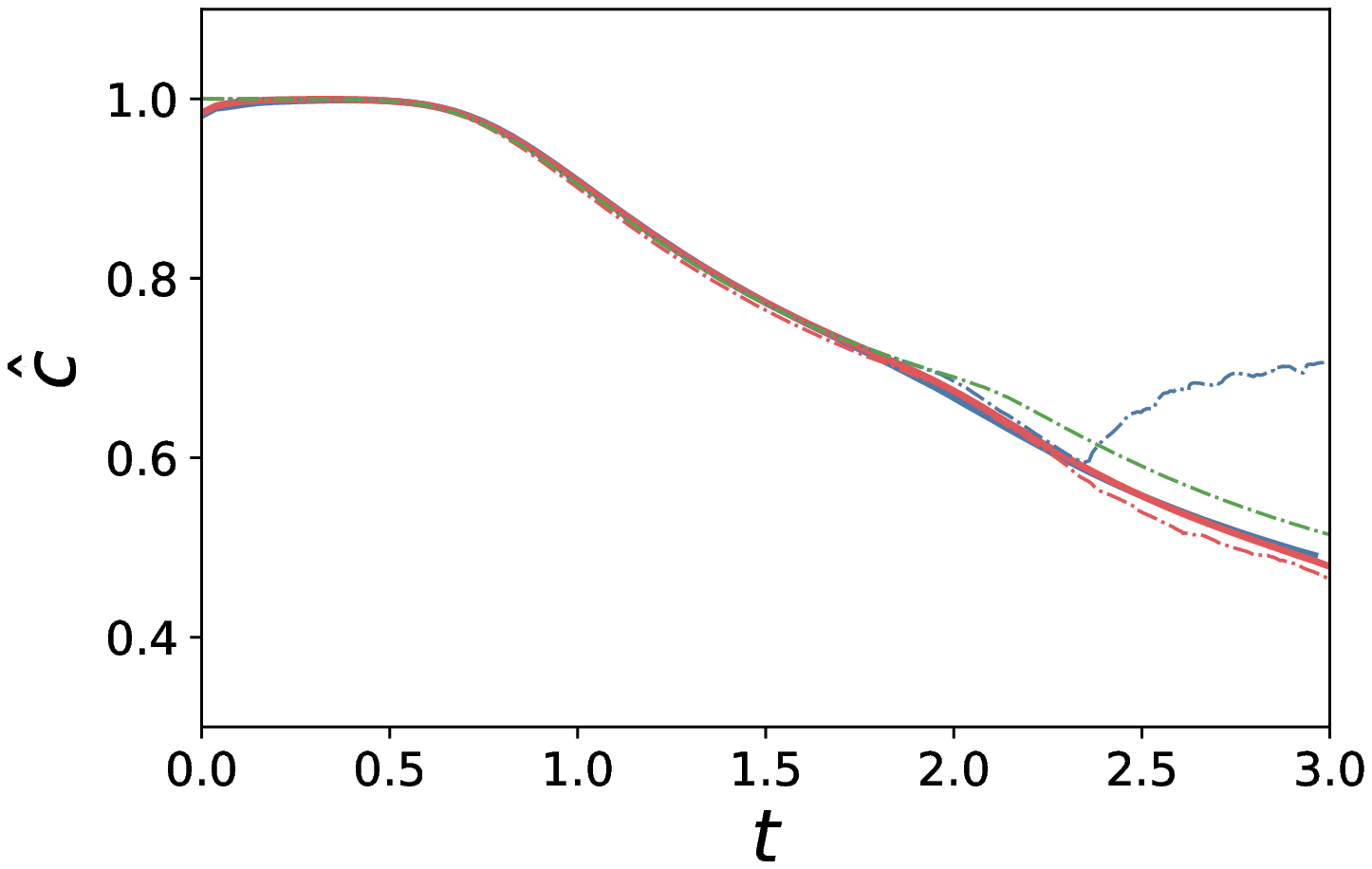}}
    \end{minipage}
\caption{Case 2. (a) Location of the $\phi = 0.5$ contour at $t = 3.0$ for the non-local model with grid spacing $\Delta x = 1/160$ compared with the front location from \cite{Hysing2009} for Group 1 (cell size $h = 1/320$), Group 2 (element mesh size $h = 1/160$), and Group 3 (900 degrees of freedom on the interface). (b-e) Comparison for Case 2 between the results of the non-local model with grid spacing $\Delta x = 1/80$ and $1/160$ and the results from \cite{Hysing2009} Group 1 (cell size $h = 1/320$), Group 2 (element mesh size $h = 1/160$), and Group 3 (900 degrees of freedom on the interface).  (b) Change in bubble area $A(t) - A(t = 0)$, (c) rise velocity $v_c$ given by eq. \ref{eq:vc}, (d) center of mass $y_c$ given by eq. \ref{eq:cm}, and (e) bubble circularity $\hat c$ given by eq. \ref{eq:hatc}. The parameters for Case 1 and 2 are given in table \ref{tab:Hysingparams}. (Color online.) 
}
\label{fig:benchmark2_front}
\end{figure}

A comparison of the front locations given by \cite{Hysing2009} and from the non-local model at time $t = 3.0$  for Case 1 and Case 2 is given in Figs. \ref{fig:benchmark1_front}a and \ref{fig:benchmark2_front}a. For Case 1 there is excellent agreement between the non-local model and the benchmark data from \cite{Hysing2009}, both in terms of the final bubble location and shape. Case 2 is computationally more challenging due to the topological changes that occur as the bubble breaks up to form satellite droplets. The results from \cite{Hysing2009} disagree on the point of breakup and the final bubble shape, as shown in Fig. \ref{fig:benchmark1_front}$c$.

Tables \ref{tab:Hysingerror} and \ref{tab:Hysingerror2} give values for Case 1 and Case 2 of $y_c(t=3)$, the maximum rise velocity $v_{c,max}$ and the time $t_{v_{c, max}}$ at which the maximum occurs, and the minimum circularity $\hat c_{min}$ and the time $t_{\hat c_{min}}$ at which the minimum occurs for the range of values presented by the three groups in \cite{Hysing2009} and the non-local model. Our results show that  the non-local model converges to the benchmark values as its resolution increases. For Case I is an excellent agreement between the non-local models and the results from \cite{Hysing2009}, with a relative error for $\Delta x  = 1/320$ of $0.02\%$ for $y_c(t = 3.0)$, $0.04\%$ for ${v}_{c, max}$, $0.96\%$ for $t_{{v}_{c, max}}$, $0.488\%$ for $\hat c_{min}$, and $1.19\%$ for $t_{\hat c_{min}}$. In each case the error is less than or on the same order of magnitude as the errors found by \cite{Strubelj2009} and \cite{Klostermann2013}. The inset of Fig. \ref{fig:benchmark1_front}$d$ shows that the final center of mass in the vertical direction $y_c$ for the non-local model with $\Delta x= 1/320$ is bounded by the benchmark values from \cite{Hysing2009}. For Case 2 the final bubble shape after breakup is inconclusive, however Table \ref{tab:Hysingerror2} shows good agreement between the non-local model and \cite{Hysing2009}. Due to computational limitations, we are unable to run higher resolution simulations of Case 2, which may improve the agreement between \cite{Hysing2009} and the non-local model.

While the numerical method used for advecting the color function $\phi$ is conservative, the bubble area is calculated as the contour corresponding to $\phi = 0.5$, and the area inside this contour is not necessarily conserved. Nevertheless, we see good conservation of the bubble area to within $0.226\% $ for  Case 1 with $\Delta x = 1/ 80$, $0.077\%$ for $\Delta x = 1/ 160$, and $0.027\%$ for $\Delta x = 1/ 320$. The change in area for Case 2 is $0.283\%$ for $\Delta x = 1/ 80$ and $0.116\%$ for $\Delta x = 1/ 160$. The change in the bubble area is plotted in Figs. \ref{fig:benchmark1_front}$b$ and \ref{fig:benchmark2_front}$b$ for Case 1 and Case 2, respectively. 

Figs. \ref{fig:benchmark1_front}$e$ and \ref{fig:benchmark2_front}$e$ show that the initial circularity is slightly under the expected value of one for a circular droplet at the beginning of the non-local simulations. This occurs because the droplet originally is fit to the square mesh. Over the first several time steps the compression algorithm for the advection of the color function rounds the edges of the droplet increasing the circularity to one.

\begin{table}
\centering
\caption{Comparison between benchmark values from the local models in \cite{Hysing2009} and the non-local model with $\Delta x = 1/80, 1/160,$ and $1/320$. The definitions for the center of mass $y_c$, rise velocity ${v}_{c}$, and circularity $\hat c$ are given in eqs. \ref{eq:cm}--\ref{eq:hatc}. }
\begin{tabular}{l c c  c  c  }
\hline 
&  $\Delta x = \frac{1}{80}$ & $\Delta x = \frac{1}{160}$ &  $\Delta x = \frac{1}{320}$ & \cite{Hysing2009}  \\  %\cite{Hysing2009} \\ 
\hline 
$y_c(t = 3.0)$     &   1.0863    &  1.0824  & 1.0807    & 1.081 $\pm$ 0.001  \\
$v_{c, max}$       &   0.2438    &  0.2426    & 0.2418  & 0.2419 $\pm$ 0.0002 \\ 
$t_{v_{c, max}}$   &   0.9199    &  0.9030  &  0.9174 & 0.9263 $\pm$ 0.005 \\
$\hat c_{min}$     &   0.8961    &  0.8967    & 0.8968 & 0.9012 $\pm$ 0.0001 \\ 
$t_{\hat c_{min}}$ &   1.9206    &  1.9314   &1.9125  & 1.89 $\pm$ 0.015   \\ 
\hline 
\end{tabular}
\label{tab:Hysingerror}
\end{table}

%Hysing = [1.081, 0.02419, 0.9263, 0.9012,1.89]
%Strubelj = [1.0679, 0.2457, 0.9235, 0.8876, 1.8915]
 %Kloster = [1.0696, 0.2348, 0.9516, 0.9044,1.9625]
% NL== [1.0807, 0.2418, 0.9174, 0.8968, 1.9125]

\begin{table}
\centering
\caption{Comparison between  values from \cite{Hysing2009} and the non-local model with $\Delta x = 1/80$ and $1/160$. The definitions for the center of mass $y_c$ and the rise velocity ${v}_{c}$ are given in eqs. \ref{eq:cm} and \ref{eq:vc}. }
\begin{tabular}{c c c  c  }
\hline 
&  $\Delta x = \frac{1}{80}$ & $\Delta x = \frac{1}{160}$  & \cite{Hysing2009}  \\  %\cite{Hysing2009} \\
\hline 
$y_c(t = 3.0)$     &   1.102   &  1.119   & 1.134 $\pm$ 0.009  \\
$v_{c,max}$       &   0.250    &  0.250    & 0.252 $\pm$ 0.002 \\ 
$t_{v_{c, max}}$   &   0.674    &  0.750 & 0.731$\pm$ 0.003  \\ 
\hline 
\end{tabular}
\label{tab:Hysingerror2}
\end{table}

\subsection{Circular bubble in shear flow}
\label{sec:shear}

A ``macroscopic'' spherical bubble in a linear shear flow $\mathbf{u} = (\dot \gamma y, 0)$ will form a steady elongated ellipsoidal shape. Denoting the major and minor axes of the ellipse as $l$ and $b$, the deformation ratio $D = \frac{l - b}{l + b}$ of an initially spherical bubble with radius $a$ was found by \cite{Taylor1932, Taylor1934} using the local model as
\begin{equation}
D  = \Ca \frac{(19 \lambda + 16)}{(16 \lambda + 16)}
\end{equation}
where $\lambda = \mu_d/\mu_f$ is the viscosity ratio between the drop and the surrounding fluid and $\Ca = \mu_f a \dot \gamma / \sigma_0$ is the capillary number, the ratio of the magnitude of the viscous forces to the capillary forces. This formula has been validated for periodic suspensions of two-dimensional macroscopic droplets in a channel \cite{Zhou1993}, and corrections have been made for macroscopic bubbles in confinement \cite{Ioannou2016}. The three-dimensional results of \cite{Taylor1934} have been shown to agree well with two-dimensional simulations for small and moderate capillary number $\Ca < 1$ \cite{Zhou1993, Guido1998}. The results of \cite{Taylor1932, Taylor1934} have been extended to second order \cite{Guido2011, Chaffey1967} to describe the angle the major axis of the deformed elliptical bubble makes with the horizontal axis, called the orientation angle  and denoted by $\theta$, to give
%update with correct citation here
\begin{equation}
    \theta = \frac{\pi}{4} - \Ca \frac{(19 \lambda + 16)(2 \lambda + 3)}{80( \lambda + 1)}. 
\end{equation}

\begin{figure}
\centering
  \sidesubfloat[]{
    \includegraphics[width=0.4\linewidth]{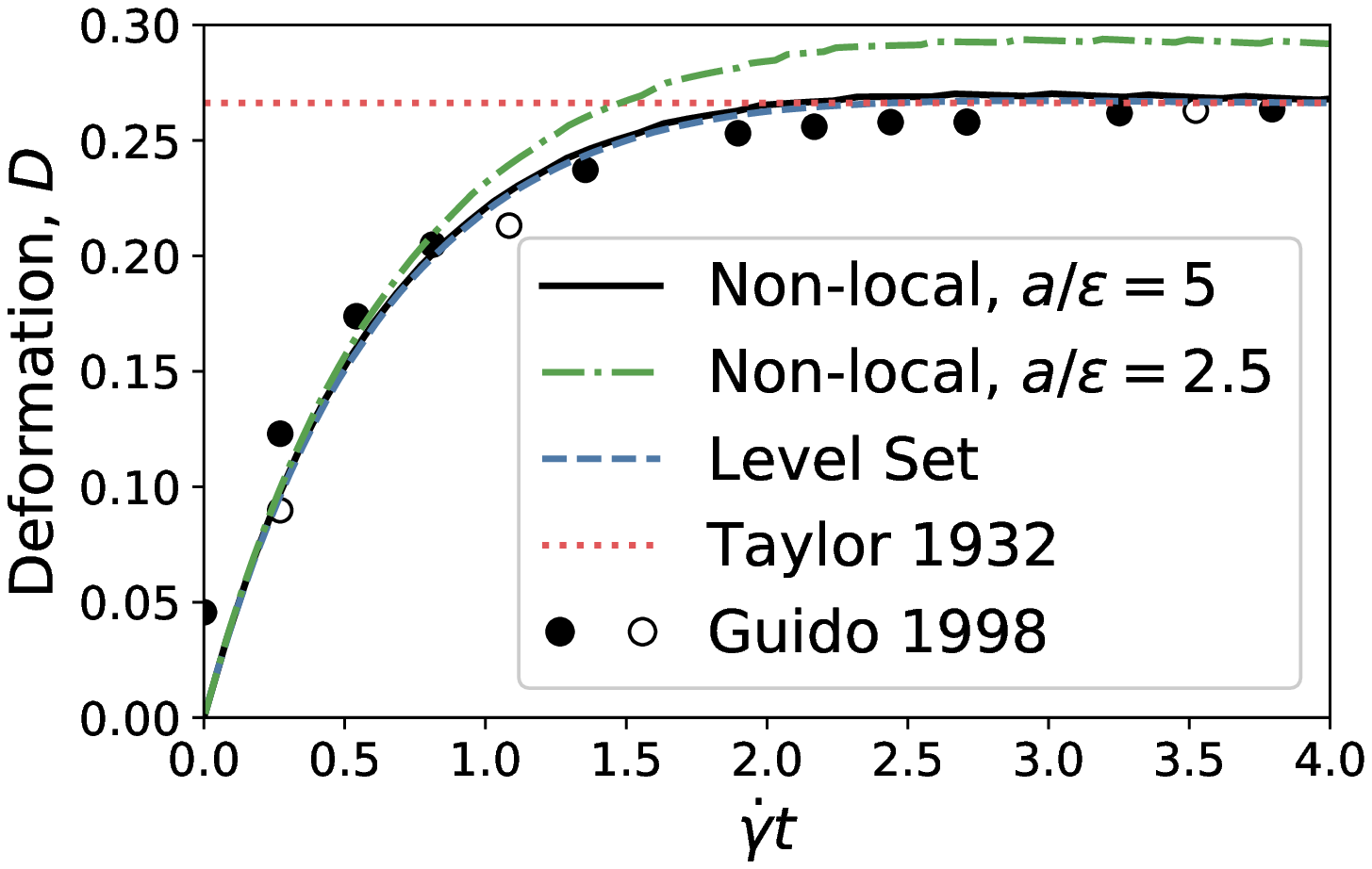}}
  \sidesubfloat[]{
    \includegraphics[width=0.4\linewidth]{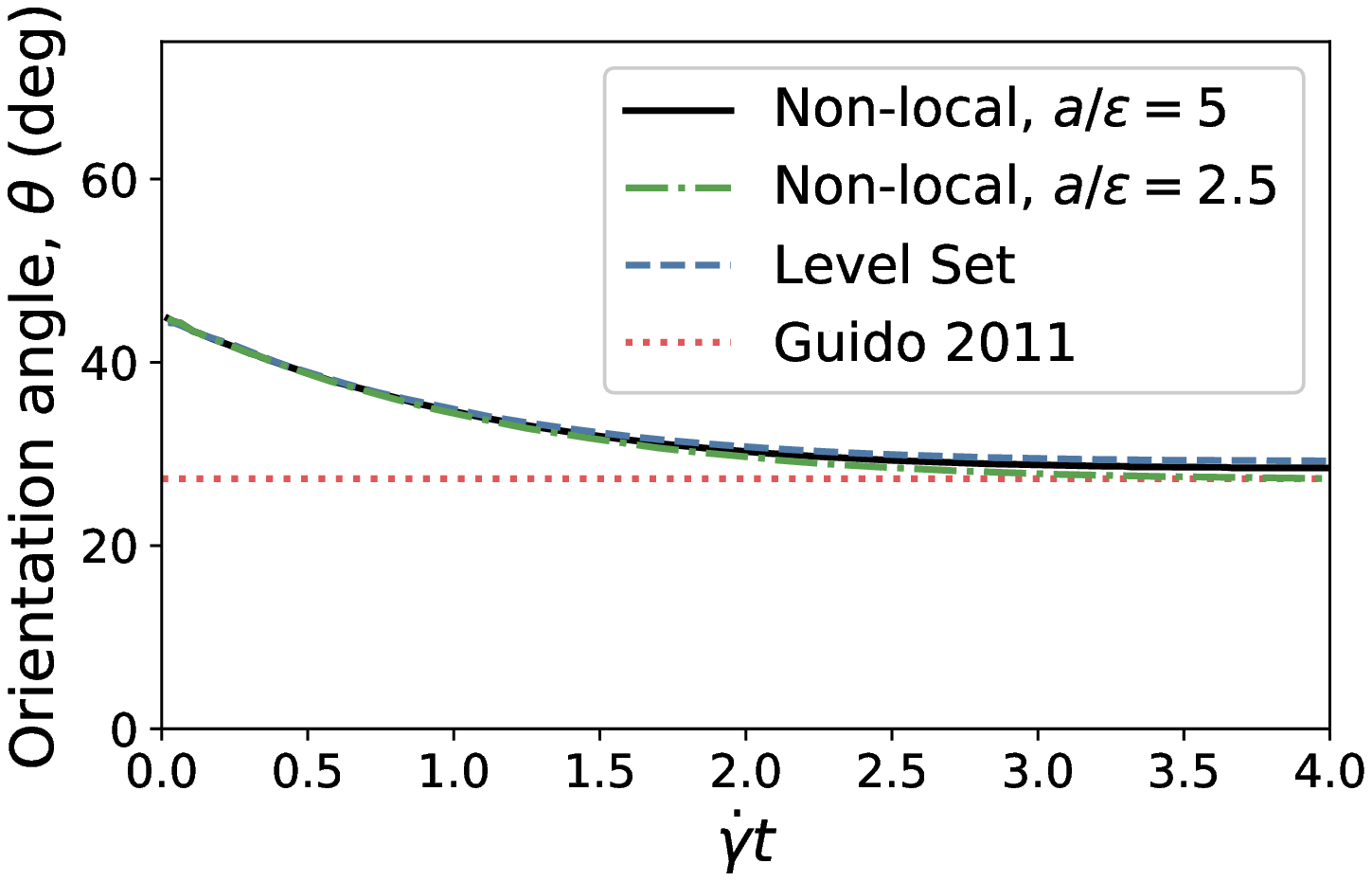} }  \\  
  \sidesubfloat[]{
 \includegraphics[width=0.45\linewidth]{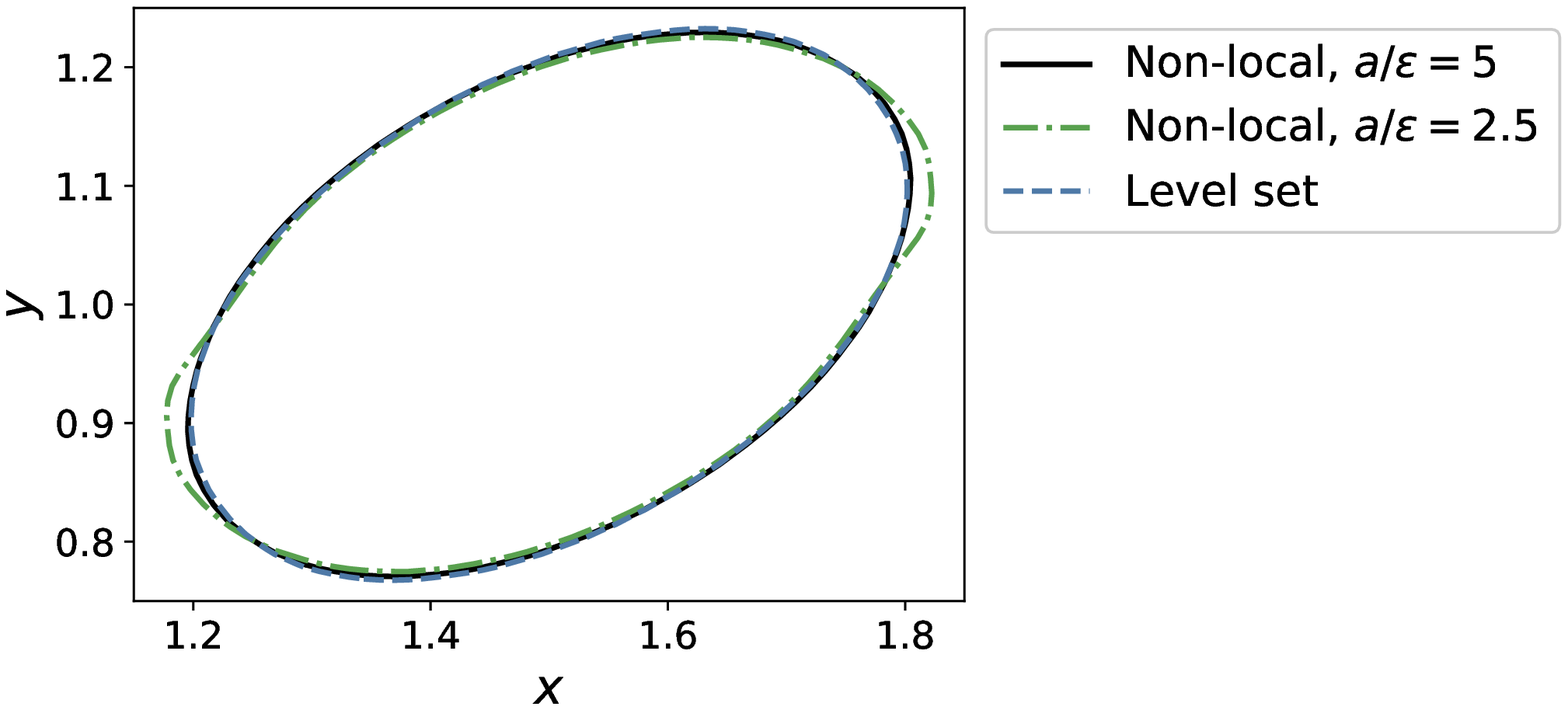}}
\caption{(a) Bubble deformation for a droplet with $\Ca = 0.24$ and $\lambda = 1.4$. Results are compared to the steady state results from \cite{Taylor1932} and experimental results from \cite{Guido1998} (open and closed symbols correspond to different digitization settings \cite{Guido1998}). {(b) Bubble orientation angle, $\theta$, compared with the analytical solution in \cite{Guido2011}. } (c) Location of the $\phi = 0.5$ contour for the non-local and CLS methods. The non-dimensional strain $\gamma$ is defined as $\gamma = \dot \gamma t$. }
\label{fig:Shear_D}
\end{figure}

We simulate a macroscale bubble ($a/\varepsilon = 5$) in shear flow with the non-local and local (CLS) models for $\Ca = 0.24$ and $\lambda = 1.4$ to provide a comparison to the experiments of \cite{Guido1998}. Results for the bubble deformation, $D$,  and the location of the fronts are given in Fig. \ref{fig:Shear_D}. At times $0 \leq t \leq 4.0$, the non-local and local models agree well with the analytical results from \cite{Taylor1932, Taylor1934} and the experimental results \cite{Guido1998}. The orientation angle $\theta$ agrees as well with the second-order analytical results from \cite{Guido2011}. The relative error between the exact deformation, $D_{exact} = 0.26625$, and the non-local model with $a/\varepsilon = 5.0$ is $0.74\%$, and the relative error for the local model is $0.39\%$. The error between the analytical value for the orientation angle $\theta_{exact}$ and the non-local model is $4.13\%$, compared with $7.15\%$ for the CLS method. 

Next, we model a ``microscopic'' bubble with the initial radius $a/\varepsilon = 2.5$,  $\Ca = 0.24$, and $\lambda = 1.4$, the same $\Ca$ and $\lambda$ values as in the simulations of the macroscopic bubble. We observe that the deformation of the microscopic bubble is significantly larger than that of the macroscopic bubble and the resulting bubble has a pronounced sigmoid shape, see Fig. \ref{fig:Shear_D}$C$. In this case, the curvature of the front is significantly lower at the rounded corners of the droplet, which in turn decreases the force due to the surface tension at those points. The droplet then continues to elongate, coming into equilibrium with a higher deformation than for macroscale droplets. Mesoscale Dissipative Particle Dynamics (DPD) simulations of a spherical microscale droplet in shear flow also predicted a larger droplet deformation for capillary number $\Ca > 0.2$ than that predicted by the Taylor theory \cite{Chen2004, Pan2014}.  
%\cite{Pan2014} attributes this difference is attributed to a higher particle Reynolds number, however, we could not determine the Reynolds number from \cite{Chen2004}. 
The deformation values found in these DPD studies correspond well with the deformation in the non-local model for microscale bubble. Note that the local model, the behavior depends only on $\Ca$ and $\lambda$ and not on the size of the bubble.   {We examined the droplet circulation for each case and found very similar results for all simulations in Fig. \ref{fig:Shear_D}.}
%\cite{Li2005} used MD to simulate nanobubbles with comparable capillary numbers to those considered here, and showed that ``clean'' nanobubbles without surfactants have finite deformation and do not extend infinitely. 

\subsection{Droplet collisions in shear flow}

\begin{figure}
%\centering
    \subfloat[][$\Ca = 0.15$: Coalescence]{ 
    \includegraphics[width=0.25\linewidth]{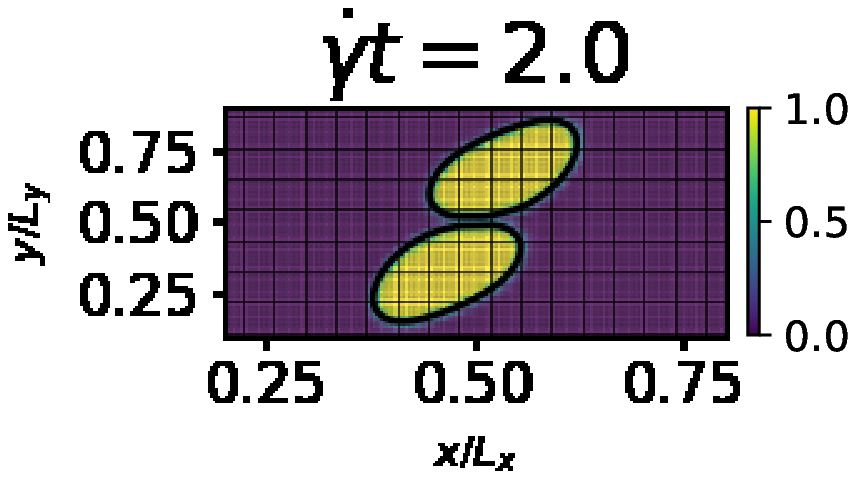}
    \includegraphics[width=0.25\linewidth]{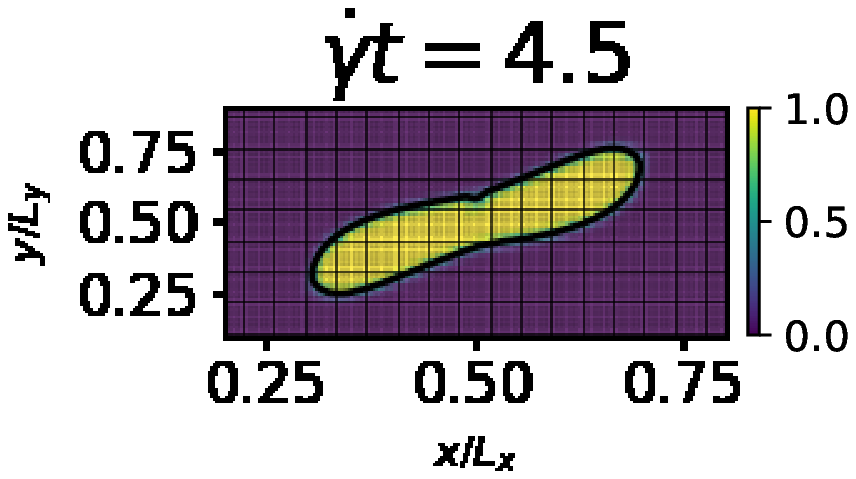}
    \includegraphics[width=0.25\linewidth]{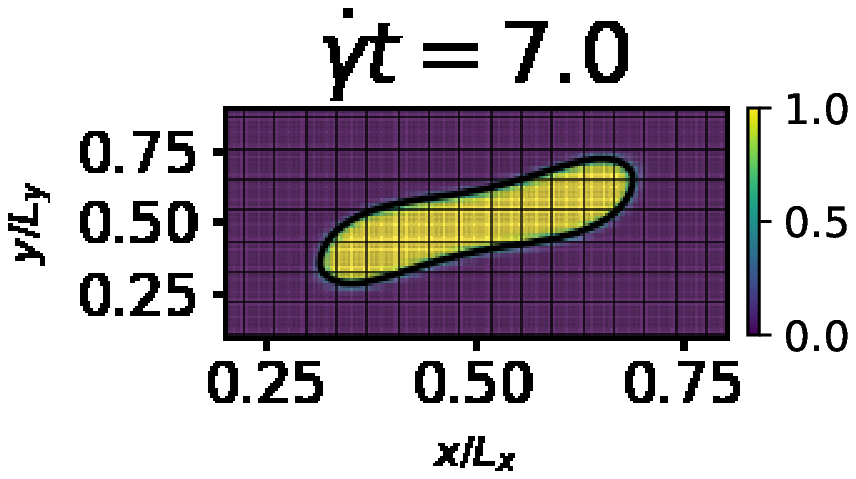}    }\\
  \subfloat[][$\Ca = 0.25$: Temporary bridge] {
    \includegraphics[width=0.25\linewidth]{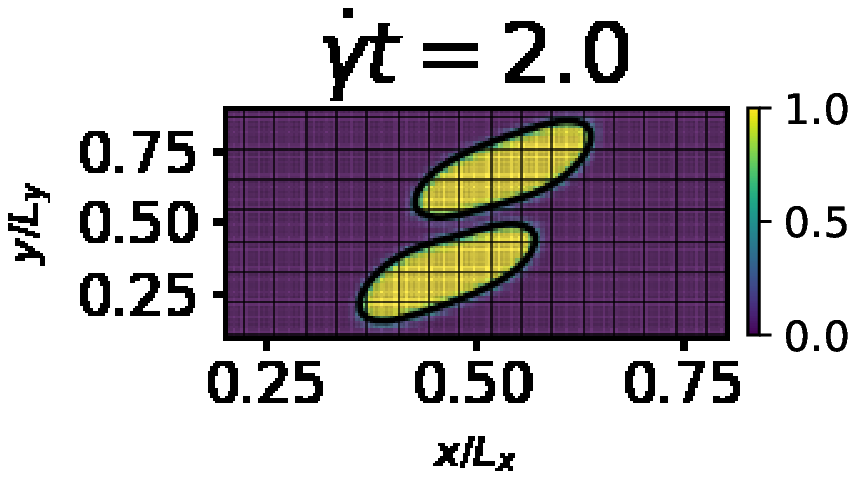}
    \includegraphics[width=0.25\linewidth]{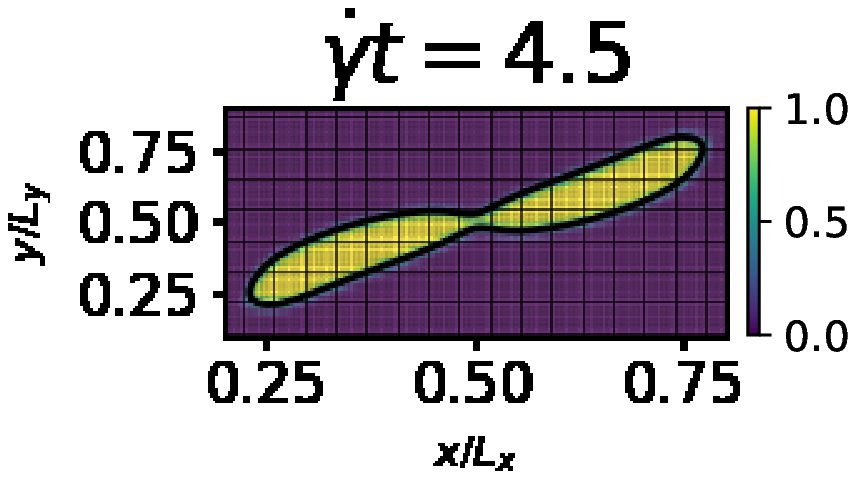}
    \includegraphics[width=0.25\linewidth]{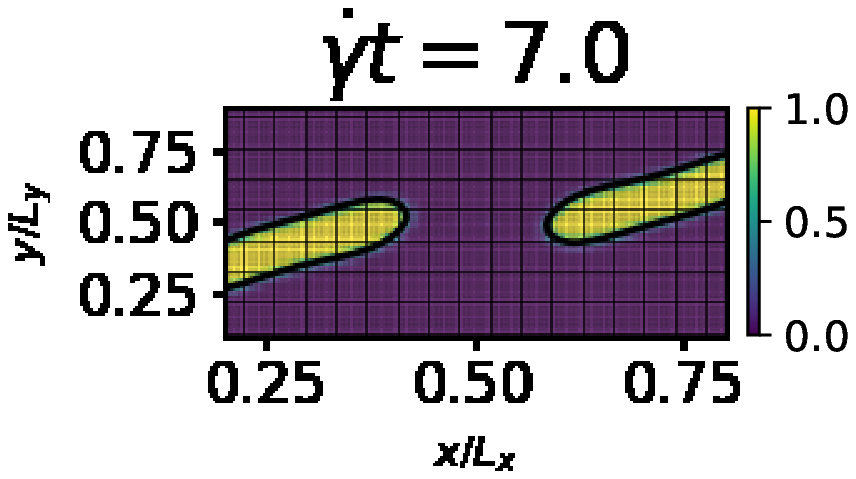}}\\
  \subfloat[][$\Ca = 0.35$: Sliding]{
    \includegraphics[width=0.25\linewidth]{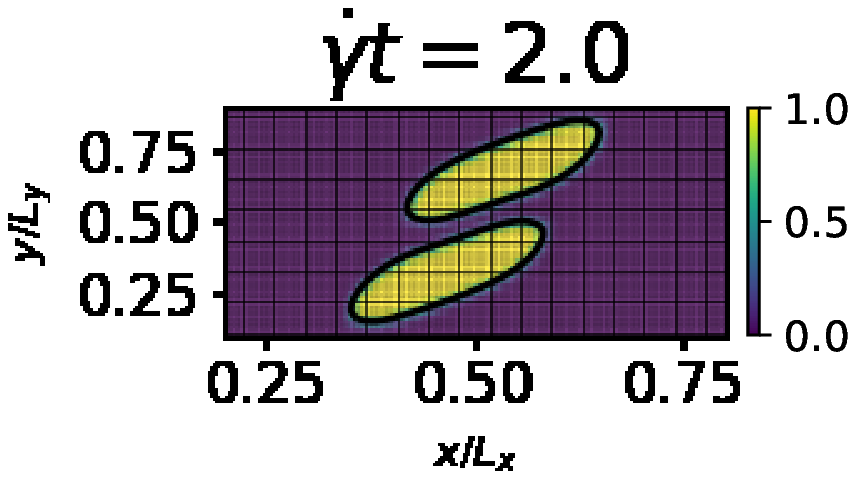}
    \includegraphics[width=0.25\linewidth]{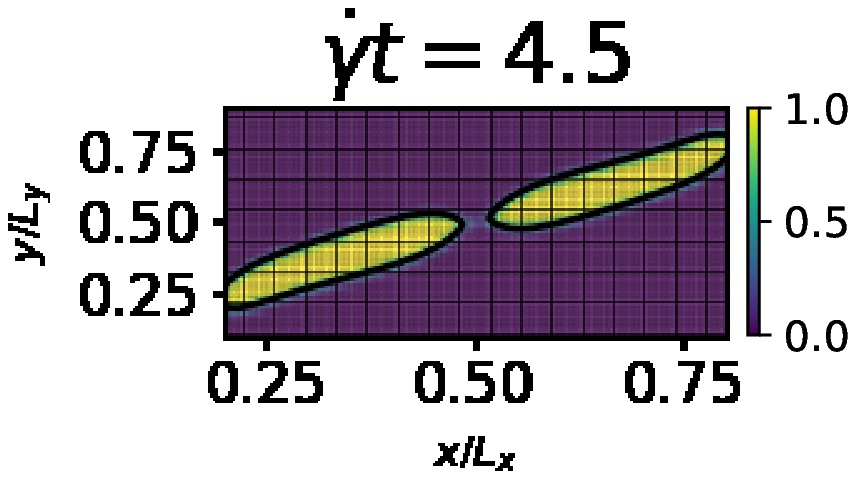}
    \includegraphics[width=0.25\linewidth]{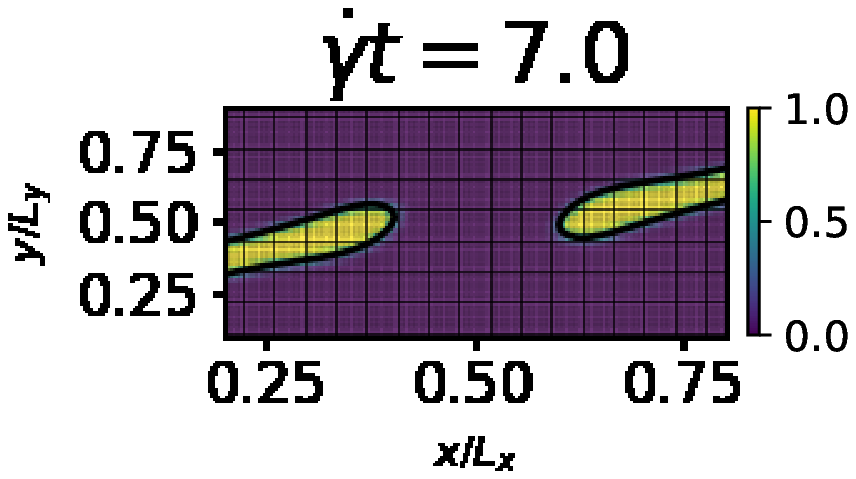}}
\caption{Example front locations illustrating the three typical trajectories of two droplets in shear flow. (a) Coalescence occurs for $\Ca = 0.15$. (b) A temporary bridge forms for $\Ca = 0.25$. (c) No coalescence occurs for $\Ca = 0.35$, and the droplets slide past each other. All simulations have resolution $\Delta x = 1/80$, $a = 0.15$, and $\varepsilon = 0.025$, which means $a/\varepsilon = 6$ and $\varepsilon/\Delta x = 2$. (Movie S1). (Color online.) }
\label{fig:two_shear_sample_traj}
\end{figure}

In this section we use the non-local model and the local CLS model to simulate the collision of two equal size droplets in shear flow. 
%This problem is computationally challenging because accurately capturing bubble coalescence requires both flexibility in modeling the bubble interface as well as the ability to capture short range interactions, on the order of nanometers, that occur between colliding bubbles \cite{Shardt2013}. 
This system has been studied previously, both experimentally and, more recently, numerically. Results have shown that the droplet behavior falls into one of three regimes: at low capillary number the droplets coalesce to form one large droplet, and at higher capillary number the droplets slide past each other \cite{Guido1998a, Chen2009, Shardt2013}. In some cases, particularly with small droplets, a third regime is possible at moderate capillary number where the droplets temporarily coalesce before breaking apart \cite{Shardt2013}. These three regimes are illustrated in Fig. \ref{fig:two_shear_sample_traj}, simulated with the non-local model for $a = 0.15$ at three capillary numbers.
% chosen to illustrate the three different regimes. 
 At the lowest capillary number, the droplets coalesce, forming one large droplet. At long times, this large coalesced droplet will find a final shape as described for a single droplet in shear flow in sec. \ref{sec:shear}. As the capillary number increases there is a transitional regime where the droplets coalesce and then separate, known as temporary bridging. Finally, at high enough capillary number the droplets slide past each other without coalescence. 
In \cite{Shardt2013},  all three regimes were simulated using a three-dimensional free-energy lattice Boltzmann method (LBM). The LBM offers an advantage over many front capturing models, including the Volume of Fluid (VoF) method and CLS method, where coalescence of droplets will depend on the grid resolution. As the grid is further refined, the film between the droplets will be better resolved, which increases the time needed for the film to fully drain and delays coalescence. \cite{Shardt2013} shows the LBM does offer grid independent results, but the behavior depends on the thickness of the diffuse droplet interface relative to the droplet size. 

Here, we choose parameters to match the results in \cite{Shardt2013}, including the Reynolds number, $Re = 1$. In comparison, the Reynolds number in the experiments in \cite{Chen2009} is  $Re < 10^{-7}.$ Further studies are needed to provide a detailed understanding of how the coalescence behavior may depend on the Reynolds number. In our simulations, the droplets have initial separation in both the $x$ and $y$ directions given by $| x_1-x_2| / a = 2.52$ and $| y_1-y_2| / a = 0.78$, where $(x_1, y_1)$ and $(x_2, y_2)$ are the centers of the two droplets. The relative channel width is $H/a =  5.1282$. The droplets have the same density and viscosity as the surrounding fluid, with $\rho = 6\mu$. We find that the transitions between the three regimes occur at capillary numbers of the same order of magnitude as presented by \cite{Shardt2013}, although we do not expect exact agreement due to the differences between two-dimensional and three-dimensional simulations. 

\begin{figure}
%\centering
   \subfloat[][CLS method, $\kappa = \Delta x/2$]{
    \includegraphics[width=0.17\linewidth]{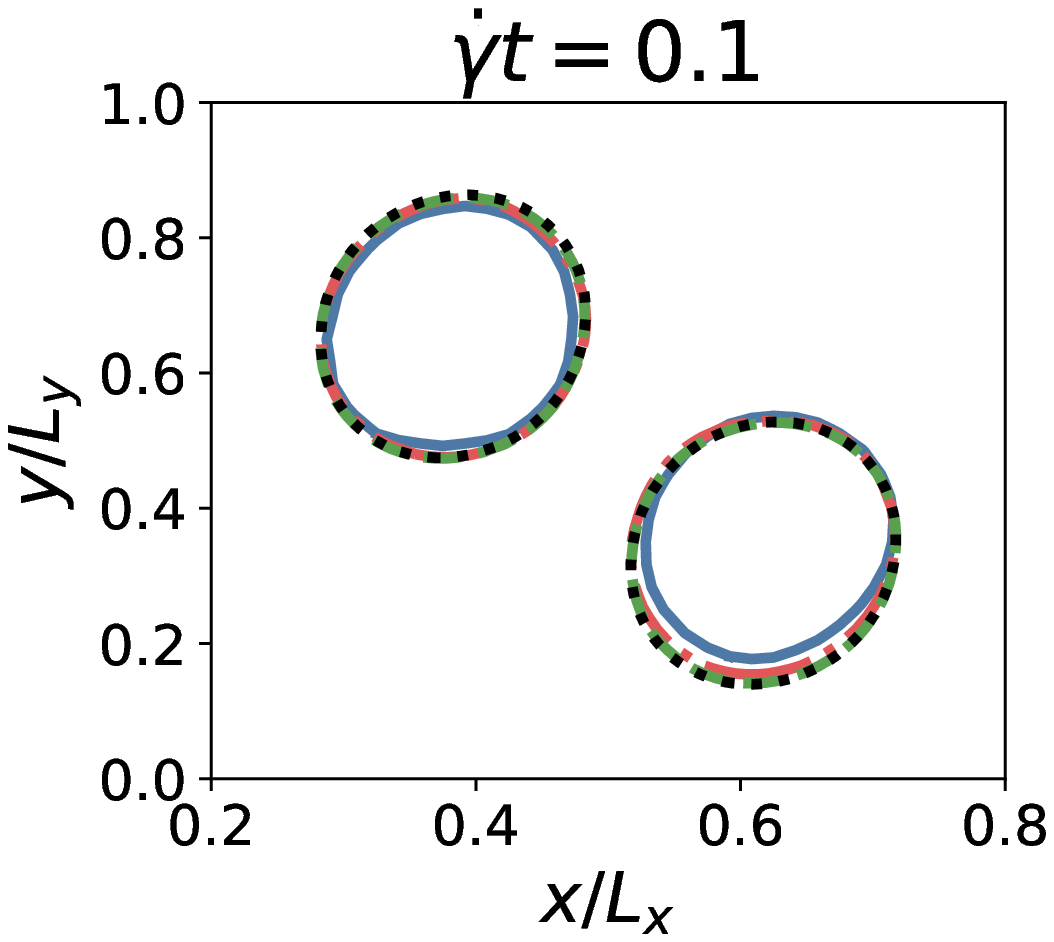}
    \includegraphics[width=0.17\linewidth]{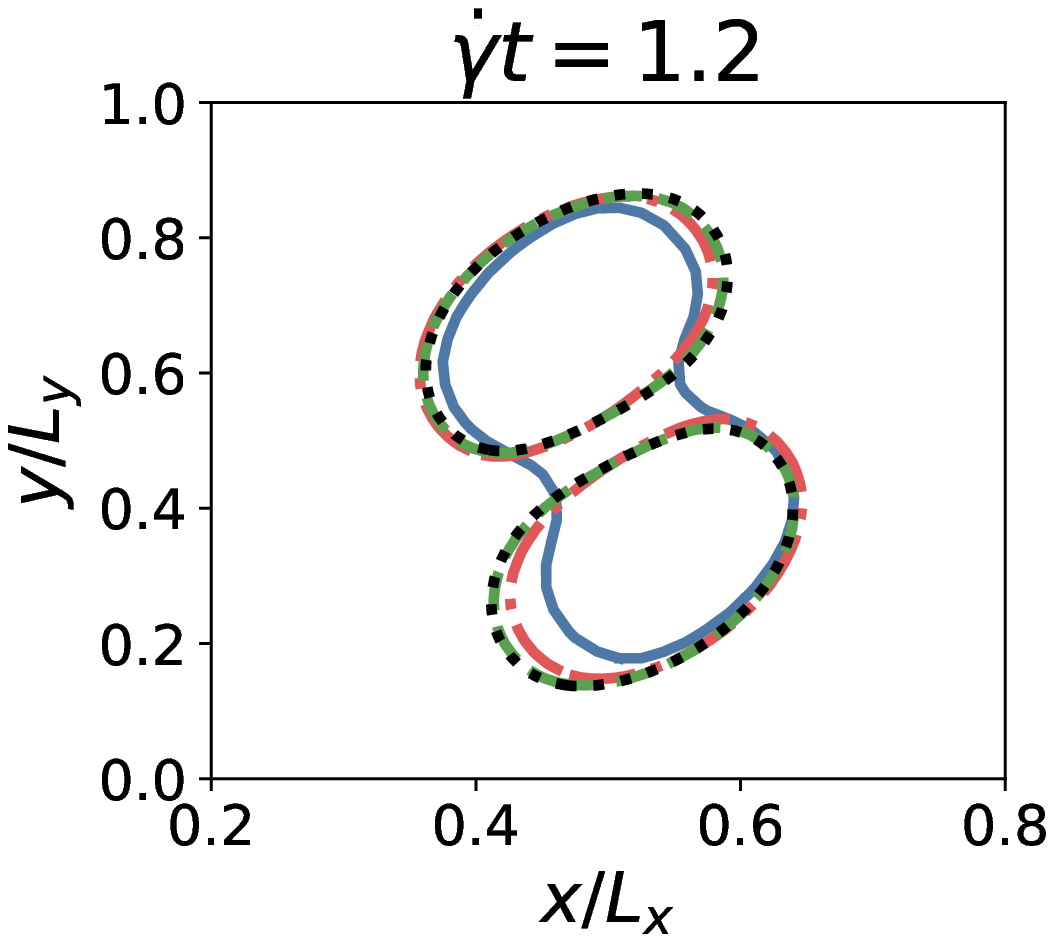}
    \includegraphics[width=0.17\linewidth]{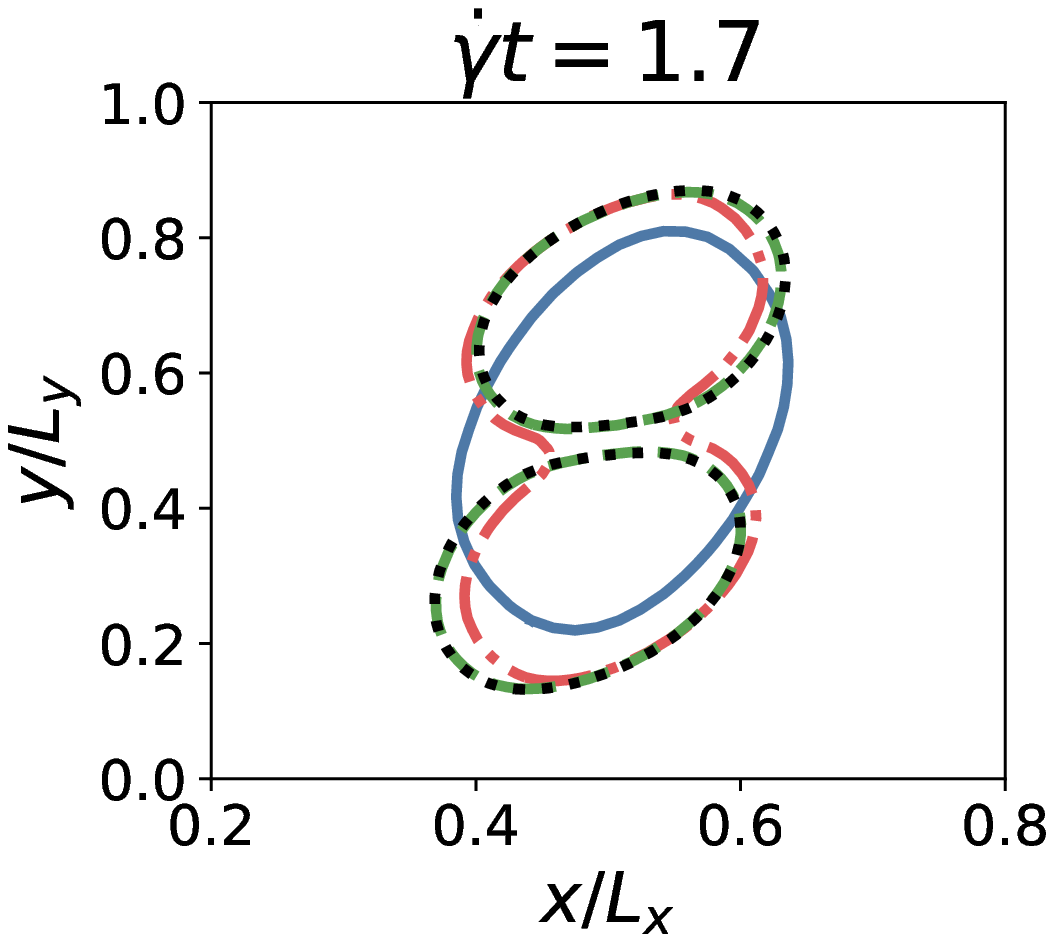}
    \includegraphics[width=0.17\linewidth]{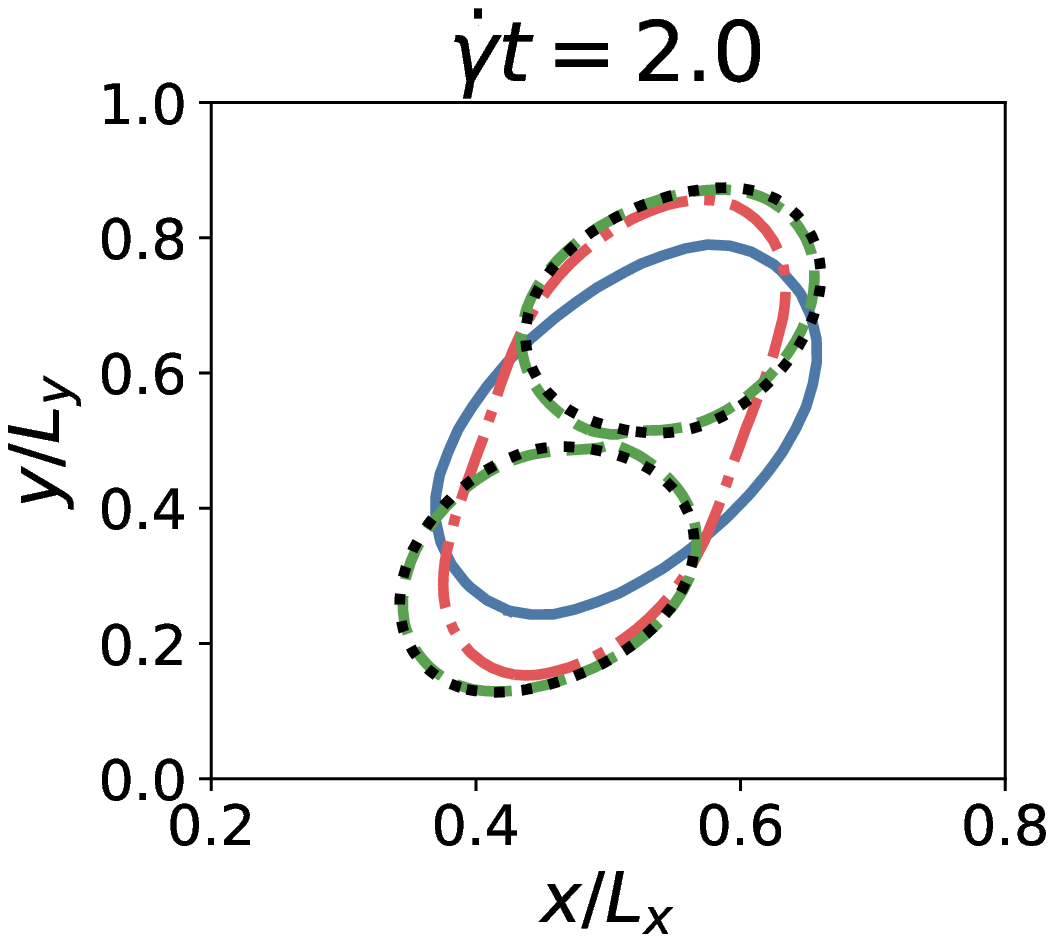}
    \includegraphics[width=0.17\linewidth]{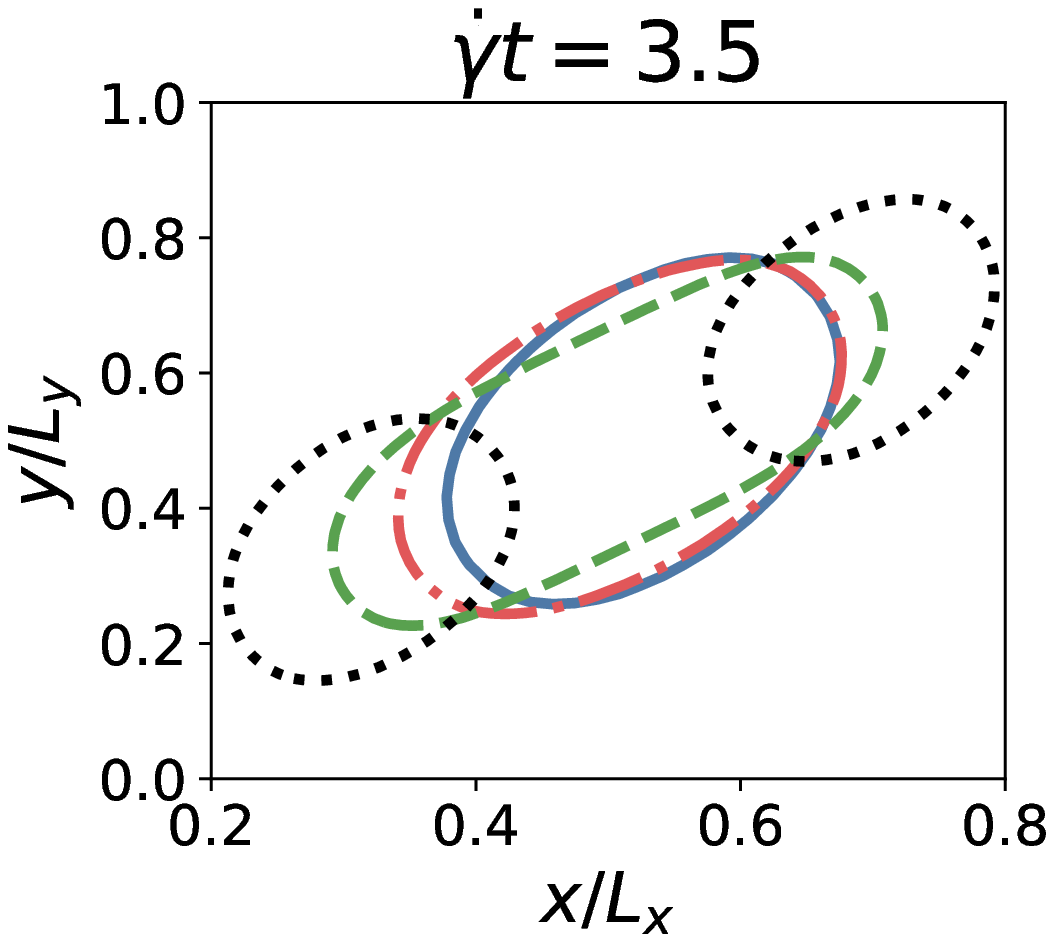}
    \includegraphics[width=0.12\linewidth]{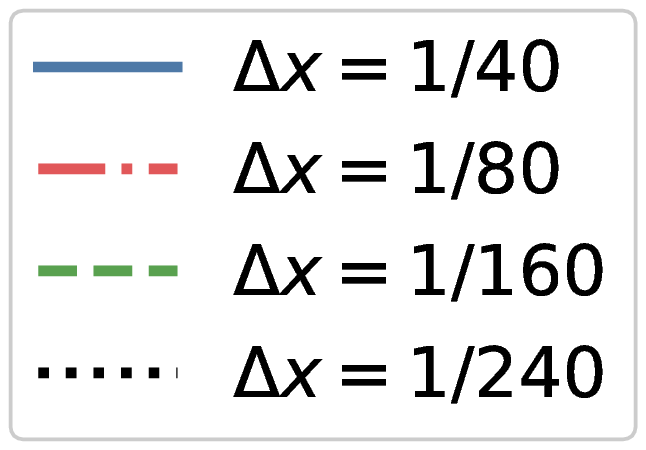}}\\
    \subfloat[][{CLS method, $\kappa = 0.0125$}]{
    \includegraphics[width=0.17\linewidth]{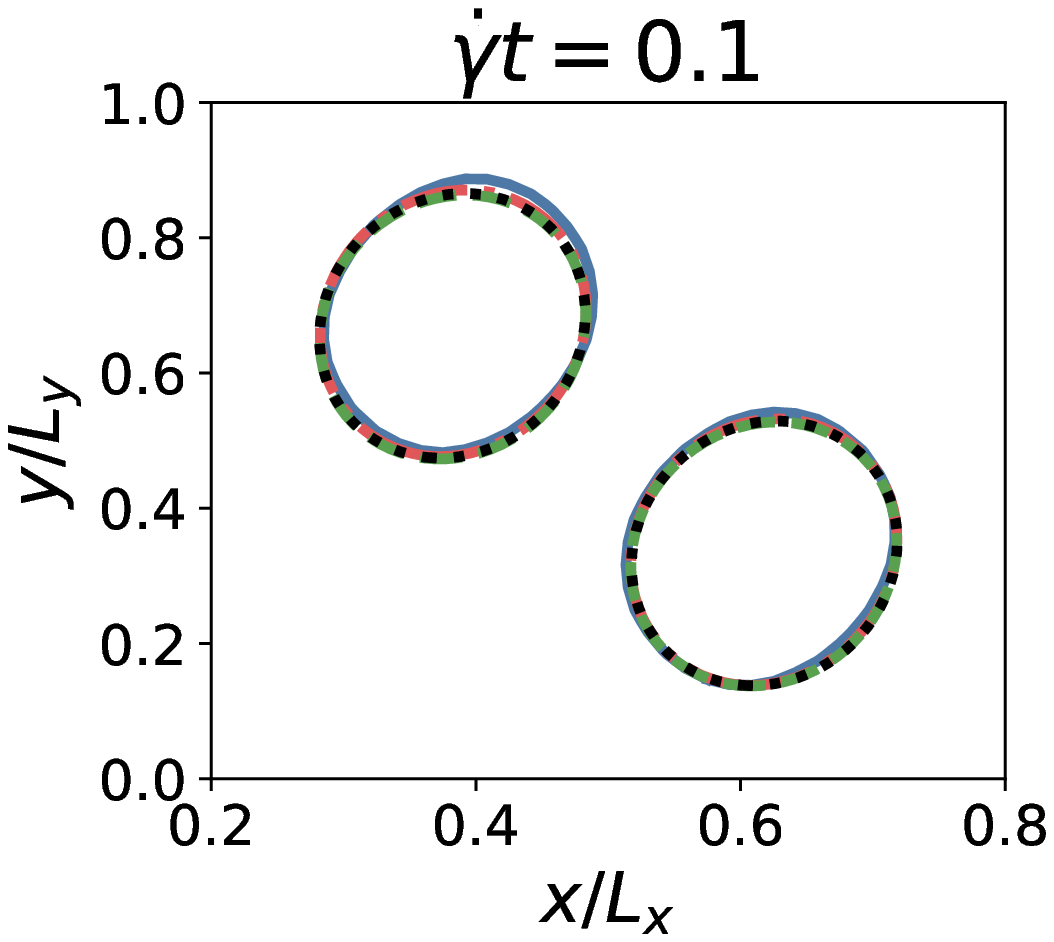}
    \includegraphics[width=0.17\linewidth]{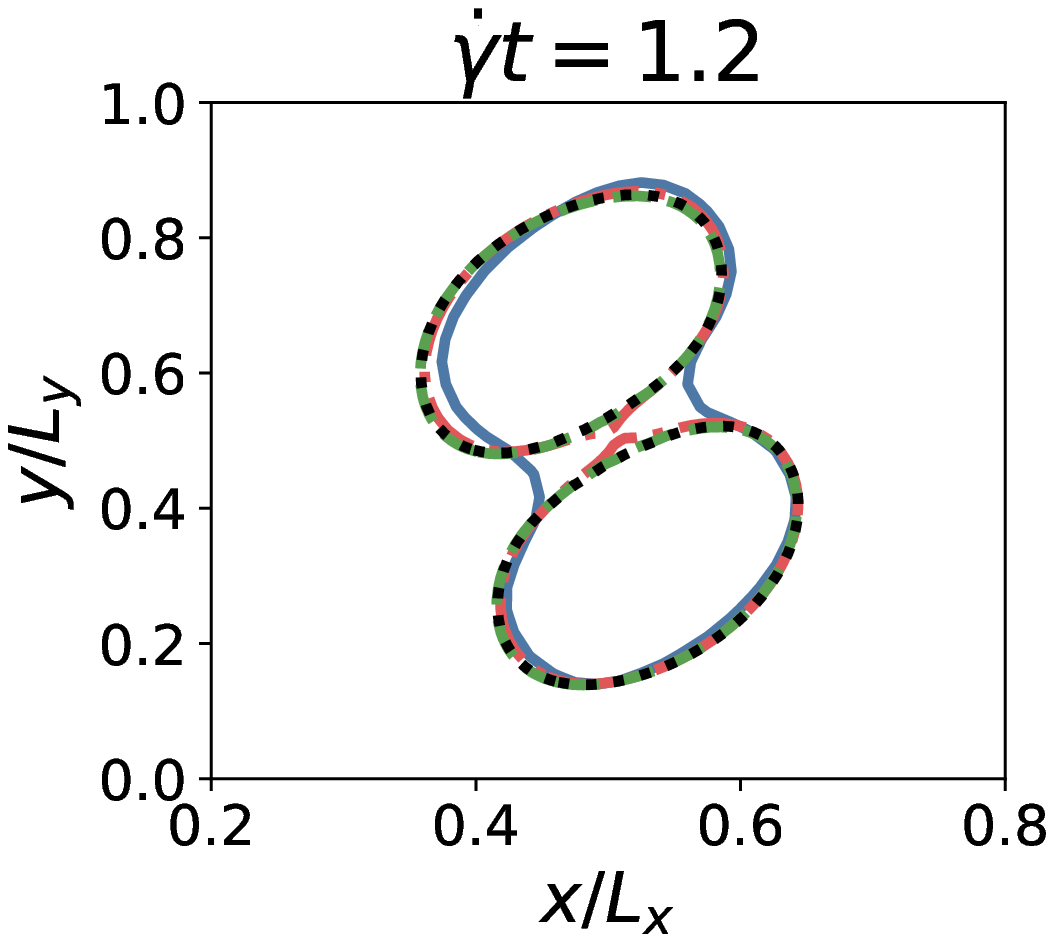}
    \includegraphics[width=0.17\linewidth]{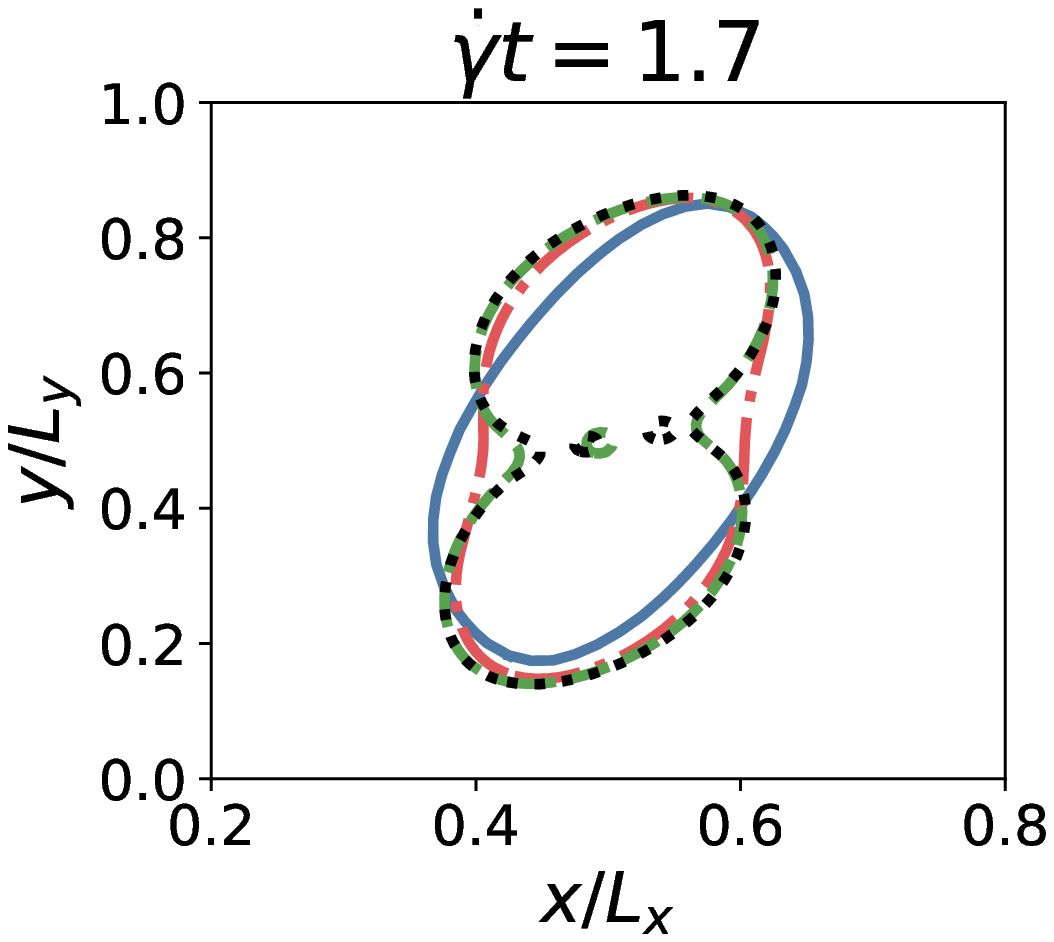}
    \includegraphics[width=0.17\linewidth]{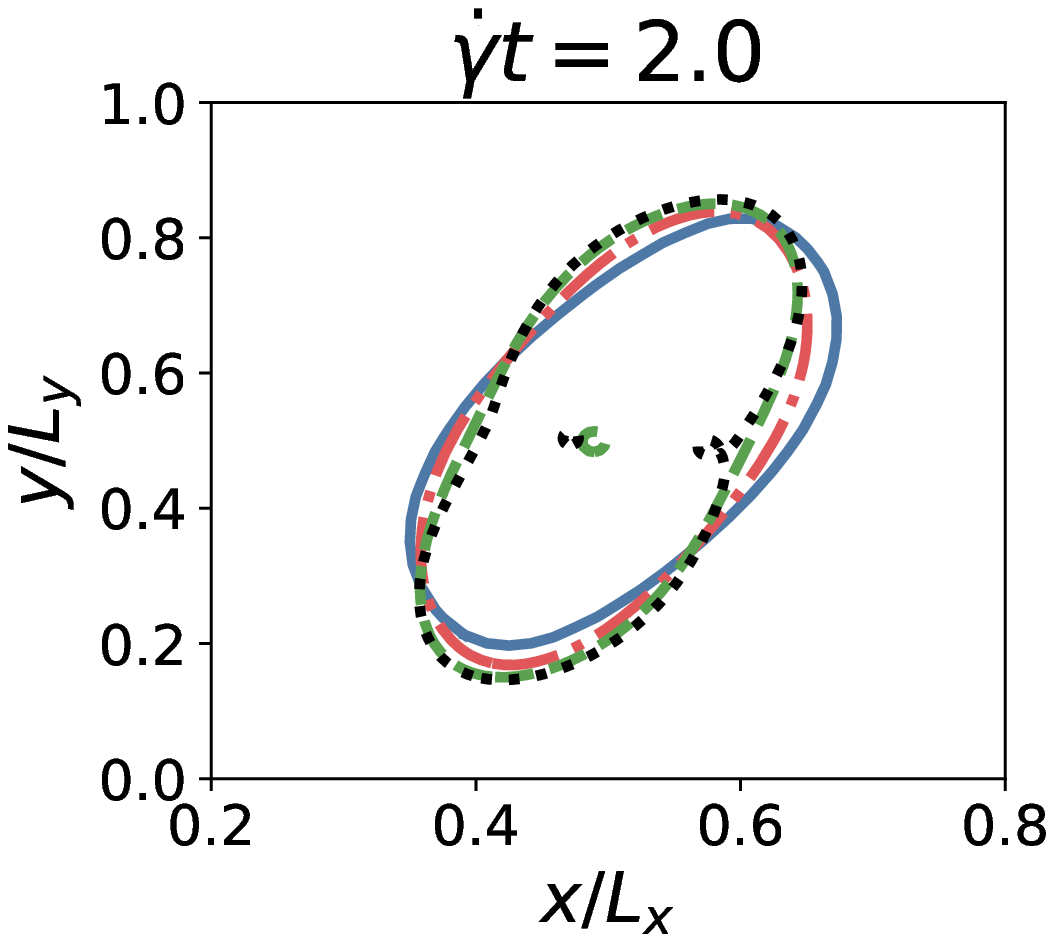}
    \includegraphics[width=0.17\linewidth]{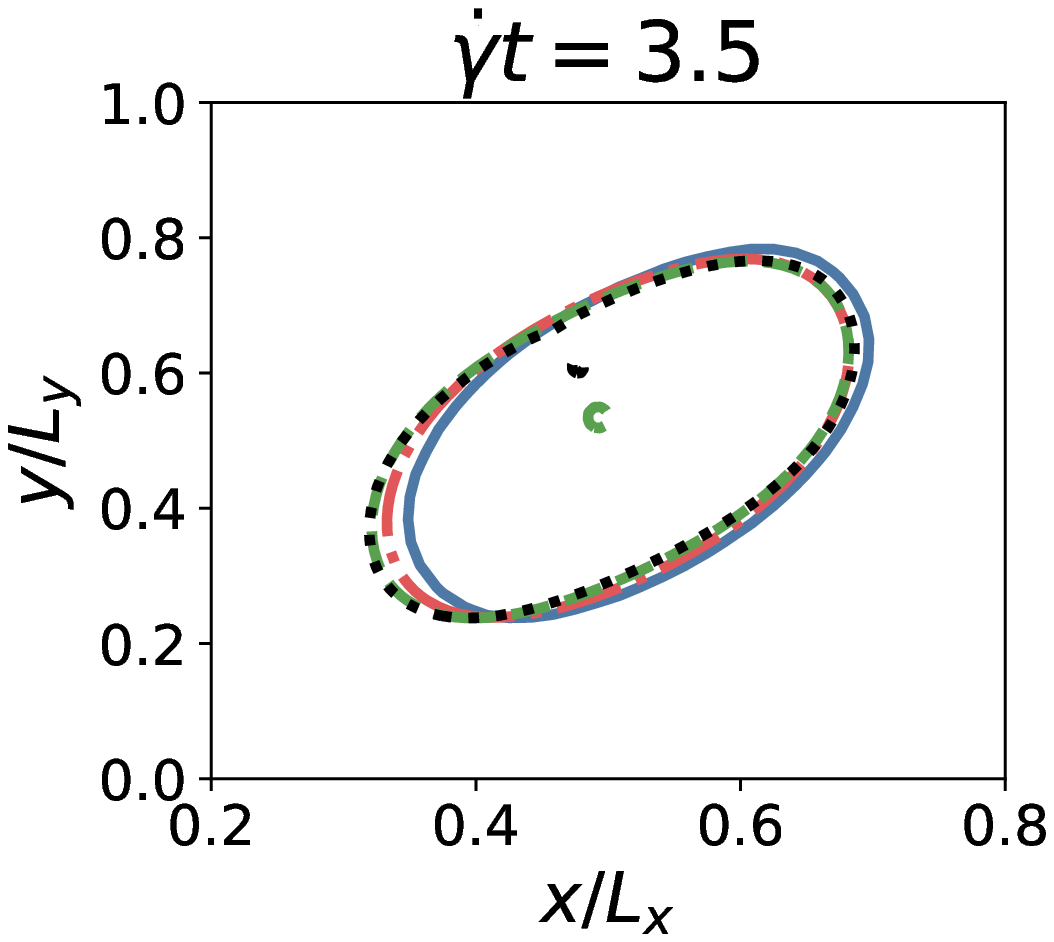}
    \includegraphics[width=0.12\linewidth]{Figures2/Figure9clegend.eps}}\\
   \subfloat[][Non-local method]{
    \includegraphics[width=0.17\linewidth]{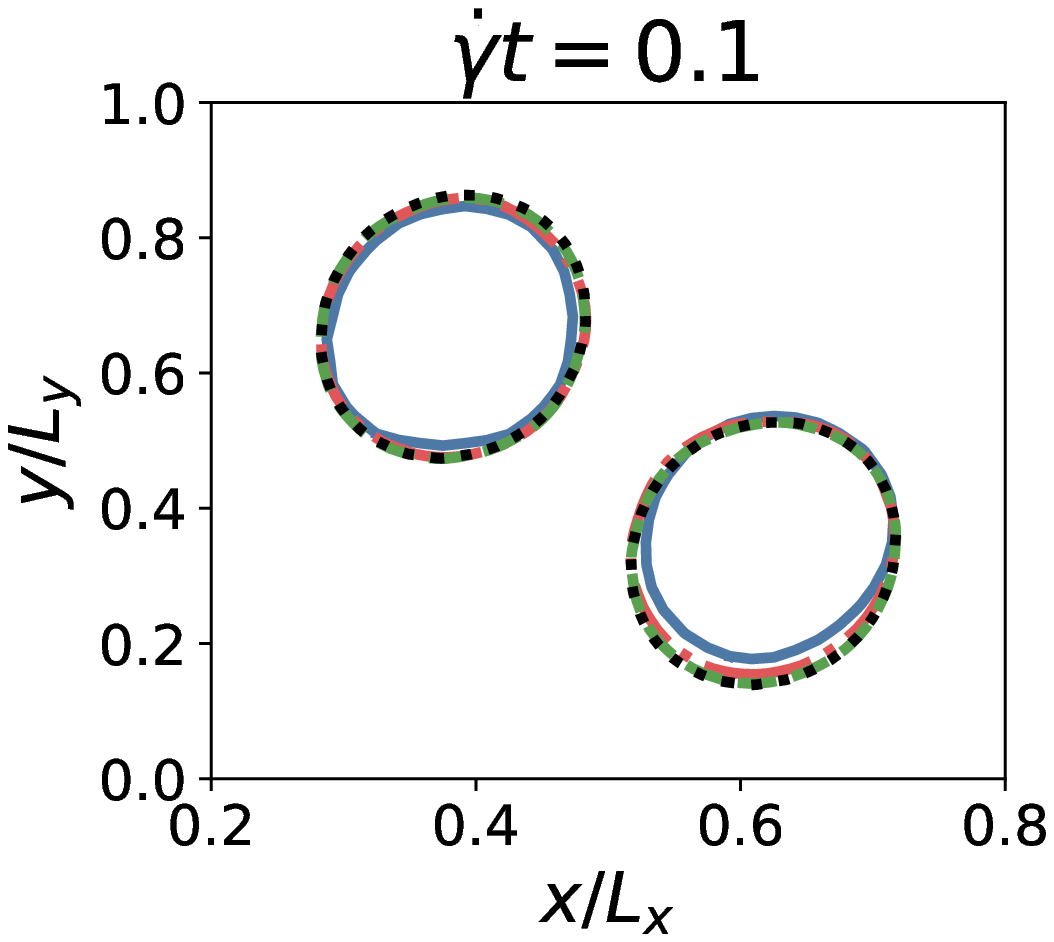}
    \includegraphics[width=0.17\linewidth]{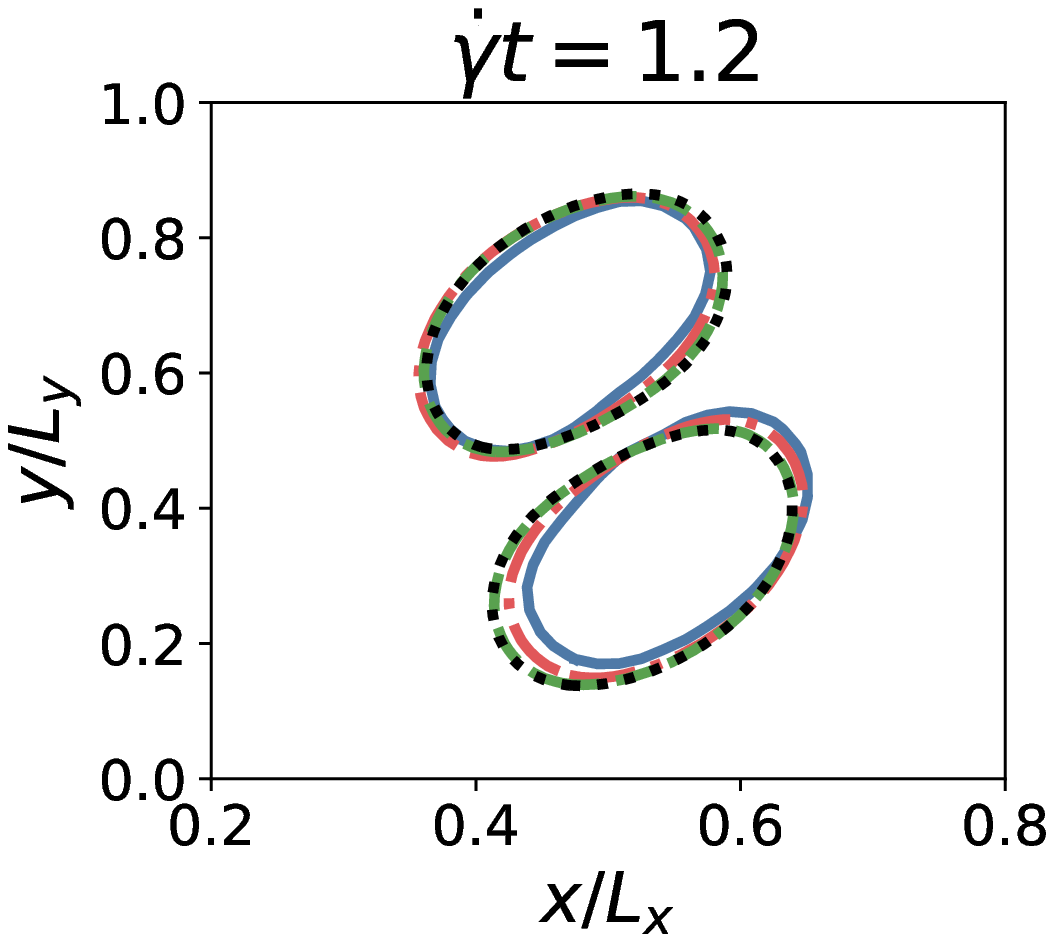}
    \includegraphics[width=0.17\linewidth]{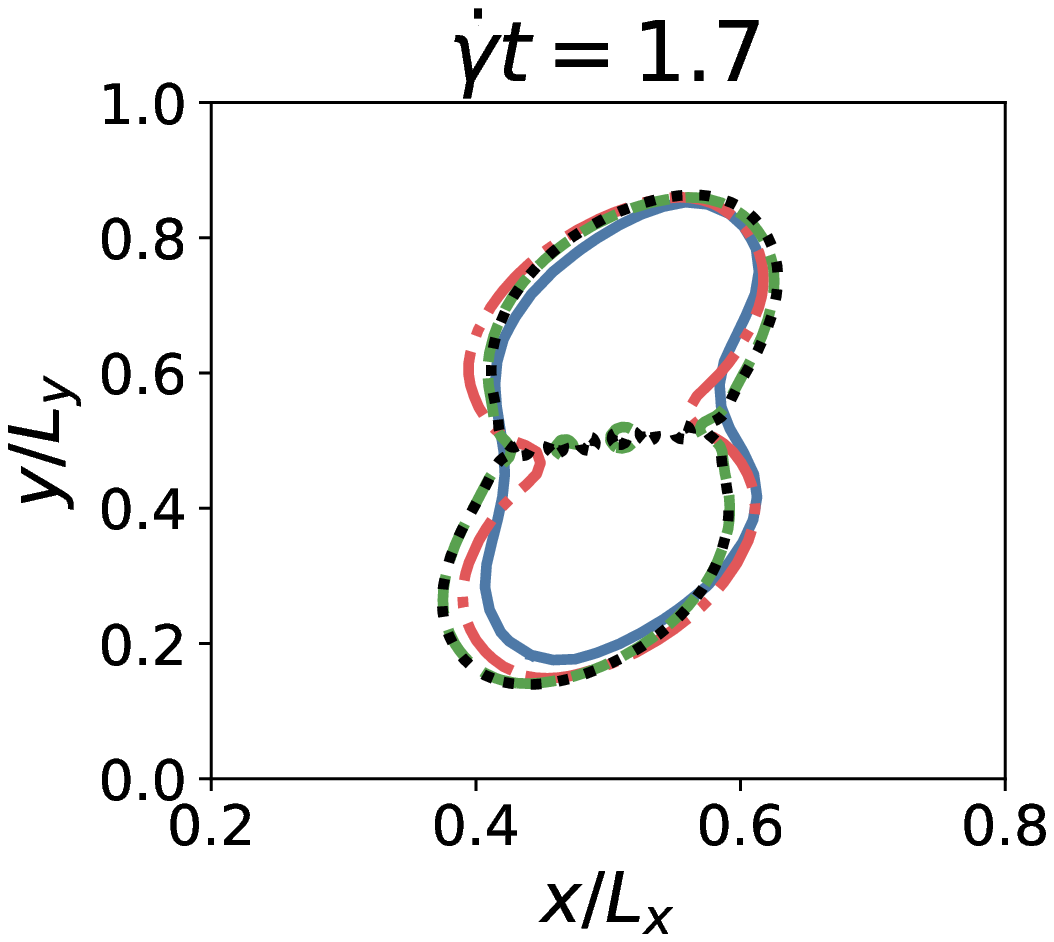}
    \includegraphics[width=0.17\linewidth]{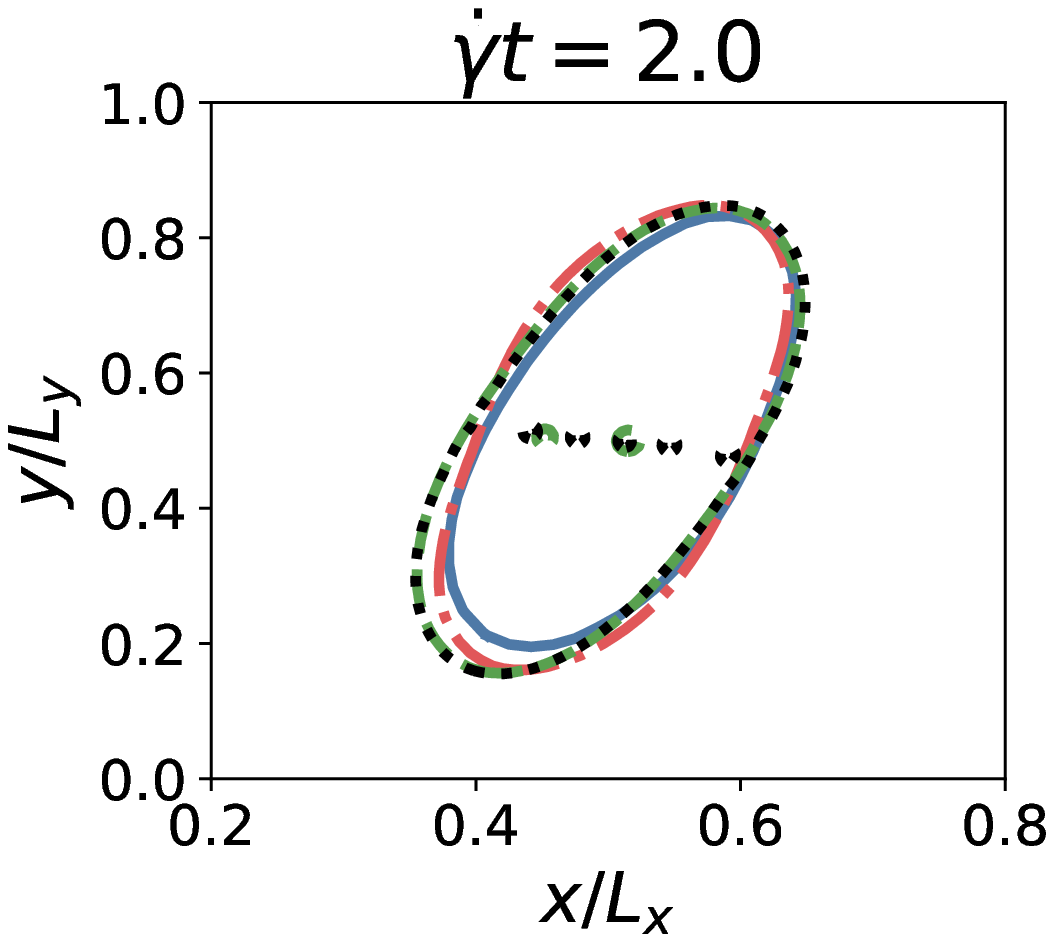}
    \includegraphics[width=0.17\linewidth]{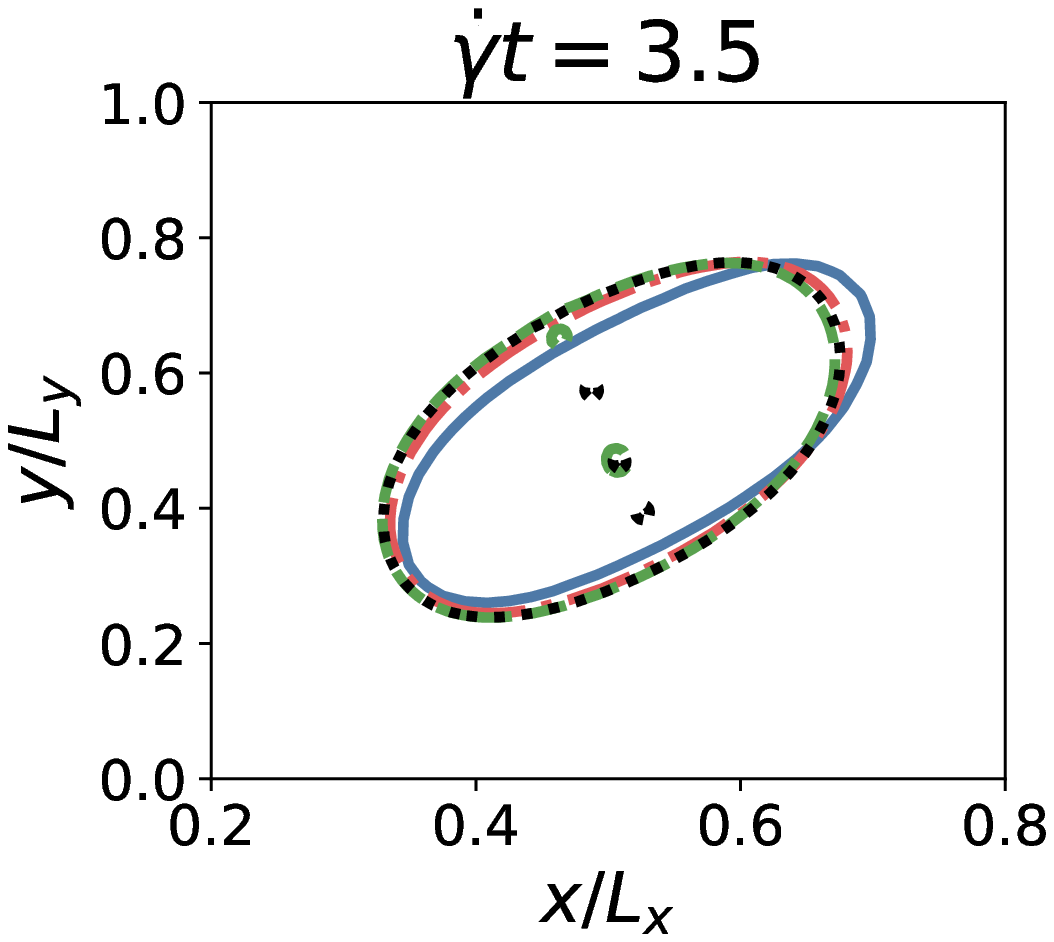}
    \includegraphics[width=0.12\linewidth]{Figures2/Figure9clegend.eps}}
\caption{Collision trajectories for two droplets with resolution $\Delta x = 1/40, 1/80, 1/160$, and $1/240$ in shear flow with $\Ca = 0.1$ calculated using the (a) CLS method with smoothing length $\kappa = \Delta x/2$, {(b) CLS method with fixed smoothing length $\kappa = 0.0125$,} and (c) non-local method with $\varepsilon = 0.025$. The times between Figures are not uniform and are chosen to be representative of the droplet behavior. The CLS method shows significantly different behavior, with simulations with finer resolution coalescing at later times, and not at all when $\Delta x = 1/240$. In contrast, the CLS simulations converge with respect to the grid size. Small satellite droplets can form at the intersection of the two droplets for the non-local model.  (Movies S2 and S3).}
\label{fig:two_shear_res10}
\end{figure}

\begin{figure}
%\centering
   \subfloat[][CLS method]{
    \includegraphics[width=0.25\linewidth]{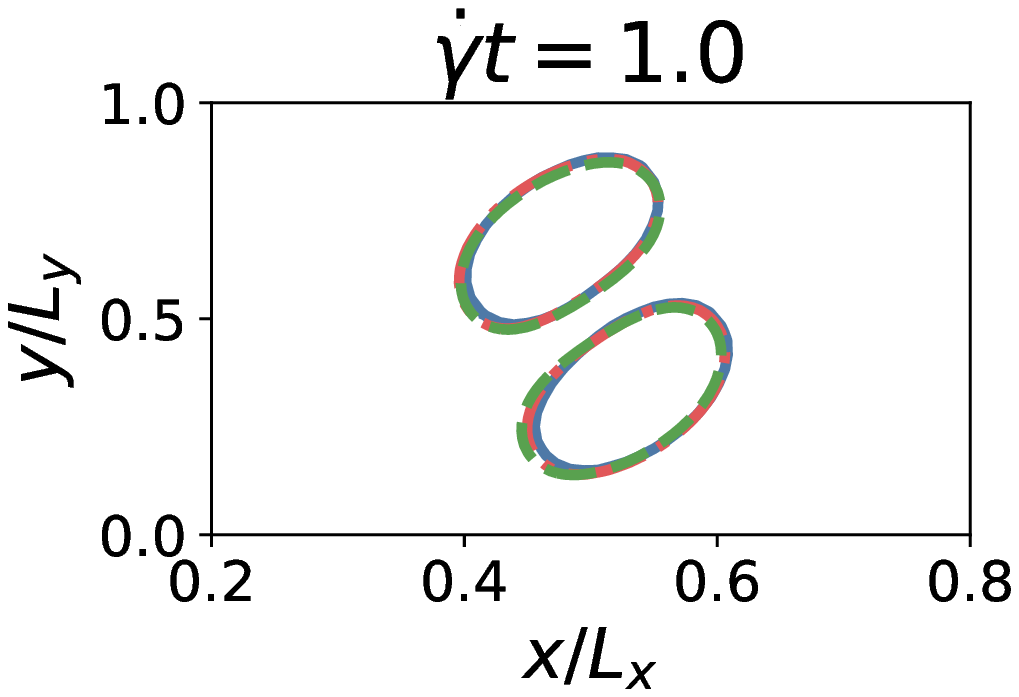}
    \includegraphics[width=0.25\linewidth]{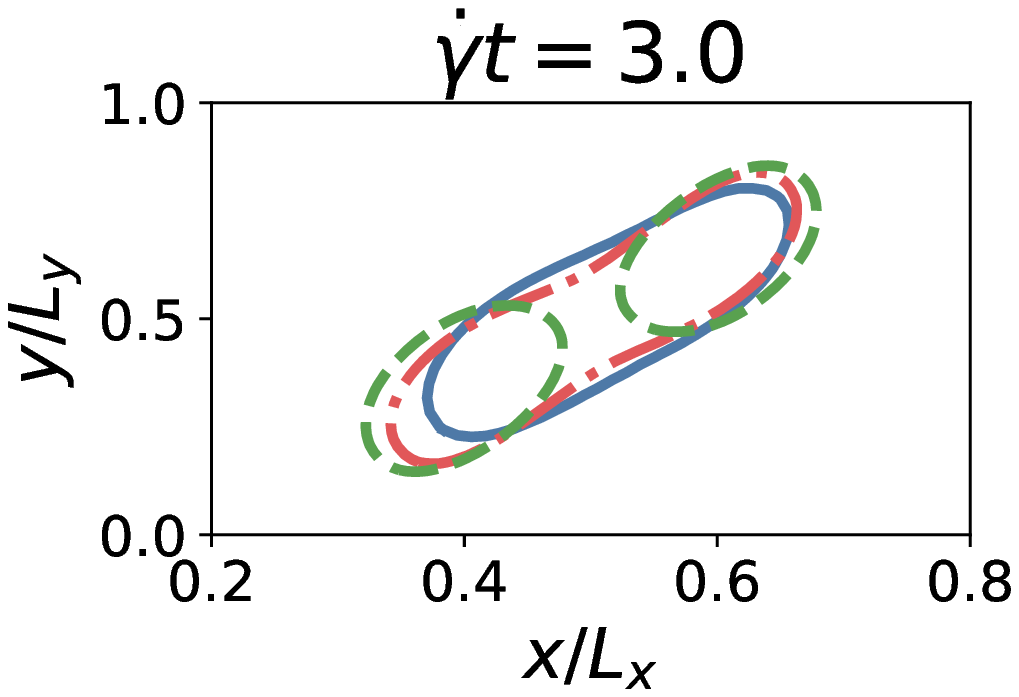}
    \includegraphics[width=0.25\linewidth]{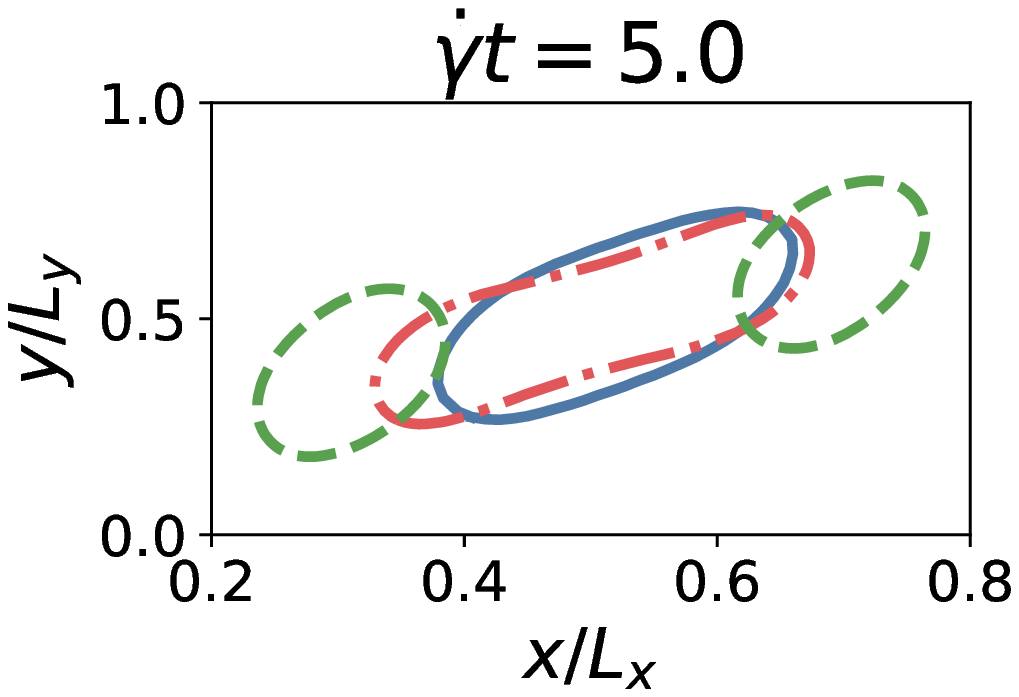}
    \includegraphics[width=0.2\linewidth]{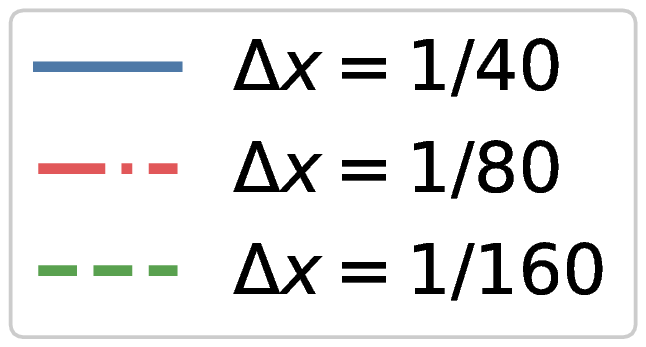}}\\   \subfloat[][Non-local method]{
    \includegraphics[width=0.25\linewidth]{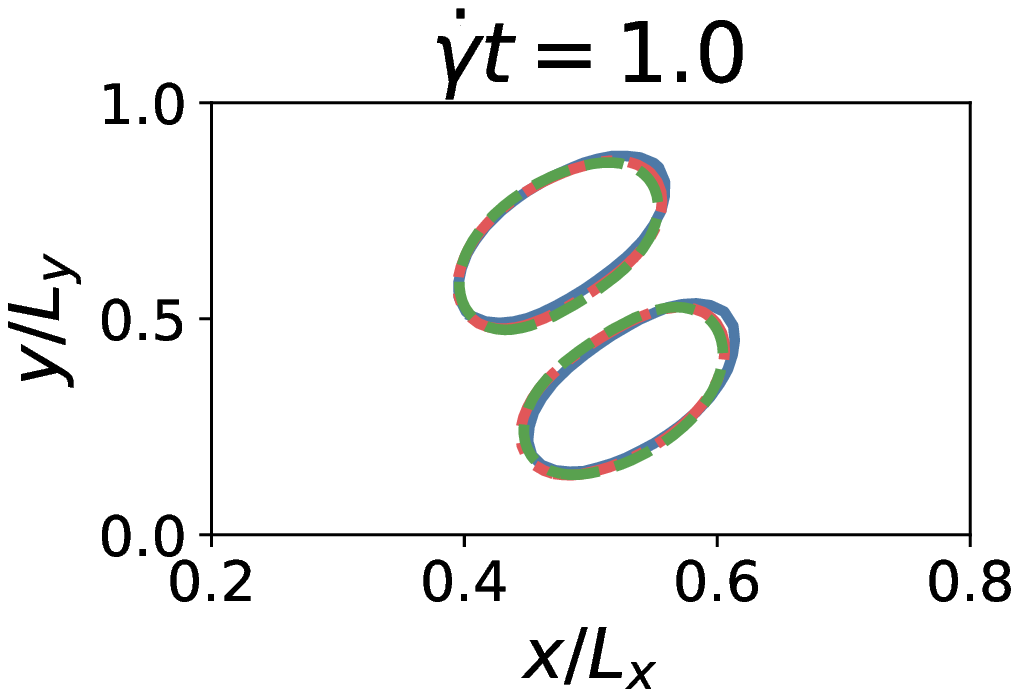}
    \includegraphics[width=0.25\linewidth]{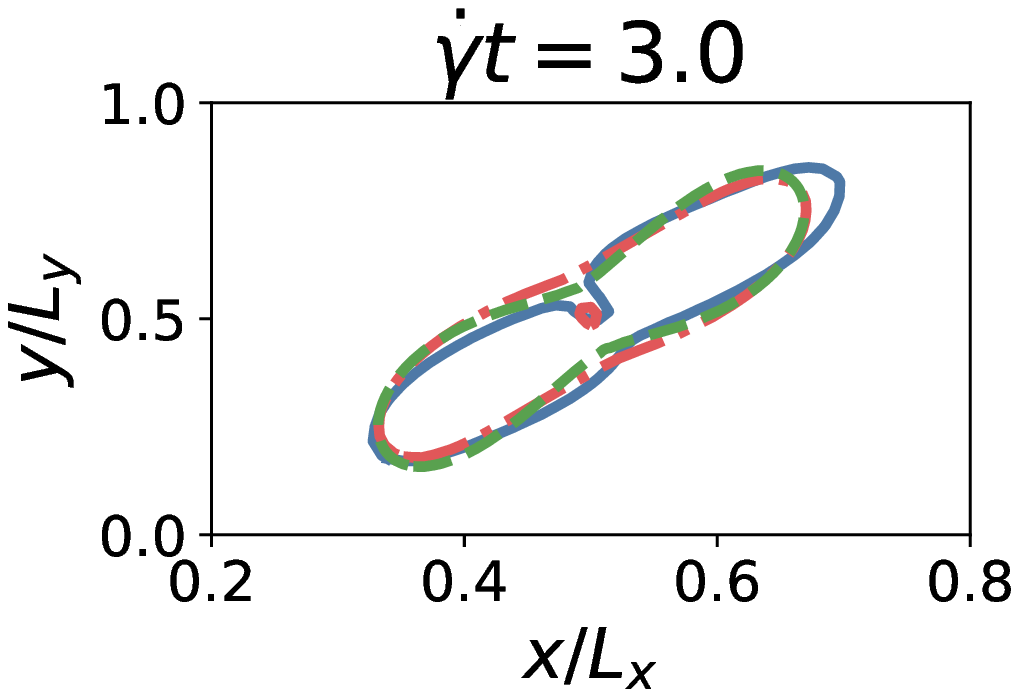}
    \includegraphics[width=0.25\linewidth]{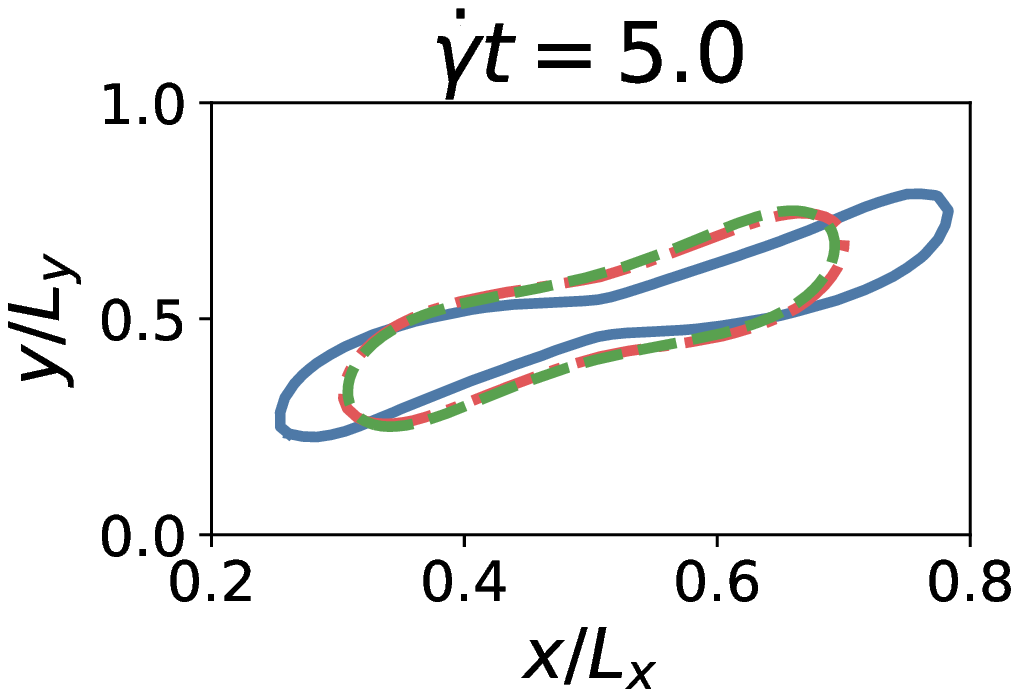}
    \includegraphics[width=0.2\linewidth]{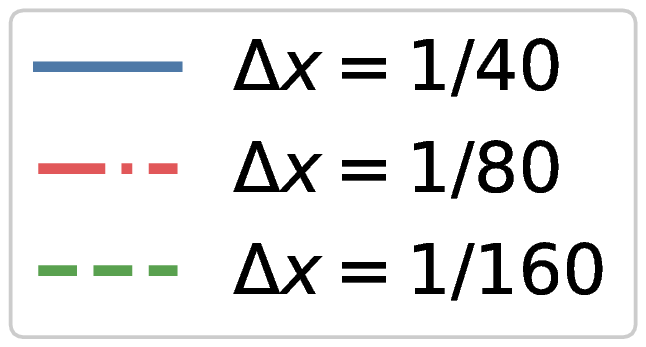}}   
    \caption{Collision trajectories for two droplets with resolution $\Delta x = 1/40, 1/80$, and $1/160$ with $\Ca = 0.15$ calculated using the (a) CLS method and (b) non-local method with $\varepsilon = 0.025$.}
\label{fig:two_shear_res15}
\end{figure}

A key advantage of the non-local model for this problem is that droplet coalescence is controlled by the parameter $\varepsilon$, not the grid resolution. In \cite{Shardt2013}, the critical capillary number for the transition from coalescing to sliding droplets was found to decrease with the droplet radius, and follow a power-law with exponent that increases with increasing P\'{e}clet  number.  In our CLS simulations, the droplets coalesce at different times depending on the resolution and do not coalesce at all at the highest resolution considered with interface smoothing length $\kappa = \Delta x/2$, as shown in Fig. \ref{fig:two_shear_res10}$A$ at $\Ca = 0.1$ and Fig. \ref{fig:two_shear_res15}$A$ at $\Ca = 0.15$. This is because in the local model, the droplets will coalesce when they are within one grid point and there are no forces interacting between the droplets to assist coalesce when the droplets are separated by one more than one grid point. {When the smoothing length is kept fixed, $\kappa = 0.0125$, the droplets continue to coalesce at slightly different times, although the final behavior is the same, as shown in Fig. \ref{fig:two_shear_res10}$B$ at $\Ca = 0.1$.}
In contrast, when the resolution is sufficient ($\Delta x \leq 1/80$), the behavior of the non-local method is similar regardless of resolution as long as $\varepsilon$ is kept constant with respect to the droplet radius as the mesh is refined. This is demonstrated in Fig. \ref{fig:two_shear_res10}$C$ at $\Ca = 0.1$ and Fig. \ref{fig:two_shear_res15}$B$ at $\Ca = 0.15$. At $\Ca = 0.1$, small satellite droplets form as two droplets coalesce. Similar phenomenon can be seen in \cite{Shardt2013}.

\begin{figure}
\centering
    \includegraphics[width=0.16\linewidth]{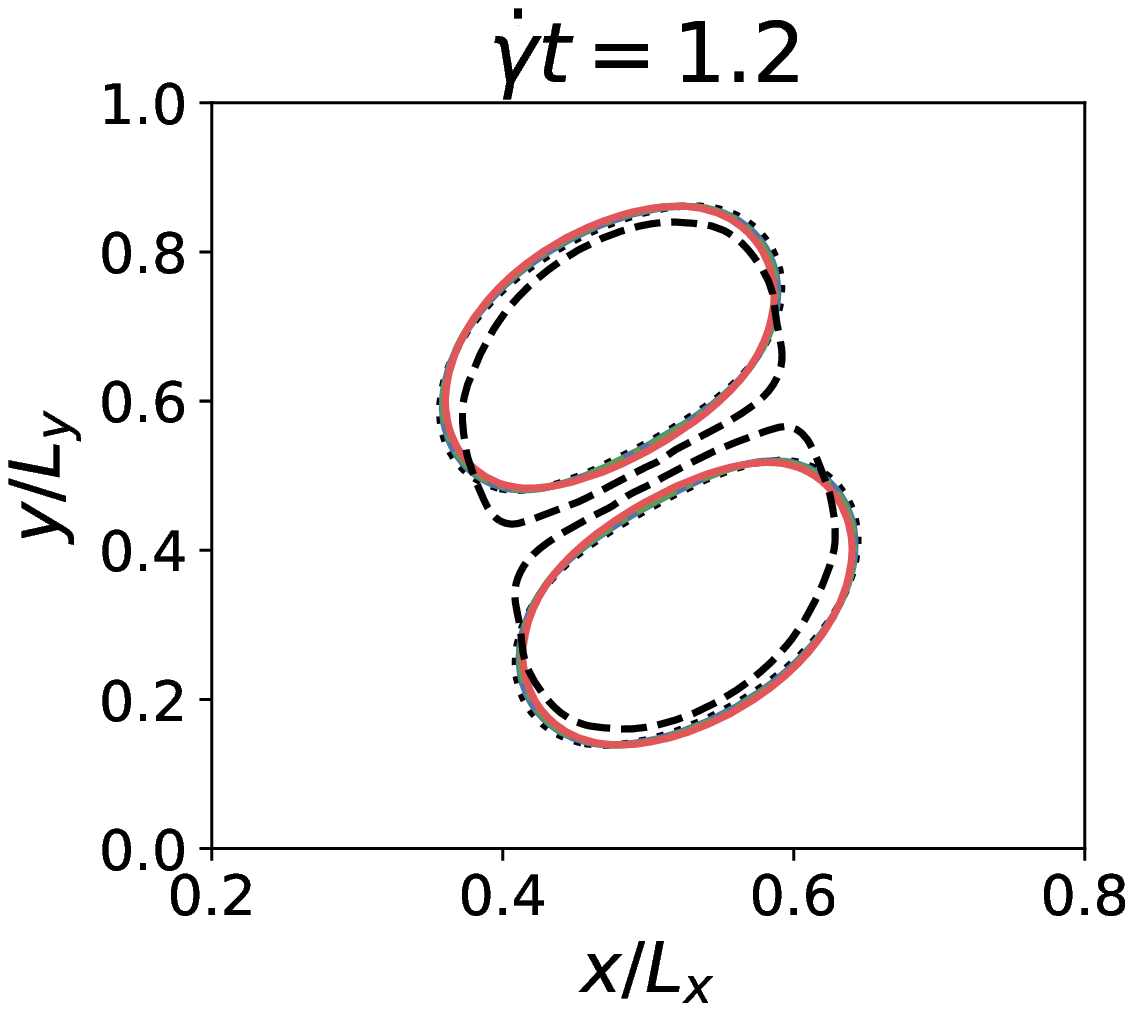}
    \includegraphics[width=0.16\linewidth]{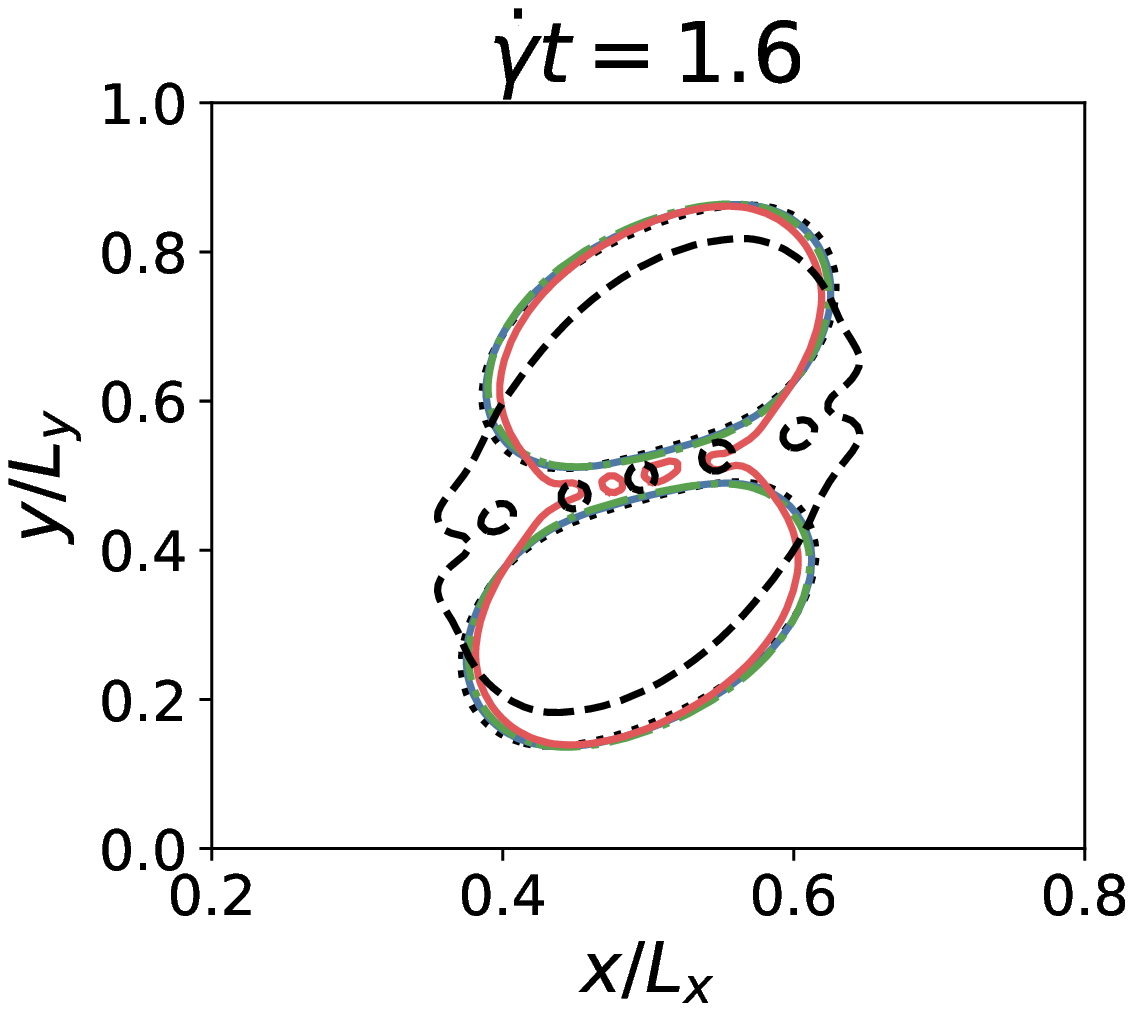}
    \includegraphics[width=0.16\linewidth]{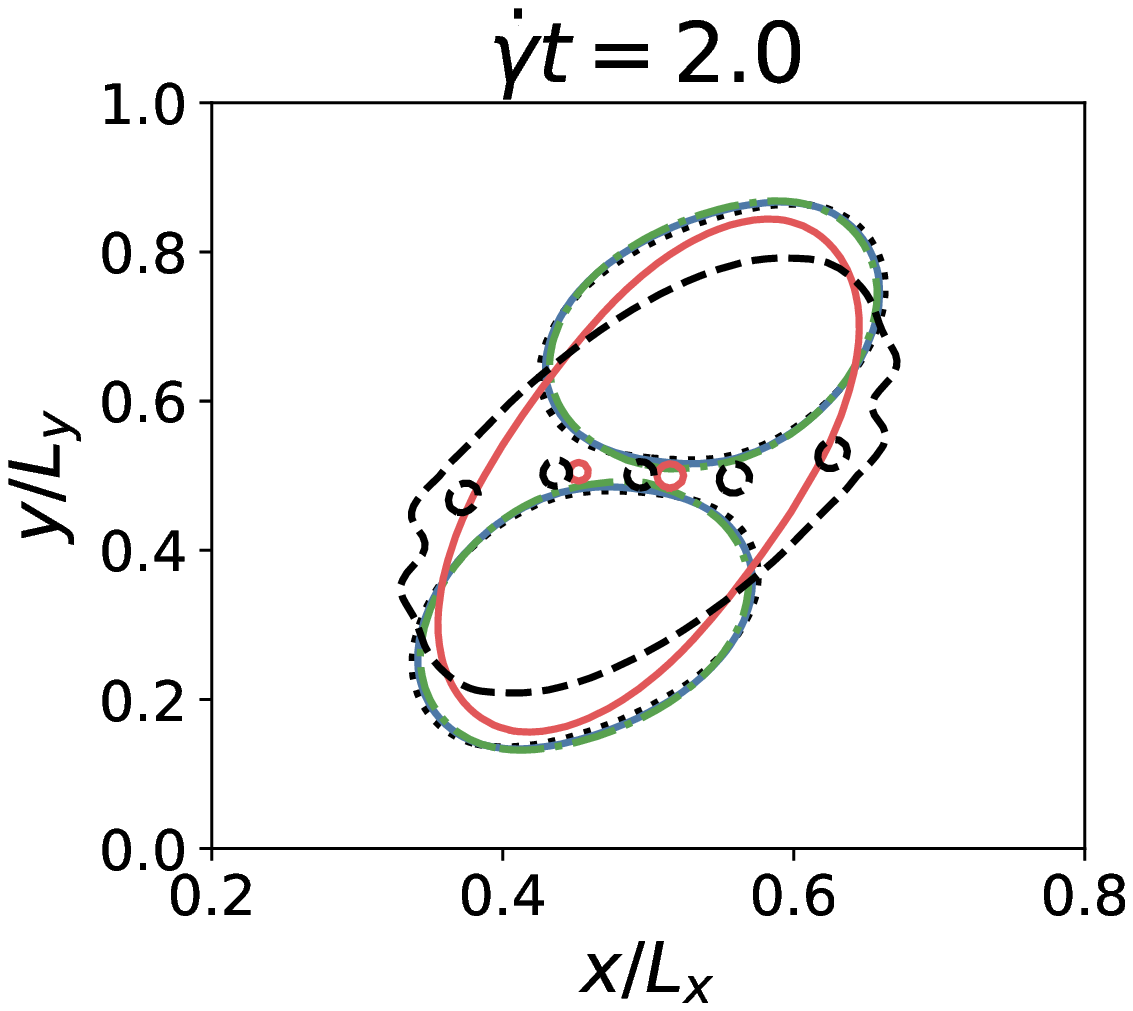}
    \includegraphics[width=0.16\linewidth]{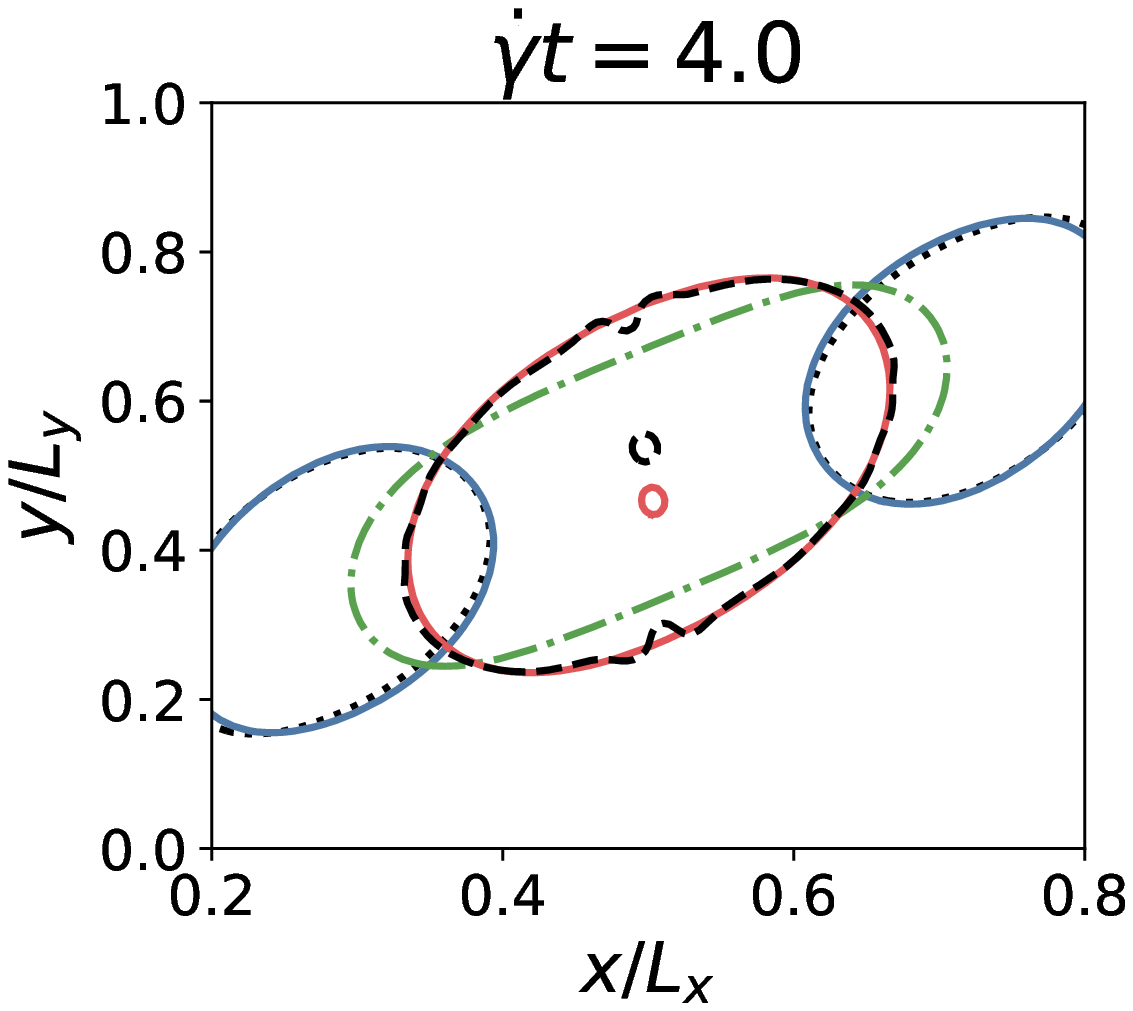}
    \includegraphics[width=0.16\linewidth]{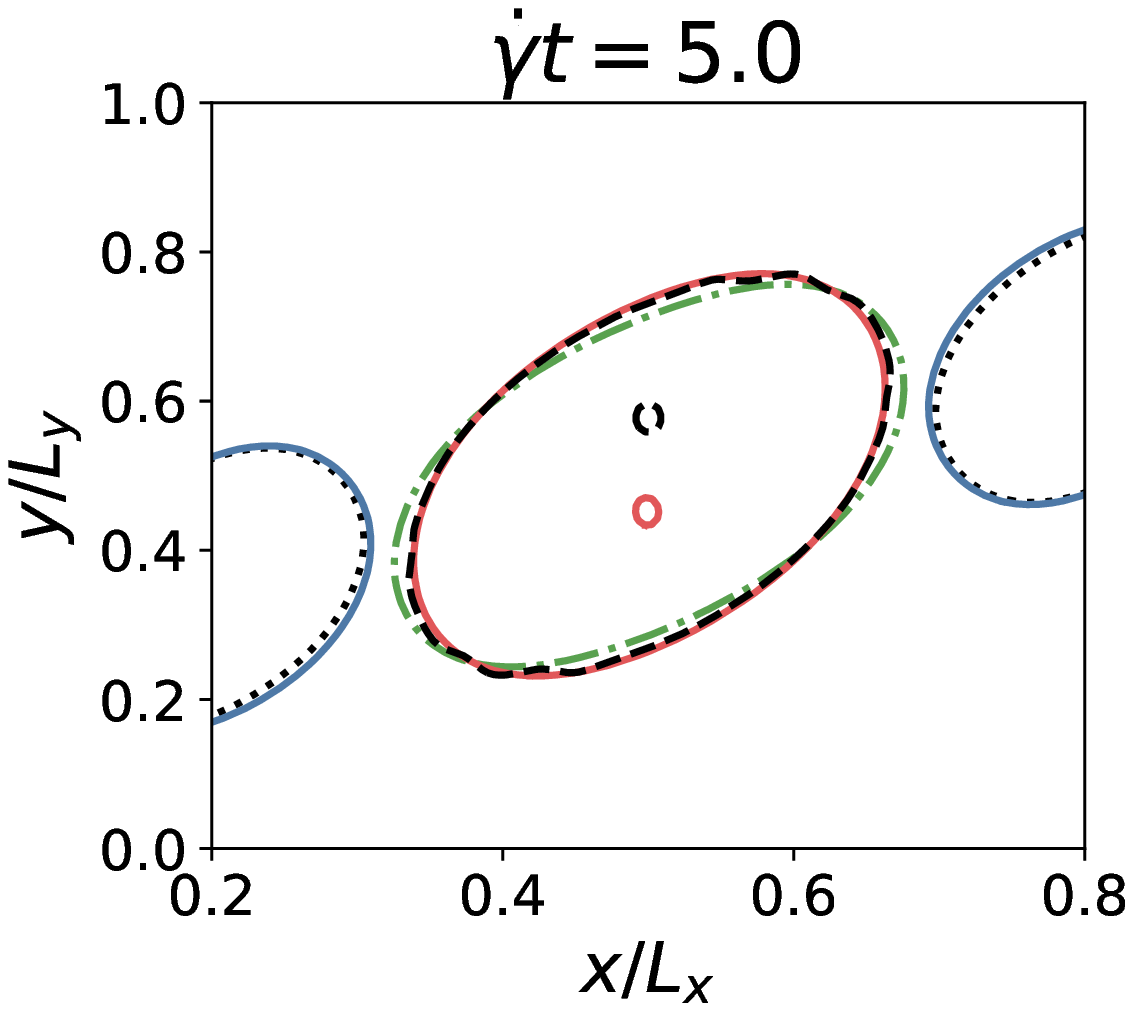}
    \includegraphics[width=0.12\linewidth]{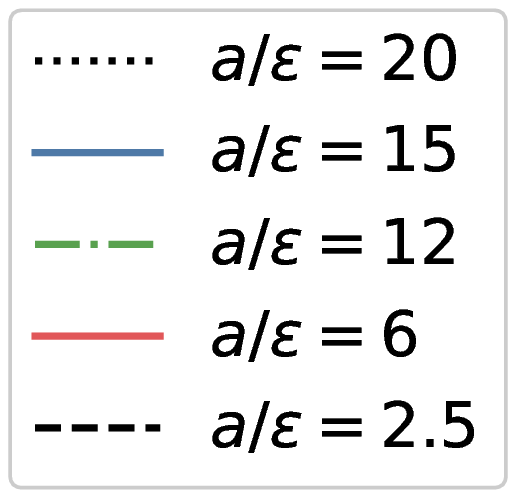}
\caption{Collision trajectories for two droplets in shear flow with $\Ca = 0.1$ calculated using the non-local method with varying values of $\varepsilon$. The times between figures is not uniform and are chosen to be representative of the droplet behavior. The values $a/\varepsilon = 20, 15, 12, 6$, and $2.5$ correspond to $\varepsilon = 0.0075, 0.01, 0.0125, 0.025$, and $0.06,$ respectively. The resolution is $\Delta x = 1/160$, so $\varepsilon/ \Delta x = 1.2, 1.6, 2.0,  4.0$, and $9.6$.  (Movie S4).}
\label{fig:two_shear_eps}
\end{figure}

\begin{figure}
\centering
 \sidesubfloat[]{
    \includegraphics[width=0.35\linewidth]{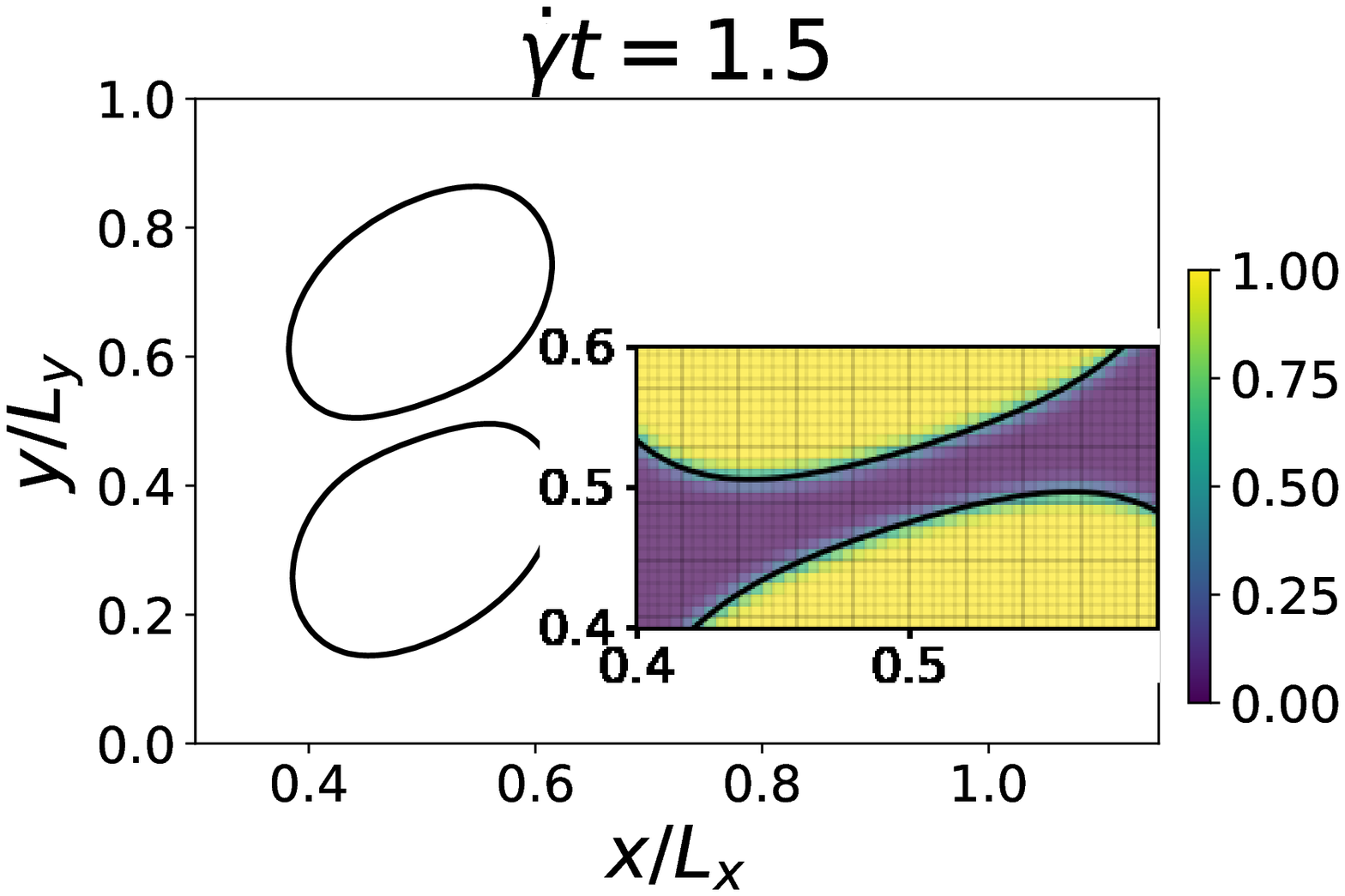}}
   \sidesubfloat[]{  \includegraphics[width=0.35\linewidth]{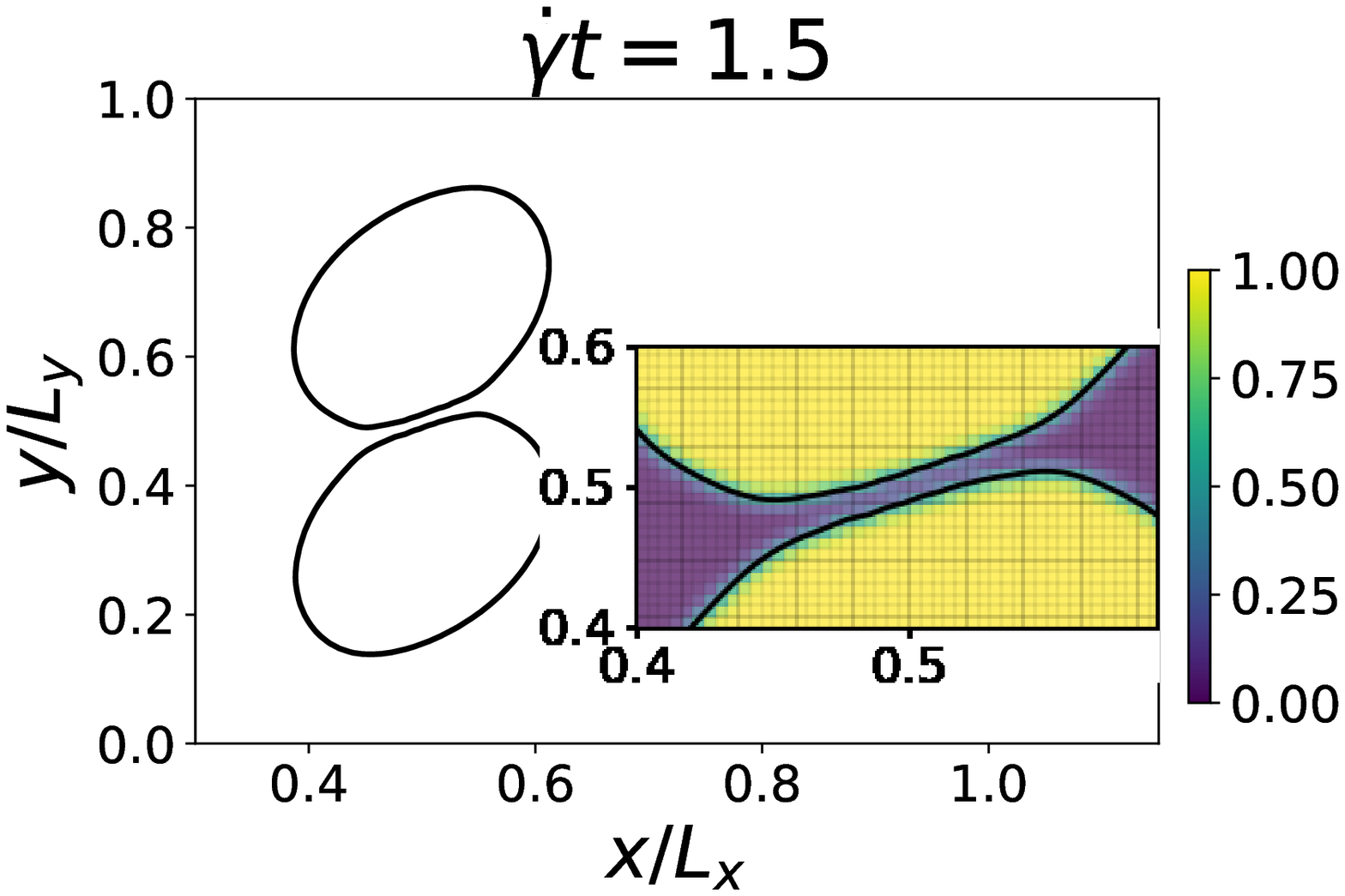}}
\caption{Front location and color function for two droplets with (a) the non-local method with $\varepsilon = 0.015$ ($a/\varepsilon = 10$), and (b) the non-local method with $\varepsilon = 0.025$ ($a/\varepsilon = 6$). At the same time, the droplets are closer together in the simulation with a higher value of $\varepsilon$. In both (a) and (b) the resolution is $\Delta x = 1/160$ and $a = 0.15$. (Color online.)}
\label{fig:two_shear_cf}
\end{figure}

The effect of $a/\varepsilon$ (the relative droplet size) is studied in Fig. \ref{fig:two_shear_eps}, which shows the droplet trajectories for five values of $\varepsilon$ at resolution $\Delta x = 1/160$ and droplet radius $a = 0.15$. The resulting droplets range from micro-droplets ($a/\varepsilon = 2.5$) to mesoscopic droplets ($a/\varepsilon = 6, 10$, and  $12$) and macroscopic droplets ($a/\varepsilon = 15$, and  $20$). 
%
%The timing of coalescence is controlled by the width of the interface given by $\varepsilon$, with larger values of $\varepsilon$ coalescing sooner. 
%
We find that the relatively smaller droplets (i.e., droplets with smaller $a/\varepsilon$) coalesce sooner. The droplets with $a/\varepsilon \geq 15$ do not coalesce during the simulation and instead slide past each other. The behavior of macroscopic bubbles does not depend on the $a/\varepsilon$ as long as $a/\varepsilon\geq15$.
The micro-scale droplets with $a/\varepsilon = 2.5$ display distinct characteristics, although the overall behavior is similar to that of the mesoscopic droplets. The fronts between these droplets flatten significantly at $\dot \gamma t = 1.2$, to a much larger degree than is seen in the other droplets. This results in many satellite droplets forming, which then migrate to the edge of the coalesced droplet over time. A zoomed-in view of the front locations and the corresponding color functions for macroscopic droplets can be seen in Fig. \ref{fig:two_shear_cf} for $a/\varepsilon = 10$ and $6$, showing that the front flattening occurs to a lesser degree than the micro-scale droplet in Fig. \ref{fig:two_shear_eps} (see Fig. \ref{fig:two_shear_cf}$B$.) The effect of increasing $\varepsilon$ relative to the droplet radius, for the same resolution and time, is to cause the two droplets to be significantly closer together leading to earlier coalescence. 

\begin{figure}
\centering    
	\includegraphics[width=0.19\linewidth]{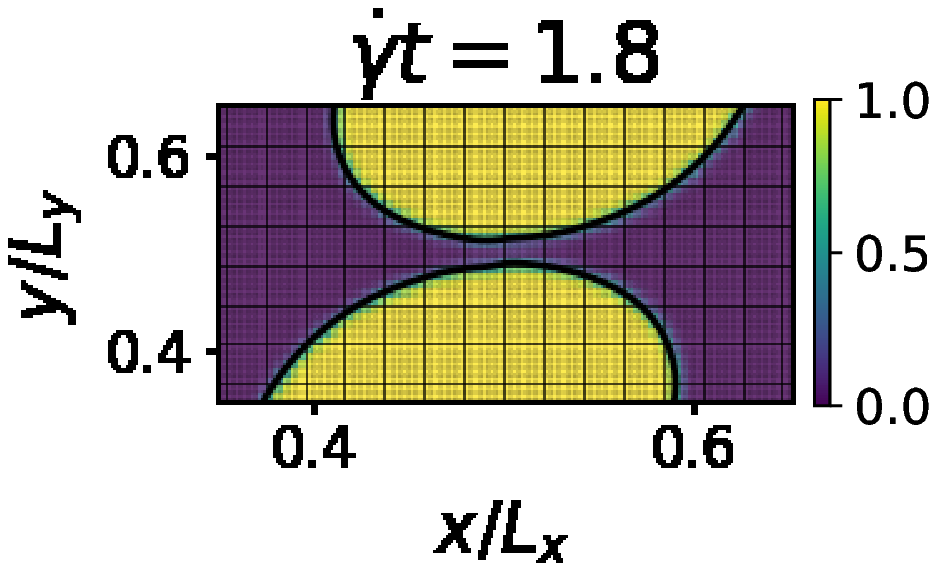}
	\includegraphics[width=0.19\linewidth]{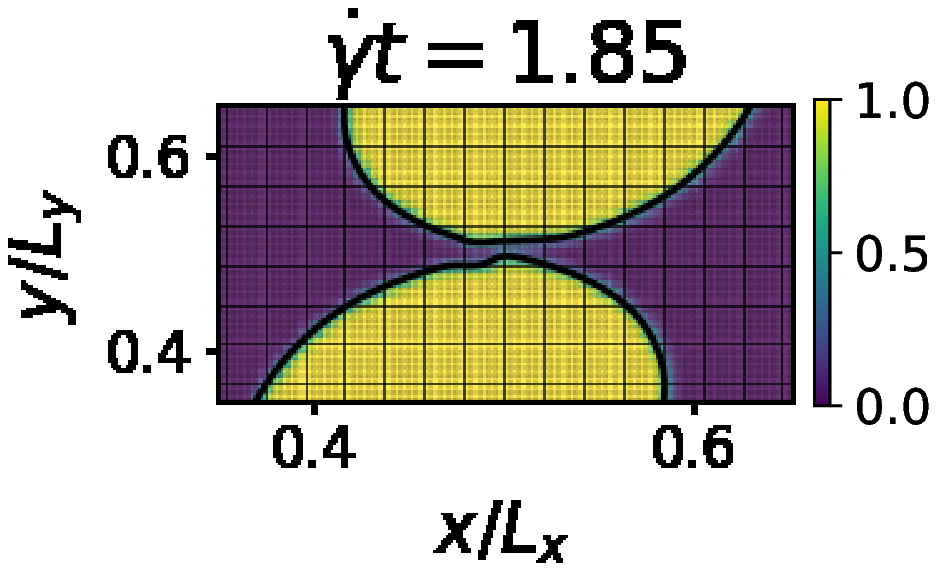}
	\includegraphics[width=0.19\linewidth]{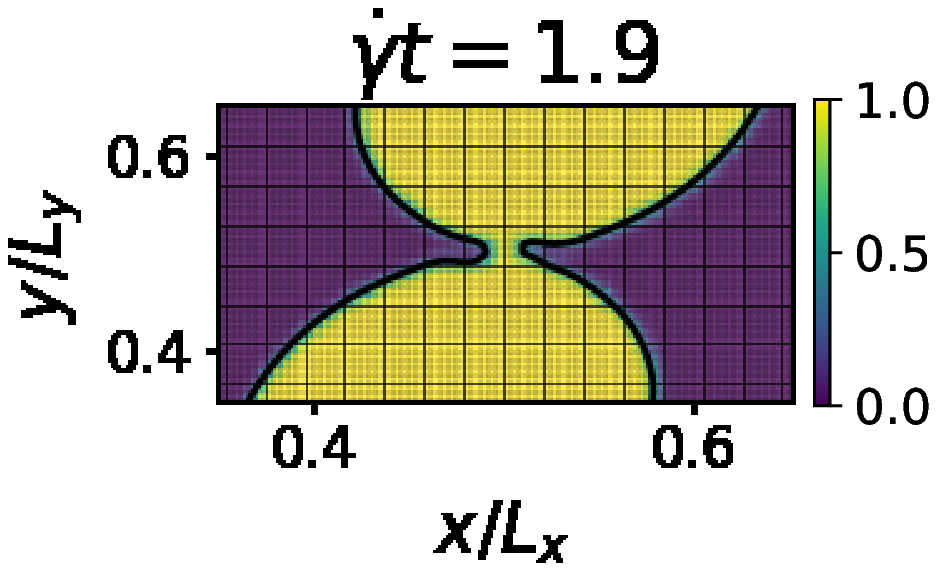}
	\includegraphics[width=0.19\linewidth]{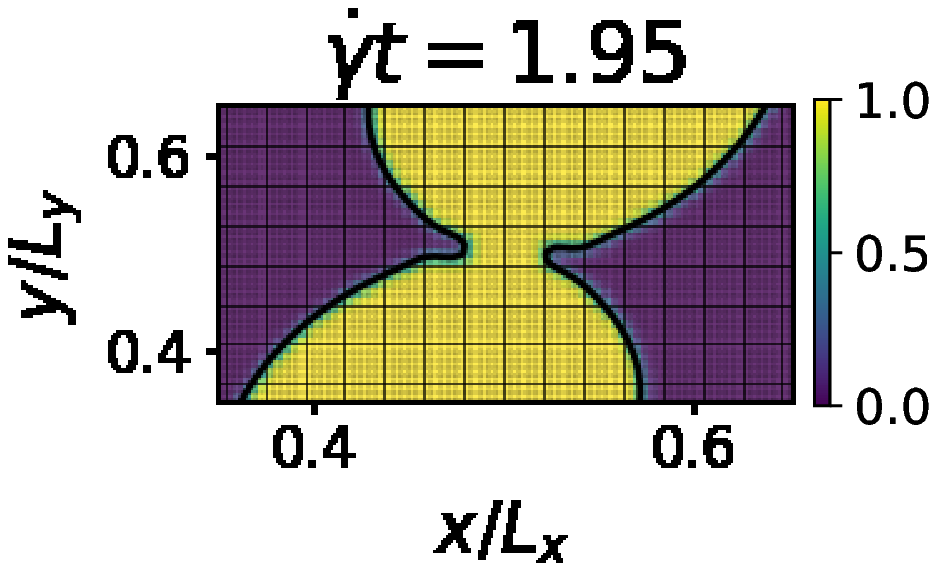}
	\includegraphics[width=0.19\linewidth]{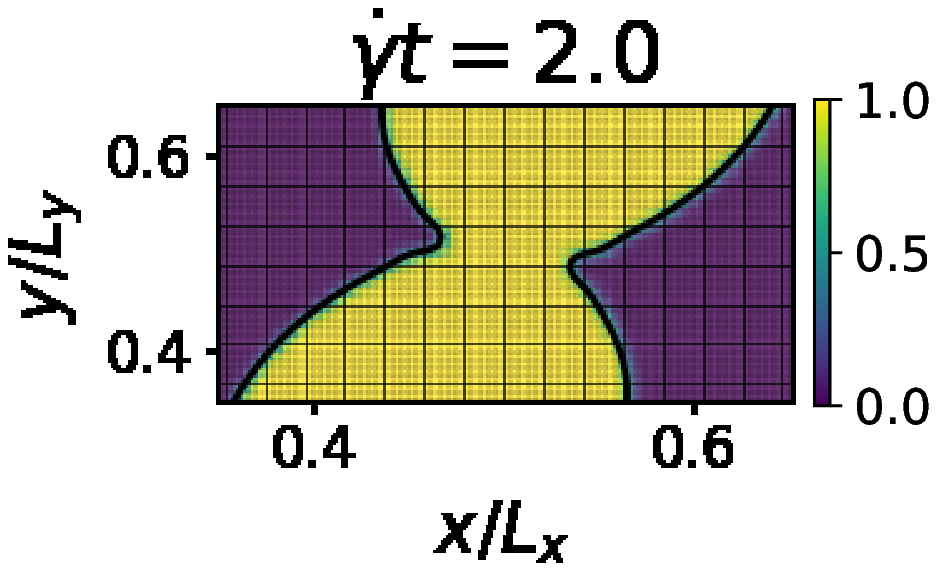}
\caption{Examples of the color function and front locations for two coalescing droplets with $\varepsilon = 0.015$, $\Delta x = 1/160$, and $a = 0.15$, giving $a/\varepsilon = 10$ and $\varepsilon/\Delta x = 2.4$. (Color online.)}
\label{fig:cf_coalescence}
\end{figure}

An example of the coalescence process with the non-local simulation is shown in detail in Fig. \ref{fig:cf_coalescence}, which plots the droplet fronts as well as the color function. In this example $ \varepsilon/\Delta x = 2.4$, and $a/\varepsilon = 10$. When the droplets are sufficiently close an asymmetry forms in the front which becomes the initial location of coalescence. As the simulation continues, this neck widens until the droplets form a single large droplet.

\section*{Conclusion}
In this paper we propose a non-local PDE model for multiphase flow that replaces the Young-Laplace law and is valid at both nano and macroscopic scales. 
%We demonstrate that our model correctly captures the anisotropic stress from MD simulations, and is simple to implement for a uniform mesh. 
Our model allows for simulations of multiscale systems with curvature radii ranging from nano to macro (micron and larger) scales, without the need to couple MD simulations with continuum NS models. Nanoscale multiphase flows, including nanobubbles, are important in biomedical applications, water and waste treatment, and membrane and surface defouling and cleaning. These applications typically include suspensions of many densely packed nanobubbles over macroscopic length scales, so the ability to use a continuum NS model will allow for efficient simulation of such systems in a way not possible with MD. 

The non-local model limits numerical error from interpolating a sharp interface across a grid. Calculating the surface tension by integrating over a local neighborhood at each point on the surface reduces the spurious, or parasitic, currents that occur in other methods. It also handles merging interfaces without the behavior depending on the grid resolution. Instead, the dynamics are controlled by the relative curvature radius $a/\varepsilon$, where $\varepsilon$ is a parameter in the non-local surface tension force. The parameter $\varepsilon$ defines the relative width of the interface. Physically, the interface width is of the order of the support of the molecular forces. Therefore, when modeling nanodroplets, $\varepsilon$ should be on the order of the support of molecular forces, i.e., on the order of several $r{A}$. The parameter $\varepsilon$ also define the resolution of numerical simulations as the grid size should be smaller than $\varepsilon$.  Therefore, using $\varepsilon$ on the order of several {\r{a}ngstr\"{o}ms} for modeling macroscopic droplets is computationally infeasible. Our results show that interfaces behave macroscopically when the radius of curvature larger than $3.5\varepsilon$ and the interface merging dynamics becomes independent of $\varepsilon$ for $a/\varepsilon >15$. Thus, it is sufficient to set $\varepsilon = a/3.5$ for modeling individual macroscopic droplets and $\varepsilon = a/15$ for modeling multiple interacting droplets.  

When compared with existing benchmarks for macroscale droplets, the proposed non-local method matches or exceeds the accuracy of local numerical methods for multiphase flow. However, the non-local method can also capture mesoscale features that cannot be resolved by local methods, as demonstrated by a droplet in shear flow. { Calculating the surface tension with the non-local model does require taking an integral at each point, which does make the surface tension calculation more computationally expensive, however each integral is independent and therefore lends itself well to massive parallelization. While not considered here, the non-local model can be easily implemented in existing codes by changing only the surface tension calculation.  Additionally, the number of points included in each integral can be truncated to those within a distance of 3.5$\varepsilon/\Delta x$ from the target point, reducing the additional amount of work per time step to be $\mathcal{O}\left(N \varepsilon^2/\Delta x^2\right)$. For a fixed mesh, the force shape function $f_\varepsilon(|\bx-\by|)$ can be precomputed before the simulation and used each time step.} Our future work will include full three-dimensional simulations. Because the force due to surface tension only needs to be calculated near the interface, fast Level Set methods such as the one proposed by \cite{Adalsteinsson1995} could be adapted to the non-local model to reduce the computational time.  In addition, simulations of multiscale systems are well fitted for adaptive mesh refinement, particularly when considering the interface between merging bubbles. 

%For consideration of merging droplets, it may be advantageous to consider a subgrid scale model coupled with the non-local model to simulation the movement of the fluid in the gap between the two droplets, in the manner of, e.g., \cite{Mason2012, Liu2018}. 

%\matmethods{
%%
%The code used for the data in this paper will be released on Github. The data sets used to generate the Figures in the paper will be released on the Pacific Northwest National Laboratory Datahub. 

%Please describe your materials and methods here. This can be more than one paragraph, and may contain subsections and equations as required. Authors should include a statement in the methods section describing how readers will be able to access the data in the paper. 

%\subsection*{Subsection for Method}
%Example text for subsection.
%}

%\showmatmethods{} % Display the Materials and Methods section
\section{Acknowledgements} 
This work was supported by the U.S. Department of Energy (DOE) Office of Science, Office of Advanced Scientific Computing Research as part of the New Dimension Reduction Methods and Scalable Algorithms for Nonlinear Phenomena project. Pacific Northwest National Laboratory is operated by Battelle for the DOE under Contract DE-AC05-76RL01830. 

The authors wish to thank Professor S. Guido for the data provided in Fig. \ref{fig:Shear_D}.

\appendix
{\section{Spurious currents}}
\label{sec:spurious_currents}
Many numerical schemes for surface tension in two-phase flows struggle to capture equilibrium solutions exactly, resulting in what are known as parasitic or spurious currents, non-zero velocity fields when the system is in a static equilibrium and thus the velocity field should be identically zero. The origin of these currents comes from errors in approximating continuous quantities by discrete operators \cite{Popinet2018}. The parasitic currents may not converge with spatial resolution \cite{Renardy2002} and scale with the surface tension and viscosity \cite{Lafaurie1994}. Some numerical methods have been proposed to limit or eliminate parasitic currents, see for example \cite{Renardy2002, Jamet2002}.

\begin{figure}
\centering
 \sidesubfloat[]{ 
    \includegraphics[width=0.4\columnwidth]{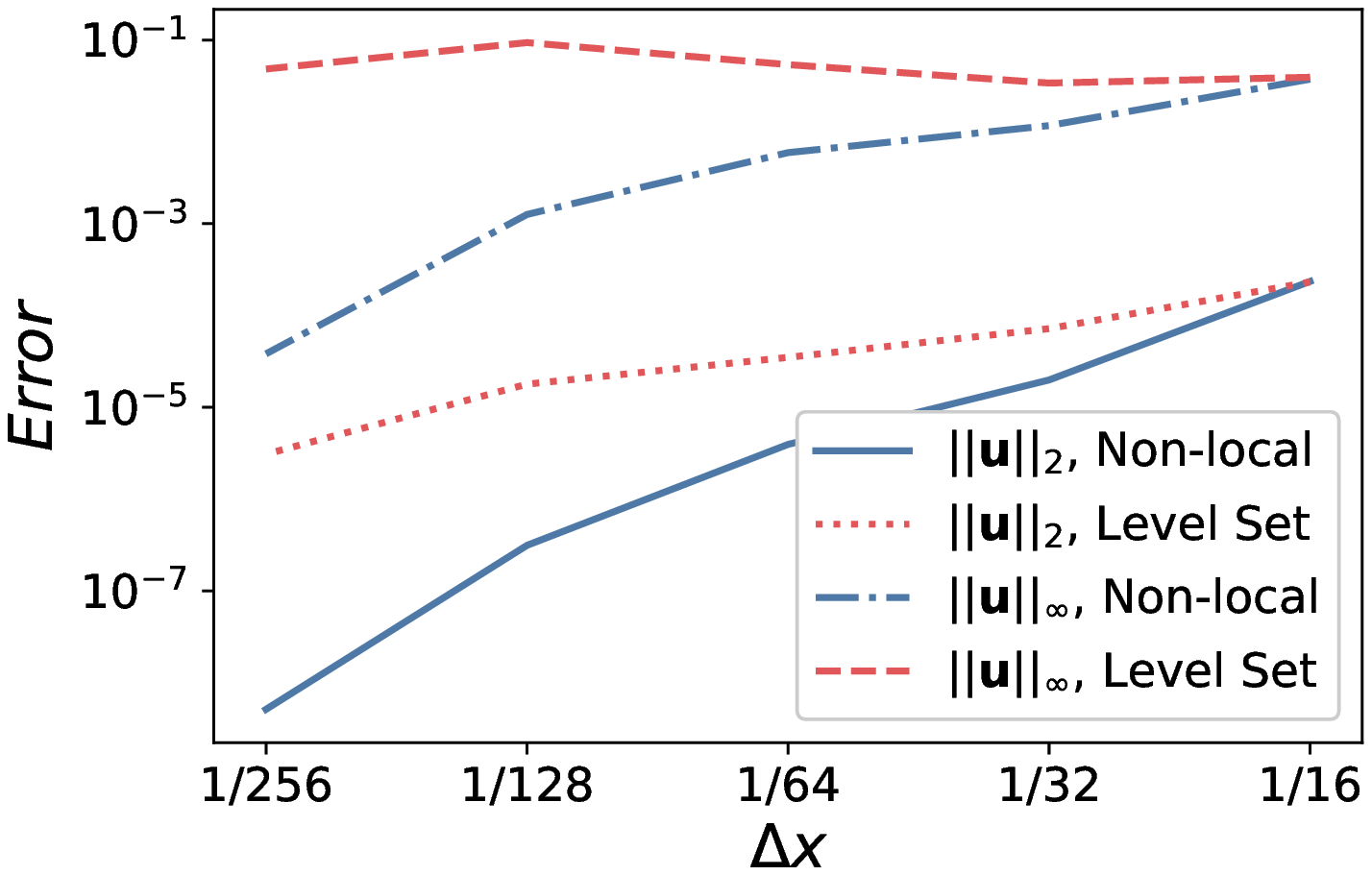} }
 \sidesubfloat[]{ 
         \includegraphics[width=0.2\columnwidth]{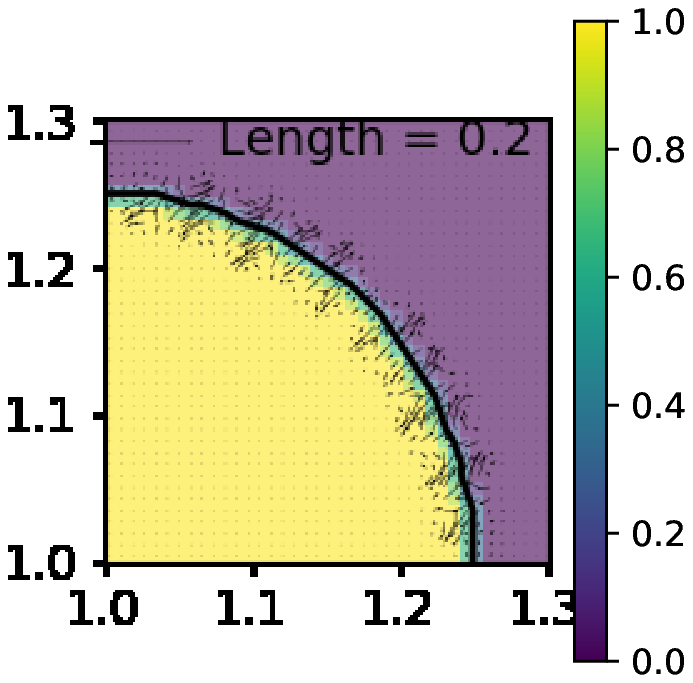}}
 \sidesubfloat[]{ 
    \includegraphics[width=0.2\columnwidth]{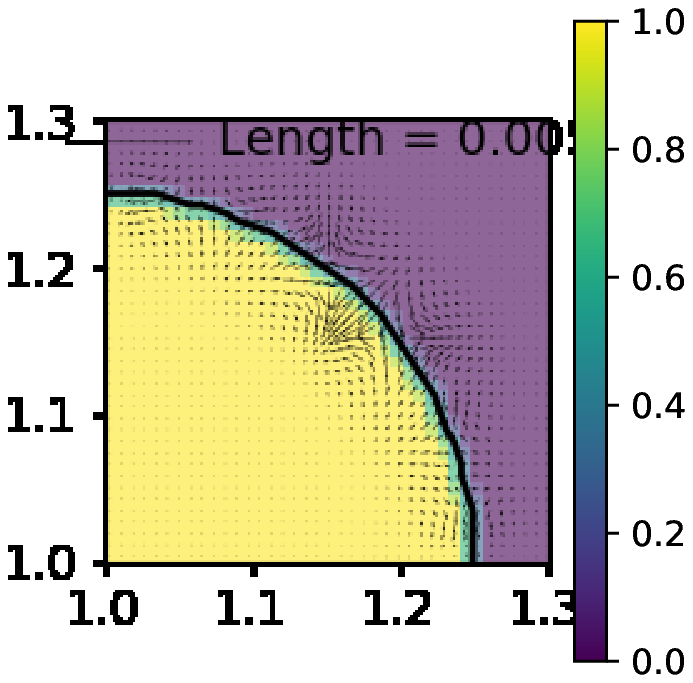}}
\caption{ {  (a) $||\mathbf{u}||_2$  and $||\mathbf{u}||_\infty$ as a function of the grid spacing for the non-local model and the local model.} (b and c) Magnitude and direction of the velocity field after 20 time steps for $\Delta x = 1/128$ for (b) the CLS method and (c) the non-local model. Note that the length scale for the velocity vectors is different in the two Figures. The color scale corresponds to the color function $\phi$. (Color online.)}
\label{fig:PC}
\end{figure}

The simplest system for studying spurious currents is that of a circular droplet with $a>3.5\varepsilon$ suspended in another fluid in the absence of gravity. In this case, the velocity field is zero and the pressure jump satisfies eq. \ref{Young-Laplace} exactly. We model a droplet at the equilibrium using the finite-volume (local) CLS and non-local models. 
The magnitude of the velocity fields are given in Table \ref{tab:PC_error} and Fig. \ref{fig:PC}$A$. Because the expected velocity field is zero, any non-zero velocity is taken to be a spurious current. The magnitude and directions of the velocity field for both the non-local and CLS model are shown in Figs. \ref{fig:PC}$B$ and \ref{fig:PC}$C$. The non-local model has spurious currents that are at least two orders of magnitude smaller than the CLS method for all quantities considered.

\begin{table}
\centering
\caption{$L_2$ and $L_\infty$ error  in the velocity field generated by the parasitic currents calculated with the non-local model and the CLS method.}
\begin{tabular}{cccccc}
     \hline
 $\Delta x$ & Non-local $||\mathbf{u}||_2$ & Non-local $||\mathbf{u}||_\infty$  &  CLS $||\mathbf{u}||_2$ &   CLS $||\mathbf{u}||_\infty$       \\ \hline 
1/16 & $2.35{\times}  10^{-4}$ & $3.74 {\times} 10^{-2}$ &  $2.31 {\times}  10^{-4}$& $3.93 {\times}  10^{-2}$ \\
1/32 & $1.98{\times}  10^{-5}$ & $1.17 {\times}  10^{-2}$ &  $7.18 {\times}  10^{-5}$& $3.40 {\times}  10^{-2}$ \\
1/64 & $3.94{\times}  10^{-6}$ & $5.93 {\times}  10^{-3}$ &  $3.49 {\times}  10^{-5}$& $5.39 {\times}  10^{-2}$ \\
1/128 & $3.14{\times}  10^{-7}$ & $1.26{\times}  10^{-3}$ &  $1.78 {\times}  10^{-5}$& $9.37 {\times}  10^{-2}$ \\
1/256 & $5.13{\times}  10^{-9}$ & $3.82{\times}  10^{-5}$ &  $3.08 {\times}  10^{-6}$& $4.83{\times}  10^{-2}$    \\   \hline
\end{tabular}
\label{tab:PC_error}
\end{table}
       
\bibliographystyle{elsarticle-num}
\bibliography{biblio2}

\end{document}